\documentclass[fleqn,usenatbib]{mnras}

\usepackage{newtxtext,newtxmath}
\usepackage{comment}
\usepackage{multirow}
\usepackage{subcaption}
\usepackage{longtable}
\usepackage{supertabular}
\usepackage{anyfontsize}
\usepackage{placeins}

\usepackage[T1]{fontenc}
\usepackage{booktabs}

\DeclareRobustCommand{\VAN}[3]{#2}
\let\VANthebibliography\thebibliography
\def\thebibliography{\DeclareRobustCommand{\VAN}[3]{##3}\VANthebibliography}

\usepackage{graphicx}	
\usepackage{amsmath}	
\usepackage{amsbsy}

\usepackage{cleveref}
\usepackage{subcaption}
\usepackage{float}
\usepackage{multirow}

\title[AGN identification at high-z]{New Methods of Identifying AGN in the Early Universe using Spectroscopy and Photometry in the JWST Era}

\author[Flor Arevalo Gonzalez \& Titanilla Braun et al.]{
Flor Arevalo Gonzalez,$^{1,2,3,4}$\thanks{The first two authors contributed equally to this work.}
Titanilla Braun,$^{1,5}$\thanks{E-mails: flordenisse9@gmail.com, titanilla.braun@gmail.com}
James Trussler,$^{1, 6}$ 
Christopher J. Conselice$^{1}$, 
\and Thomas Harvey,$^{1}$ 
 Nathan Adams,$^{1}$ 
Duncan Austin,$^{1}$ 
Qiong Li,$^{1}$
Ignas Juodžbalis,$^{7,8}$ 
Kimihiko Nakajima$^{9}$
\\
\\
$^{1}$ Jodrell Bank Centre for Astrophysics, School of Physics and Astronomy, The University of Manchester, Manchester M13 9PL, United Kingdom \\
$^{2}$  Physics Department, Tor Vergata University of Rome, Via della Ricerca Scientifica 1, 00133 Rome, Italy \\
$^{3}$ INAF – Astronomical Observatory of Rome, Via Frascati 33, 00040 Monte Porzio Catone, Italy \\
$^{4}$ Physics Department, Sapienza University of Rome, Piazzale Aldo Moro 5, 00185 Rome, Italy \\
$^{5}$ Department of Physics, University of Oxford, UK \\
$^{6}$ Center for Astrophysics $|$ Harvard \& Smithsonian, 60 Garden St., Cambridge MA 02138 USA \\
$^{7}$ Kavli Institute for Cosmology, University of Cambridge, Madingley Road, Cambridge, CB3 OHA, UK.\\
$^{8}$  Cavendish Laboratory - Astrophysics Group, University of Cambridge, 19 JJ Thomson Avenue, Cambridge, CB3 OHE, UK. \\
$^{9}$ National Astronomical Observatory of Japan, 2-21-1 Osawa, Mitaka, Tokyo 181-8588, Japan. \\
}

\date{Accepted XXX. Received YYY; in original form ZZZ}

\pubyear{2024}

\begin{document}
\label{firstpage}
\pagerange{\pageref{firstpage}--\pageref{lastpage}}
\maketitle

\begin{abstract}
We explore spectroscopic and photometric methods for identifying high-redshift galaxies containing an Active Galactic Nucleus (AGN) with JWST observations.  After demonstrating the limitations of standard optical methods, which appear ineffective in the low-metallicity environment of the early universe, we evaluate alternative diagnostic techniques using the current JWST observational capabilities.  Our analysis focuses on line ratios and equivalent widths (EWs) of UV emission lines:  \ion{C}{IV}, \mbox{\ion{He}{II} $\lambda$1640}, \mbox{\ion{O}{III}] $\lambda$1665}, and \ion{C}{III}], and the faint optical line, \mbox{\ion{He}{II} $\lambda$4686}. We find that the most valuable diagnostic quantities for finding AGN are the line ratios: \mbox{(\ion{C}{III}] + \ion{C}{IV}) / \ion{He}{II} $\lambda$1640} and \mbox{\ion{C}{III}] / \ion{He}{II} $\lambda$1640}, as well as the EW of \mbox{\ion{He}{II} $\lambda$1640}. For more reliable AGN identification, the \mbox{\ion{He}{II} $\lambda$1640} and \mbox{\ion{O}{III}] $\lambda$1665} lines would need to be detected separately.  We show that the \mbox{\ion{He}{II} $\lambda$1640/H$\beta$} ratio effectively separates AGN from star-forming galaxies, though it is contingent on a low dust content. We also show that in order to effectively use these diagnostics, future observations require longer exposure times, especially for galaxies at \mbox{z > 6}. 
Subsequently, we plot three real high-redshift sources on these diagrams which present strong UV emission lines. However, in order to classify them as strong AGN candidates, further study is needed due to the blending of \ion{He}{II} + \ion{O}{III}] and unreliable optical lines.  Lastly, we carry out a selection process using spectral energy distribution (SED) fitting with \texttt{EAZY} to identify strong AGN candidates in the JADES NIRCam photometry. One galaxy in our sample emerged as a strong AGN candidate, supported by both photometric selection and strong UV emission. We present a sample of similar AGN candidates in the JADES data based on this method.
\end{abstract}




\section{Introduction}\label{sec:intro}
Thanks to the availability of new extensive observational datasets from the James Webb Space Telescope (JWST), there is now scope for a comprehensive characterization of the formation and evolution of galaxies at high redshifts. Previously, research findings were often influenced or constrained by limitations of incomplete sampling and by limited data quality at specific redshifts. Now, due to observations in a wide range of wavelengths, JWST enables the scientific community to investigate the properties of the earliest objects in our universe (\cite{JWSTmission_Gardner2023}).

The correct classification and categorization of galaxies is the first step towards the study of their properties across time, bringing more clarity into what influences their formation and evolution. Trustworthy methods of identification between different types of galaxies can help us analyse more accurately their observational data and extract more precise conclusions about their properties, applying specific tailored techniques. 
This is even more relevant at high redshifts, given the expected different cosmic conditions at earlier stages of the universe (\cite{Kewley_2013}), and the difficulty in obtaining reliable and accurate data.

Active galactic nuclei (AGN) are very compact and highly luminous sources found in the centre of galaxies that are shown to dominate the luminosity output of their host galaxy. These differ from normal star forming galaxies (referred to as SF galaxies or SFG from here on) in which the main source of the luminosity is not concentrated to such a small region and the total output of the galaxy is significantly lower considering a broad band of wavelengths. 

Thus, one of the goals of galaxy formation and evolution research is to determine which galaxies contain an active galactic nuclei (AGN). This is important because AGN are powerful sources that play a crucial role in the evolution of their host galaxies, influencing processes such as triggering or quenching star formation rates (e.g., \cite{Vecchia_2004}). Additionally, the analysis of the abundance of SFG or AGN at high redshifts allows us to assess their relative impact on shaping the history of the universe; for example, by studying how they affect the reionization process during the reionization epoch (\cite{Robertson_2015}, \cite{Duncan_Conselice_2015}).

It is possible to differentiate between AGN and SF galaxies as their main source of radiation are different. AGN are theorized to be powered by a supermassive black hole (SMBH) surrounded by an accretion disc (\cite{accretionBH_lynden1969}), while the line emission of SF galaxies mostly originates from regions of singly ionized hydrogen \mbox{(\ion{H}{II})} where star formation occurs. AGN usually have much harder ionizing spectra than SF galaxies, making it possible to distinguish between them using emission-line intensity ratios.

In the conventional unified model, AGN can be further categorized into two groups based on the orientation of the accretion disk around the black hole relative to the observer (\cite{SantosSoltau_2024_UMAGN}, \cite{Antonucci_1993}). Type I AGN, also known as Broad-Line AGN (BL AGN), exhibit both narrow and broad emission lines since the higher velocity broad-line region (BLR), which is closer to the central SMBH, is directly detectable. In the case of type II AGN, also named Narrow-Line AGN (NL AGN), the BLR is obscured by a dusty medium around the accretion disk, hence only narrow emission lines are detected.

Although broad emission lines are a typical feature of AGN, they can also be produced by extreme stellar kinematics in compact galaxies, as noted in \cite{Baggen_2024}. This means that while broad lines can suggest AGN presence, diagnostic plots comparing emission line ratios are still essential for properly distinguishing AGN from alternative explanations such as dense stellar systems, outflows and supernova. 

Line ratio plots were first proposed by \citet*{BaldwinPhillipsTerlevich_1981}, who suggested a set of optical diagnostic diagrams, now commonly referred to as BPT diagrams, to separate emission spectra powered by different excitation mechanisms. The method was revised by \cite{VeilleuxOsterbrock_1987} creating the VO87 diagrams. These standard optical diagnostic diagrams are based on the line ratios \mbox{[\ion{O}{III}]/H$\beta$}, \mbox{[\ion{N}{II}]/H$\alpha$}, \mbox{[\ion{S}{II}]/H$\alpha$}, and \mbox{[\ion{O}{I}]/H$\alpha$}.

Using these optical diagrams, \cite{p3_kewley2001} defined a "maximum starburst line" based on the upper limit of theoretical photoionization models such that galaxies above this line have a very high probability of containing an AGN. Since this is an upper limit for the star forming model, it excludes possible AGN candidates below it. \cite{Kauffman_2003} revised this demarcation line to divide pure star forming galaxies from the composite region between their revised line and the \cite{p3_kewley2001} maximum starburst line. Later the AGN region was further divided to differentiate between Seyferts and LINERs (Low Ionization Nuclear Emission-line Regions) by \cite{p2_kewley2006}. This shows the great distinguishing power of the standard BPT diagrams.

However, in the earlier stages of the Universe, astrophysical conditions differ significantly from the local universe. The low metallicity environment that is present in the early Universe complicates the utilization of commonly used diagnostics. According to available models and observations, there is a clear correlation between cosmic time and the metallicity of galaxies (\cite{Nagamine_2001}, \cite{Maiolino_2008}), as more recent galaxies have been subject to more consecutive processes of star formation. As metallicity decreases and the ionization parameter increases, distinguishing AGN from other sources of ionization using optical emission lines becomes more challenging (\cite{Ubler_2023}).

This is supported by recent studies suggesting that the BPT and VO87 diagrams might not be effective for the identification of high redshift AGN.
A study by \cite{Backhaus_2022} stated that the VO87 diagram cannot efficiently differentiate between AGN and SF galaxies at higher redshifts due to these regions overlapping.
\cite{Ubler_2023} show empirically that the standard diagnostic methods cannot separate between AGN and SF galaxies in such low metallicity systems found in the early universe. 
\cite{Harikane2023} presented a sample of AGN with \mbox{z $ > 4$} which were all indistinguishable from SF galaxies on the BPT diagram due to being in the region occupied by star forming galaxies. 
\cite{p4_maiolino2023jades} identified AGN in the JWST data which also would be missed in the standard BPT diagram.

While Type I AGN can still be identified by their broad lines, this makes the identification of Type II narrow-line AGN much more difficult. Therefore, many new diagnostics have been proposed.
\cite{Backhaus_2022} suggested that the OHNO diagram, consisting of a plane with line ratios \mbox{[\ion{O}{III}]/H$\beta$} and \mbox{[\ion{Ne}{III}]/[\ion{O}{II}]}, can be a promising alternative. 
\cite{Ubler_2023} examined the potential of using the \mbox{\ion{He}{II} $\lambda$4686} emission line for the discrimination and proposed the diagrams showing the equivalent width (EW) of \mbox{\ion{He}{II} $\lambda$4686} vs. the ratio \mbox{\ion{He}{II} $\lambda$4686/H$\beta$} and \mbox{\ion{He}{II} $\lambda$4686/H$\beta$} vs. \mbox{[\ion{N}{II}] $\lambda$6584/H$\alpha$}, the latter of which was first suggested by \cite{Shirazi_2012}. While the \mbox{\ion{He}{II} $\lambda$4686} emission line is effective in separating AGN and SF galaxies due to being independent of metallicity, it is very faint in even powerful AGN. 

Another possibility is to use ultraviolet emission lines instead of optical ones to search for AGN. These systems might be more reliable at high redshifts as these lines primarily originate from high-ionization states that require the production of hard ionizing photons, and are therefore likely to trace AGN activity.
This was explored by \cite{p5_Feltre2016}, who find that, amongst other lines, \mbox{\ion{C}{IV} $\lambda\lambda$1548,1551}, \mbox{\ion{O}{III}] $\lambda\lambda$1661,1666}, \mbox{[\ion{C}{III}] $\lambda$1907+\ion{C}{III}] $\lambda$1909}, and \mbox{\ion{He}{II} $\lambda$1640} are individually good at separating photoionization by an AGN and star formation, and that possible diagrams using three of these lines can be highly effective.

In this paper, we explore some of the suggested diagnostic quantities and diagrams using photoionization models produced by \cite{p11_NakajimaMaiolino2022} and where possible apply them on the JADES data (\cite{p1_bunker2023jades}). Our main focus is on three sources with \mbox{NIRSpec IDs} 9422, 18846, and 10058975, which exhibit strong ultraviolet emission lines.  We also comment on the applicability and sensitivity of these diagrams and explore further diagnostic quantities, for example the possible application of UV-Optical ratio diagrams. Apart from the spectroscopic identification, we also investigate a method using photometry, combining spectral energy distribution (SED) fitting and morphology, developed by \cite{p6_Juodzbalis2023} to find high probability high-z AGN candidates that can then be followed up with NIRSpec spectroscopy.  

The structure of this paper is as follows. Section \ref{sec:data} describes the observational and modelled datasets that were used. Section \ref{sec:spectroscopy_based_diagnostics} focuses on the spectroscopic analysis of the data, discussing standard optical methods for AGN identification, along with designing alternative optical and ultraviolet diagnostic diagrams. We also detail the calculation of the equivalent widths of emission lines in the spectra of three potential AGN sources and uses these measurements to position the galaxies on the discussed diagrams. In Section \ref{sec:photometry} we carry out a photometric selection process on the JADES data. In Section \ref{subsec:sensitivity} we examine the sensitivity of the diagnostic diagrams discussed earlier. We summarise our work and its possible extensions and draw our final conclusions in Section \ref{sec:conclusions}.

\section{Data}\label{sec:data}

\subsection{Observational Data}\label{subsec:obs_data}
The observational data we use can be categorized into two main groups: spectroscopic and photometric data. The spectroscopic data used throughout this paper is from the JWST Advanced Deep Extragalactic Survey (JADES) released by \cite{Eisenstein_2023_JADESoverview} using the data taken by the JWST Near InfraRed Spectrograph (NIRSpec, \cite{NIRSpec_Jakobsen_2022}, \cite{NIRSpec_Ferruit_2022}) . In particular we concentrated on the analysis of the 253 sources with detailed spectroscopy from \cite{p1_bunker2023jades}. The latter includes a measurement of the spectroscopic redshifts (z) for 178 sources, and provides 1D spectra with a catalogue of line fluxes above a signal-to-noise ratio of 5.  

We use the 1D spectra and catalogue of the Prism/Clear data with the main focus on three of the sources with \mbox{NIRSpec IDs} 00009422, 00018846, and 10058975 , which exhibited strong ultraviolet emission lines: \mbox{\ion{C}{IV} $\lambda\lambda$1549}, \mbox{\ion{He}{II} $\lambda$1640 + \ion{O}{III}] $\lambda\lambda$1661,1666}, and \mbox{\ion{C}{III}] $\lambda\lambda$1909}, in contrast to other observed objects. These lines are \ion{C}{IV}, \mbox{\ion{He}{II} + \ion{O}{III}]} blend (or \mbox{\ion{He}{II} $\lambda$1640} and \mbox{\ion{O}{III}] $\lambda$1665} if mentioned separately), and \ion{C}{III}], respectively, from here on. \mbox{\ion{C}{IV}} and the blended \mbox{\ion{He}{II} + \ion{O}{III}]} lines are reported in all three sources with a signal-to-noise ratio \mbox{S/N > 5}, however, the \ion{C}{III}] emission line of source 00009422 is not reported in \cite{p1_bunker2023jades}. To measure the emission line flux from the spectrum of the galaxy, we use the Line Measuring package: $\mathrm{L}_{\mathrm{I}}\mathrm{M}_{\mathrm{E}}$ . It fits the continuum and line flux to calculate the integrated flux of the line (for more details, see \cite{Fern_ndez_2024}). 

\begin{figure*}
    \includegraphics[width=.95\textwidth]{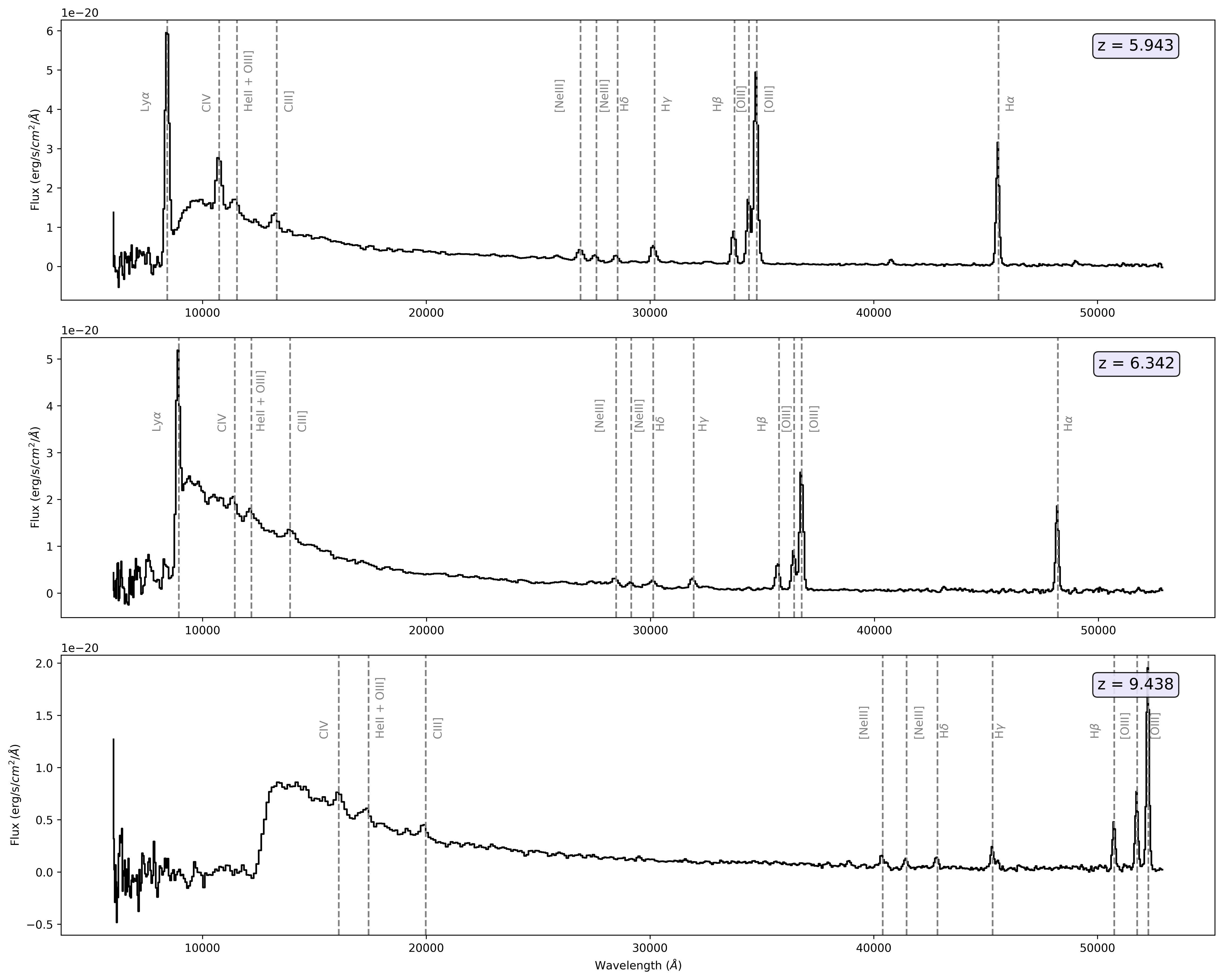}
    \caption{The NIRSpec spectra of galaxies 9422, 18846, and 10058975 are shown with spectral flux in units of $\mathrm{erg/s/cm^{2}/\text{\AA}}$ on the y axis and wavelength in units of angstroms on the x axis. The grey dashed lines are important emission lines found in the spectra.  These spectra are from the JADES survey (\protect\cite{p1_bunker2023jades}). }
\label{fig:spectra}
\end{figure*}

These galaxies are amongst those with the highest redshifts in the total sample, with 10058975 showing the highest redshift (z = 9.438) amongst the three (00018846 with a z = 6.342 and 00009422 with a  z = 5.943). The spectra of the three objects, 00009422 (or 9422), 00018846 (or 18846), and 10058975, are shown in Figure~\ref{fig:spectra} with the grey dashed lines denoting important emission lines. Source 9422 was identified spectroscopically as an AGN by \cite{p14_scholtz2023jades},\cite{Tacchella_2024_GS9422} and \cite{Li_2024_GS9422}), but rejected as an AGN in \cite{p15_cameron2023}. 10058975 was also identified as an AGN through spectroscopy by \cite{p14_scholtz2023jades}. Galaxy 18846, to the best of our knowledge, has still previously not been identified as an AGN. 

The various identified AGN sources displayed on the optical diagrams in Sections \ref{subsec:BPT_VO87} and \ref{subsubsec:OHNO} were collected from the literature. They are mostly confirmed as AGN  based on their broad line emission (\cite{p4_maiolino2023jades}, \cite{Harikane2023}, \cite{Ubler_2023}, \cite{Kocevski2023}, \cite{Larson_2023}, \cite{Kokorev_2023}). Note that on these diagrams, we also display JADES data points from the Prism/Clear data catalogue, as well as from the R=1000 grating catalogue.

The photometric data is from the JWST Near Infrared Camera (NIRCam, second data release, \cite{NIRCam_Rieke_2023}, \cite{Eisenstein_2023_JADESoverview}, \cite{Eisenstein_2023_OriginsField_DR2_NIRCam}).
We used the following filters from the NIRCam photometric catalogues for our objects: F090W, F115W, F150W, F200W, F277W, F356W, F444W, F335M, and F410M. All of these use the CIRC2 measurement (associated with a 0.15 arcsec radius circular aperture).

\subsection{Modelled Data}\label{subsec:mod_data}
The theoretically modelled sample for the AGN and SF galaxies is from a version based on \cite{p11_NakajimaMaiolino2022} using \texttt{CLOUDY} photoionization model calculations (version 13.05; \cite{Ferland1998, Ferland2013}). This data is available in spectral form, and also in a tabulated form containing values of emission line fluxes and some equivalent widths.

\begin{table*}
    \centering  
    \begin{tabular}{|l|l|l|l|} \hline 
         \multicolumn{2}{|c|}{\textbf{AGN models}} &  \multicolumn{2}{|c|}{\textbf{SFG models}}\\ \hline \hline  
         \textbf{$\alpha$} & $-1.2$, $-1.6$, $-2.0$ &  \textbf{Mass upper cut} & 100 $\mathrm{M_{\odot}}$, 300 $\mathrm{M_{\odot}}$ \\ \hline 
         \textbf{$\mathrm{T_{bb}}$} & $5 \times 10^{4} \, \mathrm{K}$, $1 \times 10^{5} \, \mathrm{K}$, $2 \times 10^{5} \, \mathrm{K}$ &  \textbf{Stellar age} & 1 Myr, 10 Myr\\ \hline 
         \textbf{Z} & \shortstack[l]{0.0014, 0.0028, 0.007, 0.014, 0.028 \\ 
         or in solar metallicity (Z$_{\odot}$ = 0.014): 0.1, 0.2, 0.5, 1, 2} & 
         \textbf{Z} & \shortstack[l]{
         $1 \times 10^{-5}$, $1 \times 10^{-4}$, $1 \times 10^{-3}$, 0.0014, 0.0028, 0.007, 0.014  \\ 
         or in solar metallicity (Z$_{\odot}$ = 0.014): \char`\~ 0.0007, \char`\~ 0.007, \char`\~ 0.07, 0.1, 0.2, 0.5, 1} \\ \hline 
         \textbf{log U} & \shortstack[l]{table: $-0.5$, $-1$, $-1.5$, $-2$, $-2.5$ \\ 
         spectra: $-0.5$, $-1.5$, $-2.5$} & 
         \textbf{log U} & \shortstack[l]{table: $-1$, $-1.5$, $-2$, $-2.5$, $-3$ \\ 
         spectra: same as table} \\ \hline
    \end{tabular}
    \caption{Parameters from \protect\cite{p11_NakajimaMaiolino2022} for the AGN and SF galaxy (SFG) models (with $\alpha$ denoting the power-law index, $\mathrm{T_{bb}}$ the temperature of the bump component, Z the metallicity, and log U the ionization parameter). Note that for the AGN models, there are a few ionization values that are included in the table data, but not in the spectral data. If a method uses both types of the data, then only values included in both the tabular and the spectral data are used.}
    \label{tab:model_parameters}
\end{table*}

For the SF galaxies the spectral energy distributions (SEDs) are from Binary Population and Spectral Synthesis stellar evolution models (BPASS v2.2.1; \cite{Eldridge2017}\cite{Stanway_Eldridge_2018}) which are based on the \cite{Kroupa2001} initial mass function (IMF) with the upper cut on the mass at both 100 and 300 solar masses, used in various diagnostic plots. These adopt the stellar ages \mbox{1 Myr} and \mbox{10 Myr}. The SF galaxy sample used in this project contains metallicities (Z) ranging from about 0.0007 Z$_{\odot}$ to 1 Z$_{\odot}$ and ionization parameters (logU) from $-1$ to $-3$.

\cite{p11_NakajimaMaiolino2022} use AGN models with the \texttt{CLOUDY} "AGN" continuum command consisting of the ``Big Bump'' as well as power-law components. These components are parameterised by the temperature of the bump ($\mathrm{T_{bb}}$), ranging from $5 \times 10^{4}$ to $2 \times 10^{5}$, and the power-law index ($\alpha$) which varies between $-1.2$ and $-2$. In the AGN sample used in this project, metallicities range from 0.1 Z$_{\odot}$ to 2 Z$_{\odot}$, and ionization parameters range from $-0.5$ to $-2.5$.  The full list of parameters, metallicities and ionization values of the models we use are listed in Table \ref{tab:model_parameters}. These models are specifically applicable to Type II AGN or the narrow-line component observed in Type I AGN.

\section{Spectroscopy-based diagnostics}\label{sec:spectroscopy_based_diagnostics} 
In this section we discuss diagnostic methods relying on spectroscopy, starting with the standard BPT and VO87 diagrams that have been widely used to identify galaxies containing an active galactic nucleus. After the standard diagnostic methods, we discuss the OHNO diagram and the use of the \mbox{[\ion{O}{III}] $\lambda$4363} auroral line, which have been suggested as optical alternatives to the BPT and VO87 diagrams for high redshift systems. 

We then explore the use of ultraviolet line fluxes and equivalent widths for separating AGN and SFG, and apply some of the plots on our sources of interest. This is one of the main points of this paper --- developing a new method for locating systems which contain AGN. Furthermore, after discussing the sensitivity of some of these possible diagnostics, we explore the possibility of using combined optical--UV line flux ratios considering the magnitude of the dust correction factor.

\subsection{The weakness of standard optical diagrams}\label{subsec:BPT_VO87} 
As previously mentioned, the BPT \citet*{BaldwinPhillipsTerlevich_1981} and VO87 \cite{VeilleuxOsterbrock_1987} diagrams have been the standard methods for identifying  AGN. In this section, we research  the N2-BPT, and S2-VO87 diagrams, which display the line flux ratio \mbox{[\ion{O}{III}] $\lambda$5007/H$\beta$} against the ratios \mbox{[\ion{N}{II}] $\lambda$6584/H$\alpha$} and \mbox{[\ion{S}{II}] $\lambda\lambda$6716,6731/H$\alpha$}, respectively.

Ratios for the models were calculated using the tabulated data of the models. The parameters of these models are as described previously with the exception that we do not include the lowest metallicity (0.0007 Z$_{\odot}$ and 0.007 Z$_{\odot}$) SFG models, since they fall outside of the relevant region for the data presented.

\begin{figure*}
    \begin{subfigure}{.5\textwidth}
        \centering
        \includegraphics[width=\linewidth]{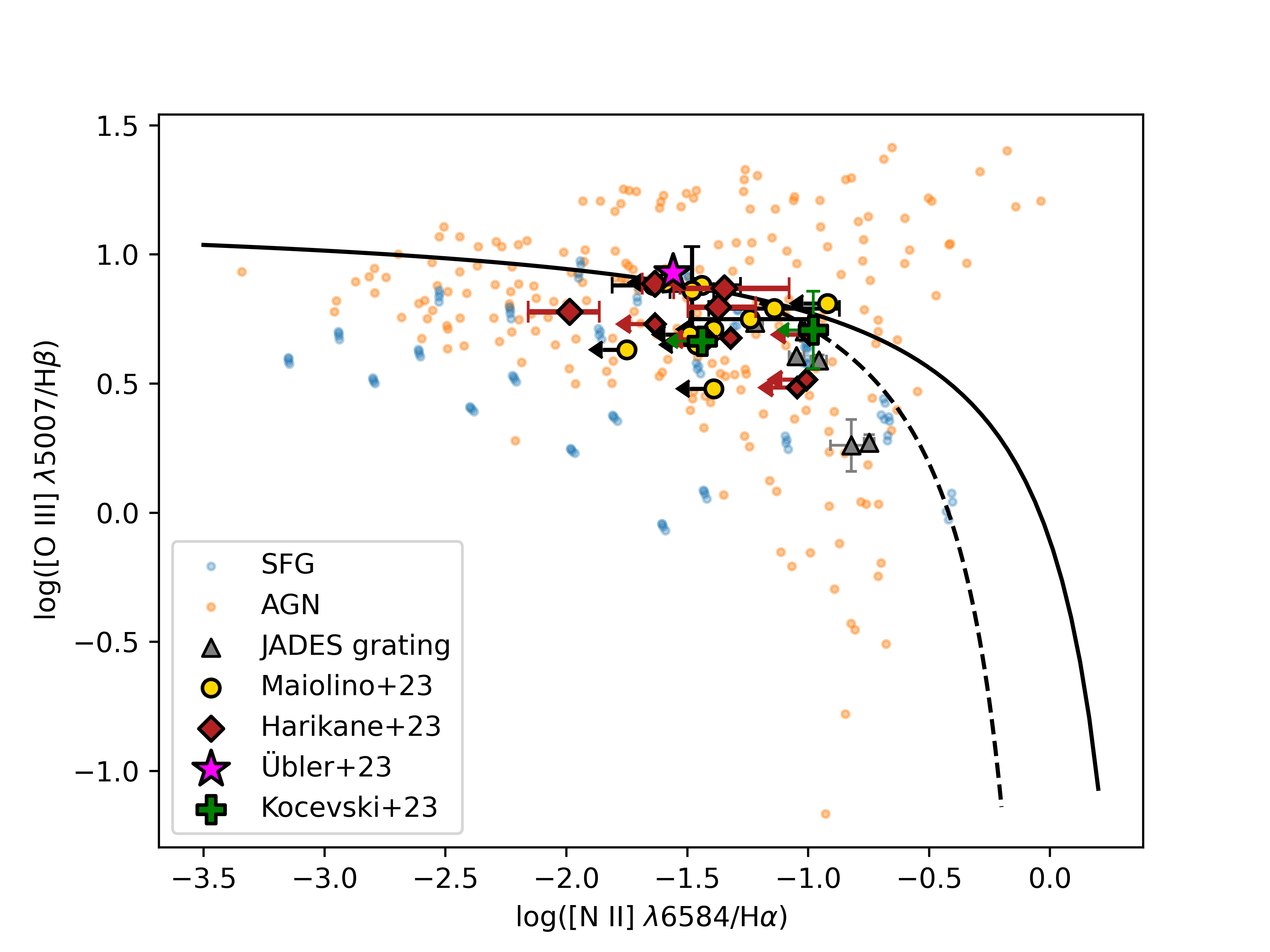}
    \end{subfigure}%
    \begin{subfigure}{.5\textwidth}
        \centering
        \includegraphics[width=\linewidth]{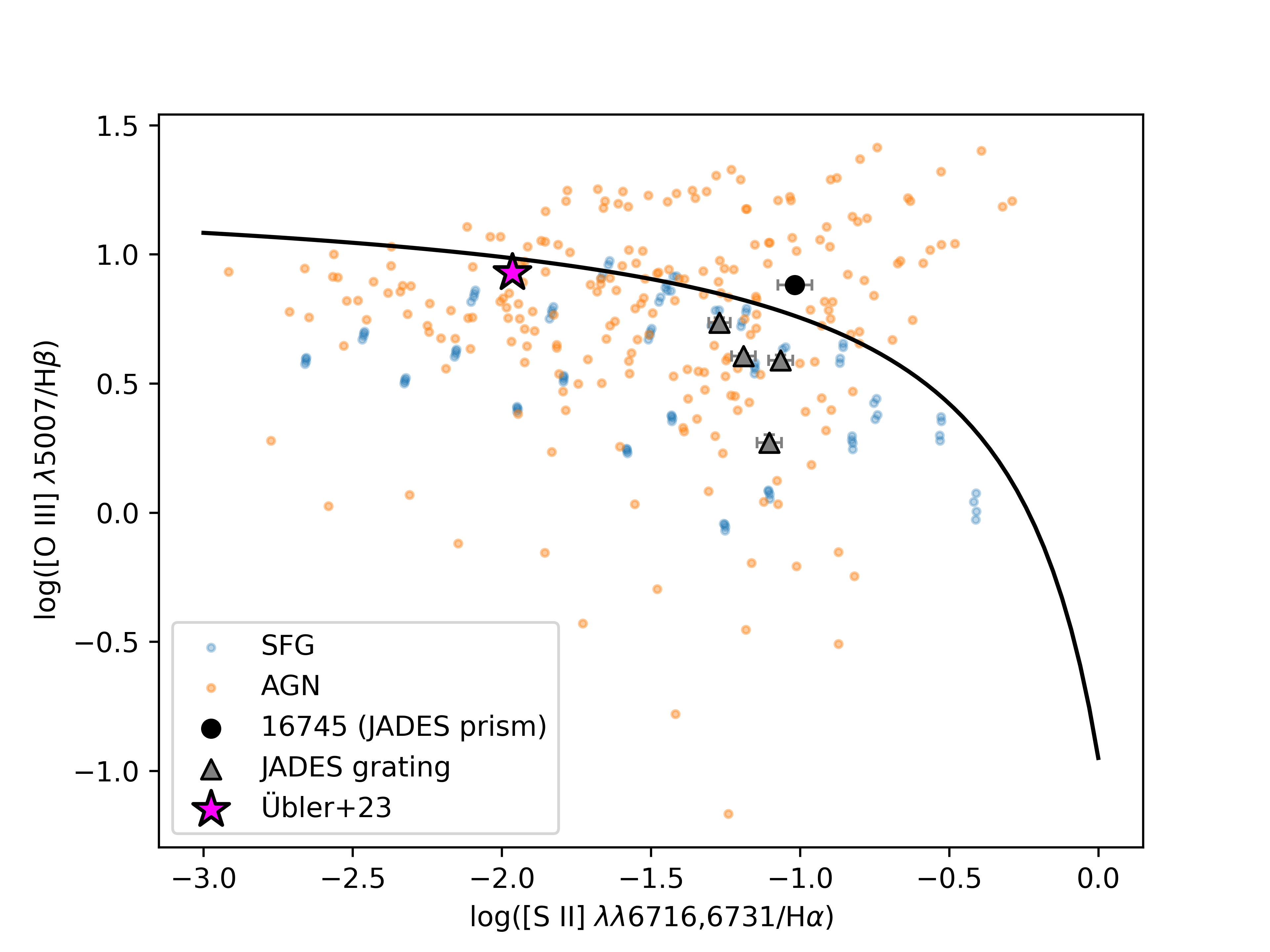}
    \end{subfigure}\\
\caption{N2-BPT (on the left) and S2-VO87 (on the right) diagrams displaying photoionization models of active galactic nuclei (AGN, orange) and star-forming galaxies (SFG, blue) created by \protect\cite{p11_NakajimaMaiolino2022} and various AGN sources from the literature (\protect\cite{p4_maiolino2023jades}, \protect\cite{Harikane2023}, \protect\cite{Ubler_2023}, \protect\cite{Kocevski2023}). The solid lines are extreme starburst lines defined by \protect\cite{p3_kewley2001} and the dashed classification line is from \protect\cite{Kauffman_2003}. Additionally, the grey triangles present JADES data points from the R=1000 grating catalogue (\protect\cite{p1_bunker2023jades}).  AGN are identified above the solid and dashed lines. A source in the Prism/Clear data catalogue with NIRSpec ID 16745, which contains all the necessary lines with a minimum SNR of 5, is located above the maximum starburst line on the S2-VO87 diagram. \protect\cite{p14_scholtz2023jades} selected this source as an AGN using similar considerations.Narrow-line fluxes were used to create these plots for all sources, except for source 16745, where the total flux was used.}
\label{fig:standard_diagrams}
\end{figure*}

The N2-BPT diagram is shown on the left side of \mbox{Figure \ref{fig:standard_diagrams}}. The photoionization models of star-forming galaxies and AGN are indicated with blue and orange points, respectively. The solid line is the extreme starburst line defined by \cite{p3_kewley2001}. 
We can see that this line agrees well with the AGN and SFG models displayed on the plot. All the modelled data points above the line are AGN, whereas both AGN and SFG points can be seen below the line. The dashed classification line was defined by \cite{Kauffman_2003} and it has been used to describe a composite region in the parameter space between the two lines (\cite{p2_kewley2006}). However, a mixed region can be seen even below the \cite{Kauffman_2003} line. In this mixed region, the location of AGN models is consistent with what is expected for AGN hosted in metal poor galaxies (\cite{p4_maiolino2023jades}).

The actual AGN sources displayed on the N2-BPT diagram are from \cite{p4_maiolino2023jades}, \cite{Harikane2023}, \cite{Ubler_2023}, and \cite{Kocevski2023}. \cite{p4_maiolino2023jades} presents 12 AGN at redshifts \mbox{4 < z < 7} from the JADES survey identified by their broad Balmer lines. The 10 sources from \cite{Harikane2023} are also broad-line AGN at \mbox{z > 4}. \cite{Ubler_2023} identifies galaxy \mbox{GS 3073} at \mbox{z = 5.55} as an AGN not only from its broad lines, but also from its high \mbox{He II $\lambda$4686} equivalent width. The position of the data point was calculated using the narrow line fluxes. Finally, \cite{Kocevski2023} reports 2 broad line AGN, CEERS 1670 and CEERS 3210, from the Cosmic Evolution Early Release Science (CEERS) Survey at \mbox{z > 5}. We use the narrow line flux values of the emission lines to position these sources on our plots.

We can see that all these sources would be missed as AGN by the use of the N2-BPT diagram if they did not display broad lines. They are located in a region of the parameter space which has been thought to be populated by star-forming galaxies and which is shown to be a mixed region of SFG and low metallicity AGN models.

Since the H$\alpha$ and \mbox{[\ion{N}{II}] $\lambda$6584} lines are blended in the prism spectra, it was not possible to determine the line ratio, and hence we could not display sources from the Prism/Clear data catalogue \citep{p1_bunker2023jades} in the N2-BPT diagram. We were able to display sources from the R=1000 grating catalogue, however, they could not be identified by this diagram.

On the right side of \mbox{Figure \ref{fig:standard_diagrams}}, we plot the S2-VO87 diagram. As before, the AGN and SFG models are represented with orange and blue points, respectively. The solid black line is the extreme starburst line from \cite{p3_kewley2001} and we can see that, similarly to the N2-BPT diagram, this line agrees well with the modelled data. Above the line we only find AGN models, while below it we can see both AGN and SFG models.

The galaxy from \cite{Ubler_2023} is represented the same way as on the N2-BPT diagram and is placed using the narrow line fluxes. We can see that this AGN would be missed by both the N2-BPT and S2-VO87 diagrams if it did not exhibit broad lines. The black data point is from the JADES dataset and was the only system with all the necessary lines in the prism catalogue to display on the plot. It uses the total flux values of the emission lines. It is located above the demarcation line, meaning that it can be identified as an AGN.  Its NIRSpec ID is 16745, and it was selected as an AGN by \cite{p14_scholtz2023jades} using this diagram. The grating data points are below the demarcation line, so they would not be identified as AGN by this diagram either.

Figure \ref{fig:standard_diagrams} shows why new identification methods are needed if we want to be able to effectively characterise high redshift galaxies. The BPT and VO87 diagrams, while still very useful in some circumstances, will miss many potential high redshift active galaxies amongst the plethora of data the JWST provides.

\subsection{Alternative optical line flux ratio diagrams} \label{subsec:optical_alternatives_OHNO_and_auroral_line}
This section explores two optical diagnostic methods which are suggested in recent previous work as alternatives of the standard methods. These are the OHNO diagram and the use of the \mbox{[\ion{O}{III}] $\lambda$4363} auroral line.

\subsubsection{OHNO diagram}\label{subsubsec:OHNO}
In this section we examine newer methods proposed to find AGN, one of which is the so-called OHNO method. The OHNO diagram and demarcation line was suggested by \cite{Backhaus_2022} and uses the line ratios \mbox{[\ion{O}{III}] $\lambda$5007/H$\beta$} and \mbox{[\ion{Ne}{III}] $\lambda$3869/[\ion{O}{II}] $\lambda\lambda$3727,3729}.

\begin{figure}
    \includegraphics[width=\linewidth]{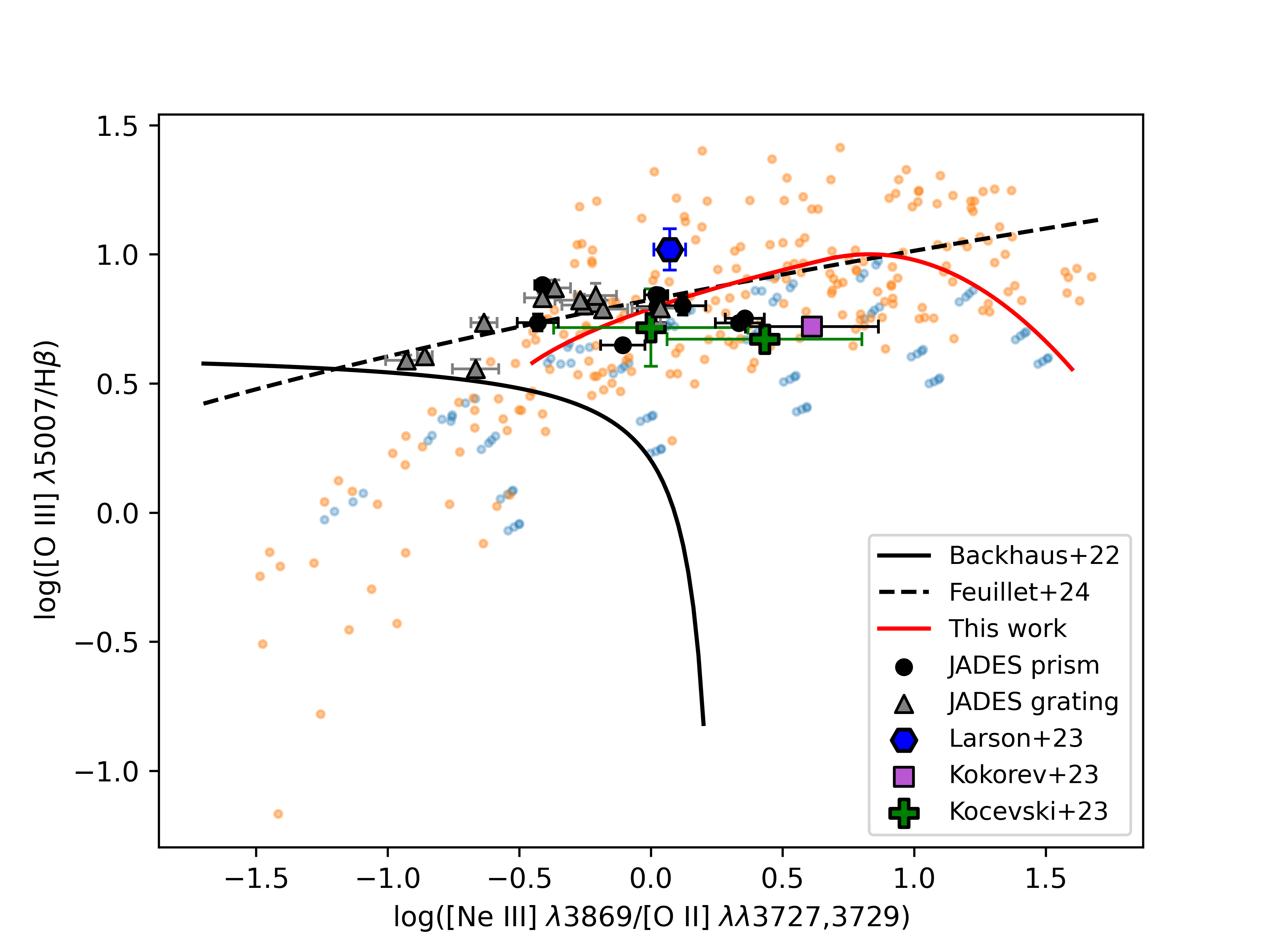}
    \caption{The OHNO diagram displaying SFG and AGN models from \protect\cite{p11_NakajimaMaiolino2022} (coloured as previously), the solid and dashed black demarcation lines are based on the work of \protect\cite{Backhaus_2022} and \protect\cite{Feuillet_2024}, respectively. Our newly defined solid red separation curve is defined by Equations \ref{eq:OHNO_1} and \ref{eq:OHNO_2}, and it separates a mixed region from a region only occupied by AGN models. Broad line AGN collected from literature (\protect\cite{Larson_2023}, \protect\cite{Kokorev_2023}, \protect\cite{Kocevski2023}), as well as sources from the JADES Prism/Clear and R=1000 grating data catalogues can be seen. Narrow line fluxes were used to create this plot with the exceptions of the data points from \protect\cite{Larson_2023} and the JADES PRISM where the total line flux was used.}
\label{fig:OHNO_diagram}
\end{figure}

The OHNO diagram can be seen plotted on Figure~\mbox{\ref{fig:OHNO_diagram}}. \cite{Backhaus_2022} created the OHNO AGN/SF line to empirically produce similar separation between AGN and SF galaxies to another diagram called unVO87 displaying the line ratios \mbox{[\ion{O}{III}] $\lambda$5007/H$\beta$} and \mbox{[\ion{S}{II}]/H$\alpha$ + [\ion{N}{II}]}. This solid black demarcation line is meant to divide the parameter space, such that AGN sources are above it. However, it does not agree with the displayed theoretical models. Based on the models, the demarcation line does not seem to show a separation between AGN and the mixed region as the ones on the NII-BPT and SII-VO87 diagrams do.

The dashed black line is another demarcation line defined by \cite{Feuillet_2024}. This shows a better separation between the models, though it still fails to perfectly distinguish the regions.Our separation curve, represented by a solid red line, was determined by visually inspecting the regions. It separates a mixed region from a region exclusively occupied by AGN models. It follows the equation

\begin{equation}
\log \left( \frac{[\ion{O}{III}]}{H\beta} \right) = \sqrt{\frac{\log \left( \frac{[\ion{Ne}{III}]}{[\ion{O}{II}]} \right) + 0.6}{3.4}} + 0.37  
\label{eq:OHNO_1}
\end{equation}
when  $ -0.45 < \log \left( [\ion{Ne}{III}] / [\ion{O}{II}] \right) < 0.7, $
and the equation
\begin{equation}
\log \left( \frac{[\ion{O}{III}]}{H\beta} \right) = 1 - 0.75 \left( \left( \frac{[\ion{Ne}{III}]}{[\ion{O}{II}]} \right) - 0.83 \right)^2
\label{eq:OHNO_2}
\end{equation}
when $ 0.7 < \log \left( [\ion{Ne}{III}] / [\ion{O}{II}] \right) < 1.6. $

\vspace{10pt}
Four previously identified AGN from \cite{Larson_2023}, \cite{Kokorev_2023}, and \cite{Kocevski2023} are displayed on Figure~\mbox{\ref{fig:OHNO_diagram}}.
\cite{Larson_2023} identified \mbox{CEERS 1019} as an AGN based on its broad lines, the
presence of weak high-ionization lines, and a spatial point-source component.
\cite{Kokorev_2023} identifies another broad line AGN at \mbox{z = 8.5} from the JWST Cycle 1 UNCOVER Treasury survey with high ionization lines and a
point-source morphology. Amongst these, only the galaxy from \cite{Larson_2023} can be identified as an AGN based on this diagnostic diagram.

Furthermore, we can see nine sources from the JADES prism catalogue displayed as black data points. Amongst them is one of our galaxies of interest with NIRSpec ID 9422, however, it is located in the mixed region. Two of these data points are located above our demarcation line: 16745 and 10013620. The source 16745 has been selected as an AGN candidate by \cite{p14_scholtz2023jades}, but galaxy 10013620 has not been identified as an AGN candidate to the best of our knowledge. This is the black data point which is above our separation curve and located on the black dashed demarcation line produced by \cite{Feuillet_2024}.

Lastly, we also present data points from the JADES R=1000 grating catalogue. These are displayed as grey triangles and most of them are outside of the mixed region, therefore could be potential AGN candidates. The NIRSpec IDs of the sources in the region of interest are as follows: 22251, 10008071, 4270, 7938, 18090, 3892, 10009506, 10011849, 8891, 3753.

Narrow line fluxes were used to create this graph, except for the data points from \protect\cite{Larson_2023} and the JADES PRISM, where the total line flux was used.

\subsubsection{The \mbox{[\ion{O}{III}] $\lambda$4363} auroral line}\label{subsubsec:auroral_line}
In a recent study, the use of the \mbox{[\ion{O}{III}] $\lambda$4363} auroral line, which has been observed in several JWST sources, was presented as a possible alternative method to locate AGN by \cite{MazzolariUbler2024_OIII4363_auroral_line} (see also \cite{Brinchmann_2023} and \cite{Ubler_2023_offsetAGN}).

\cite{MazzolariUbler2024_OIII4363_auroral_line} proposed that a high \mbox{[\ion{O}{III}] $\lambda$4363/H$\gamma$} ratio is sufficient to identify the presence of an AGN, and suggested three diagnostic diagrams displaying this ratio against other ratios: \mbox{[\ion{O}{III}] $\lambda$5007/[\ion{O}{II}] $\lambda\lambda$3727} (O32),  \mbox{[\ion{Ne}{III}] $\lambda$3869/[\ion{O}{II}] $\lambda\lambda$3727} (Ne3O2), and \mbox{[\ion{O}{III}] $\lambda$5007/[\ion{O}{III}] $\lambda$4363} (O3O3).
They supported these suggestions by not only empirical calibrations, but with a wide range of theoretical models. The usefulness of these diagnostics is based on the intensity of the [\ion{O}{III}] $\lambda$4363 line relative to the H$\gamma$ line. This is influenced by the temperature and metallicity of the interstellar medium (ISM), as well as the ionization parameter (see \cite{MazzolariUbler2024_OIII4363_auroral_line} for more detail).

The \mbox{[\ion{O}{III}] $\lambda$4363} auroral line is a very promising possible alternative to the standard diagnostic methods, however, we note that this line is quite hard to observe in the data we are considering. Only a few sources had the necessary emission lines recorded in the line flux catalogue and these sources were under the demarcation lines, meaning they could be either AGN or SFG.

\subsection{UV alternatives}\label{subsec:UV_alternatives}

\subsubsection{Using UV emission line ratios} \label{subsubsec:UV_ratio_plots}

Lines such as \mbox{\ion{C}{IV}} and \mbox{\ion{He}{II} $\lambda$1640} have high potential for separating stellar population and AGN since they require extremely high energy ionizing photons. In particular, the \mbox{\ion{He}{II} $\lambda$1640} line is very sensitive to this ionization which is produced more dominantly by AGN than by stars, and thus line ratios using \mbox{\ion{He}{II} $\lambda$1640} were suggested to be useful in separating SF galaxies and AGN (\cite{p5_Feltre2016, Gutkin2016}).  

\cite{p5_Feltre2016} stated that the line ratios \mbox{\ion{C}{IV}/\ion{C}{III}]}, \mbox{\ion{C}{IV}/\ion{He}{II} $\lambda$1640}, and \mbox{\ion{C}{III}]/\ion{He}{II} $\lambda$1640} can help separate AGN and SF galaxies very well. In their model spectra, the \mbox{\ion{C}{IV}/\ion{He}{II} $\lambda$1640} and \mbox{\ion{C}{III}]/\ion{He}{II} $\lambda$1640} ratios distinguish between active and inactive galaxies even on their own, while the \mbox{\ion{C}{IV}/\ion{C}{III}]} ratio could do the separation with the help of an additional ratio.

A paper by \cite{p12_nakajima2018} also discusses possible UV diagnostic diagrams involving the lines \ion{C}{III}], \ion{C}{IV}, and \mbox{\ion{He}{II} $\lambda$1640}.
A diagnostic diagram using the line ratios \mbox{\ion{C}{IV}/\ion{C}{III}]} and \mbox{\ion{C}{III}] + \ion{C}{IV} / \ion{He}{II} $\lambda$1640} presented by them is called the C4C3-C34 diagram. They showed that AGN and SFG models (\cite{p5_Feltre2016, Gutkin2016}) show a remarkable difference in the \mbox{\ion{C}{III}] + \ion{C}{IV} / \ion{He}{II} $\lambda$1640} parameter (C34), because for AGN this value is smaller than for SFG considering a fixed value of \mbox{\ion{C}{IV}/\ion{C}{III}]} (C4C3).

Though these line ratios are very promising, in the observational data from the NIRSpec prism, \mbox{\ion{He}{II} $\lambda$1640} and \mbox{\ion{O}{III}] $\lambda$1665} are not present on their own, only as a blended line. We therefore considered whether the blended line could still provide some information and we explored these UV line ratios using the tabulated modelled data presented in Section \ref{subsec:mod_data}.

\begin{figure*}
    \begin{subfigure}{.5\textwidth}
        \centering
        \includegraphics[width=\linewidth]{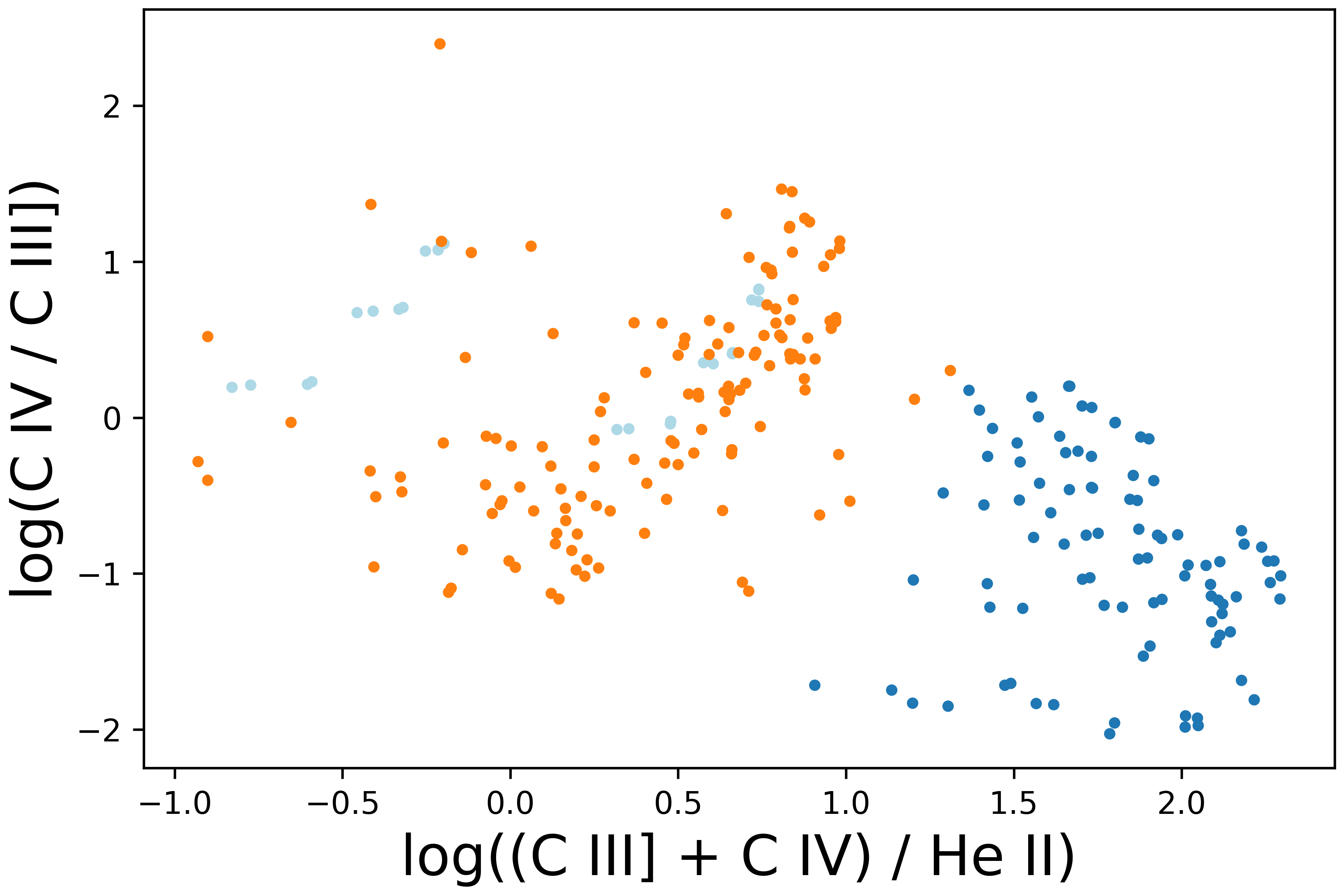}
        \label{fig:ratios_a}
    \end{subfigure}%
    \begin{subfigure}{.5\textwidth}
        \centering
        \includegraphics[width=\linewidth]{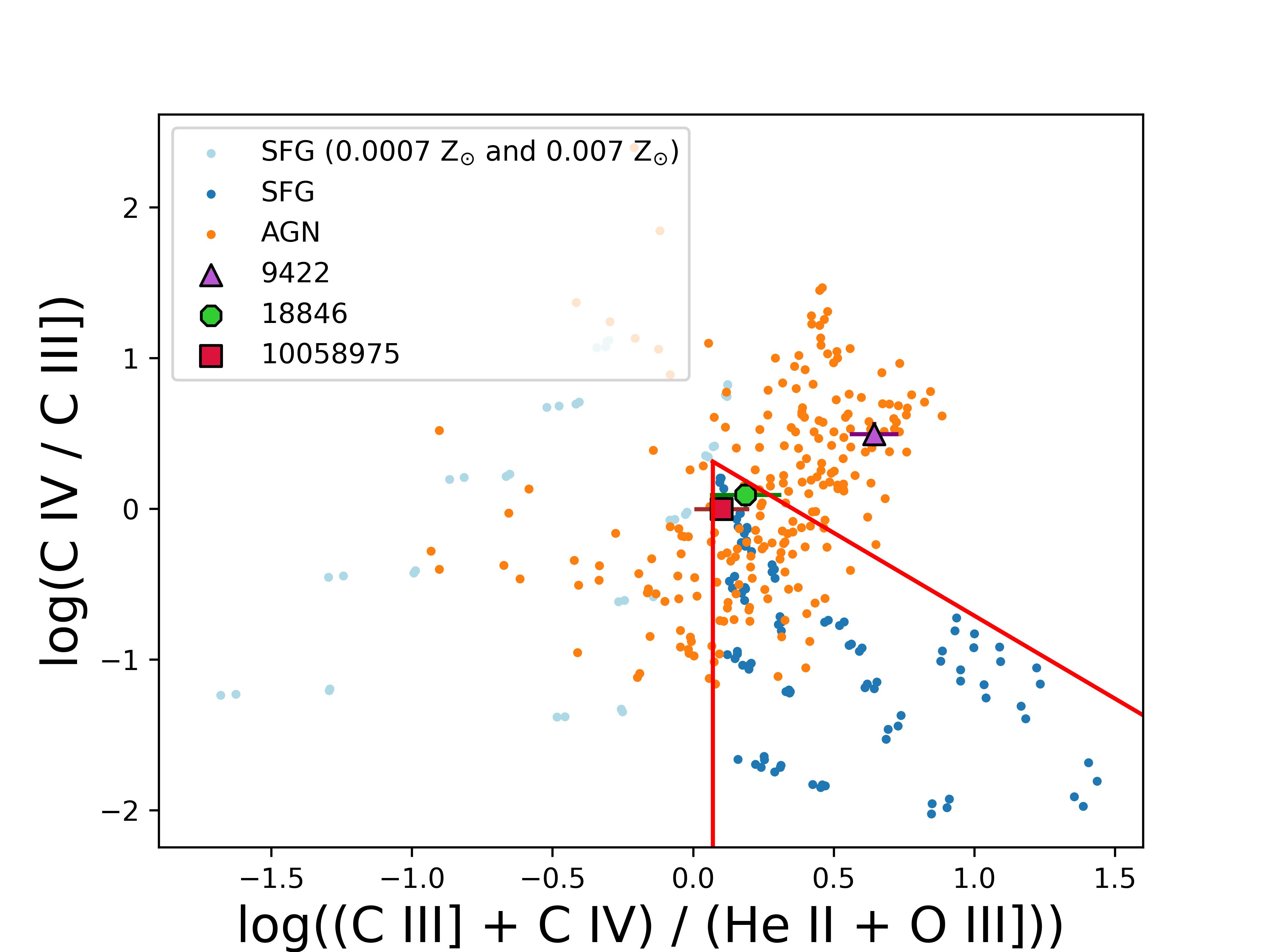}
        \label{fig:ratios_e}
    \end{subfigure}\\%
    \caption{Ultraviolet emission line flux ratio diagrams created based on the models from \protect\cite{p11_NakajimaMaiolino2022} presented in Table \ref{tab:model_parameters}. As in Fig. \ref{fig:standard_diagrams}, the AGN models are represented with orange dots and the SF galaxy models with blue dots. The lighter blue points represent the SF galaxies models with very low metallicities (0.0007 Z$_{\odot}$ and 0.007 Z$_{\odot}$). These metallicities are 2--3 orders of magnitude lower than metallicities of AGN models in the same region. Our galaxies of interest (with NIRSpec IDs 9422, 18846, and 10058975) are represented using three different colors (purple, green and red).}
    \label{fig:UV_line_ratios_paper}
\end{figure*}

As before, the plots show the AGN and SFG models in orange and blue, respectively, while the lighter blue points are the SFG models with very low metallicities (0.0007 Z$_{\odot}$ and 0.007 Z$_{\odot}$). It can be seen on the left side of Figure  \ref{fig:UV_line_ratios_paper} that if the lowest metallicity SF galaxies are disregarded, then the \mbox{(\ion{C}{III}] + \ion{C}{IV}) / \ion{He}{II} $\lambda$1640} ratio indeed shows a complete separation between AGN and SF galaxies, even on their own. The low metallicity SF galaxies exhibit similar ratios to AGN due to both the metal and the \mbox{\ion{He}{II} $\lambda$1640} emission being very weak, while in active galaxies, the \mbox{\ion{He}{II} $\lambda$1640} emission is exceptionally strong.

Replacing \mbox{\ion{He}{II} $\lambda$1640} in the ratio presented on the left side of Figure \ref{fig:UV_line_ratios_paper} with the blended line, an additional plot was created. This can be seen on the right of Figure \ref{fig:UV_line_ratios_paper}, where the ratio \mbox{(\ion{C}{III}] + \ion{C}{IV}) / (\ion{He}{II} + \ion{O}{III}])}(CHeO) is presented on the x-axes. We call this new diagnostic diagram C4C3-CHeO . It is clear that on this plot the region of parameter space that is uniquely occupied by AGN is reduced. This means that the range of AGN and star forming galaxies that can be identified from this diagram is more limited. There can still be many cases, however, where we can make an SFG--AGN separations as two distinct regions remain visible. Galaxies with:
\begin{equation}
\log \left( \frac{\text{\ion{C}{III}] + \ion{C}{IV}}}{\text{\ion{He}{II} + \ion{O}{III}]}} \right) < 0.07
\label{eq:Line_ratio_separation_left_side}
\end{equation}
and
\begin{equation}
\log \left( \frac{\text{\ion{C}{IV}}}{\text{\ion{C}{III}]}} \right)  
> -1.1 \log \left( \frac{\text{\ion{C}{III}] + \ion{C}{IV}}}{\text{\ion{He}{II} + \ion{O}{III}]}} \right) + 0.39
\label{eq:Line_ratio_separation_right_side}
\end{equation}
when $ \log(({\text{\ion{C}{III}] + \ion{C}{IV}}})/({\text{\ion{He}{II} + \ion{O}{III}]}})) > 0.07 $  can be reliably identified as AGN, while those with values outside these thresholds fall into a mixed region that can be either star-forming galaxies or AGN. The AGN in the upper right region primarily consist of those with the highest ionization parameters in the model (logU = $-0.5$ and $-1.0$), with most AGN with logU = $-0.5$ falling entirely within the region. The AGN in the left region primarily consist of those with the lowest ionization parameters in the model (logU = $-2.5$).

This figure also displays the JWST/NIRSpec data sample on the diagram. As can be seen, the only three galaxies from our sample that have the needed UV lines are our galaxies of interest. The remaining galaxies did not achieve a sufficient signal-to-noise (S/N) ratio, and therefore, do not appear in the emission line flux catalogue. \ion{C}{IV} and the blended \mbox{\ion{He}{II} + \ion{O}{III}]} lines were detected in all three sources with a signal-to-noise ratio \mbox{S/N > 5}. However, the \ion{C}{III}] emission line was not reported for source 9422 so we use $\mathrm{L}_{\mathrm{I}}\mathrm{M}_{\mathrm{E}}$ to calculate the emission line flux. Unfortunately, galaxies 18846 and 10058975 fell within areas of the diagnostic plot where clear identification was not possible. In contrast, galaxy 9422 was located in the AGN-only region, suggesting it could be classified as an AGN.

While this diagnostic offers a promising alternative to standard diagnostic methods, we note that the emission lines constituting these ratios are difficult to observe in the existing JWST data and would require additional observations, such as longer exposure time data, to be used effectively.

\subsubsection{Using UV EWs and line ratios}\label{subsubsec:UV_EW_and_ratio_plots}

\cite{p12_nakajima2018} not only discusses UV line ratios, but also suggests that the equivalent widths of these UV lines could potentially be used as diagnostic parameters. This paper shows that the EWs of \mbox{\ion{C}{IV}}  and \mbox{\ion{C}{III}]} plotted against the ratios \mbox{\ion{C}{IV}/\ion{He}{II} $\lambda$1640} and \mbox{\ion{C}{III}]/\ion{He}{II} $\lambda$1640}, respectively, provide a very good separation between AGN and SF galaxies.

\cite{p11_NakajimaMaiolino2022} then complemented this study by using the EW of \ion{He}{II} $\lambda$1640. They found that two distinct regions could be seen when using the EW of \ion{He}{II} against the ratios of either \ion{C}{III}]/\ion{He}{II} or \ion{O}{III}]/\ion{He}{II}. Unfortunately, we are unable to use these graphs with the prism data because of the blended \mbox{\ion{He}{II} + \ion{O}{III}]} line. In this section, we show how the plots change when using the blended \mbox{\ion{He}{II} + \ion{O}{III}]} line instead.

The line ratios were calculated using the tabular form of the modelled data. The EWs were calculated from a mix of the modelled spectra and the table data. The continuum was fitted by applying the $\mathrm{fit\_generic\_continuum}$ function of the specutils Python package on manually selected regions around the emission lines in the spectral models.

\begin{figure*}
    \begin{subfigure}{.5\textwidth}
        \centering
        \includegraphics[width=\linewidth]{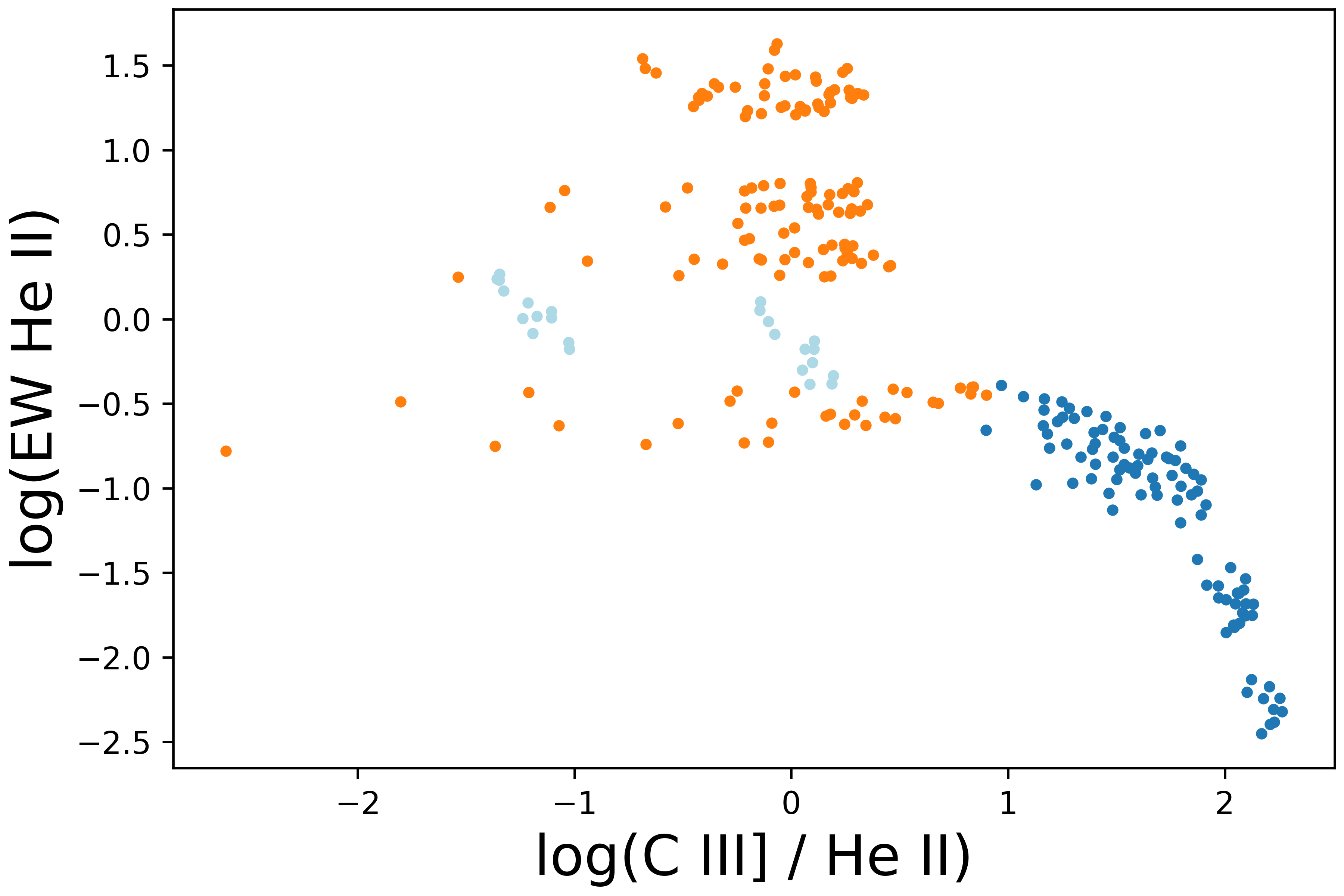}
        \label{fig:bl_a}
    \end{subfigure}%
    \begin{subfigure}{.5\textwidth}
        \centering
        \includegraphics[width=\linewidth]{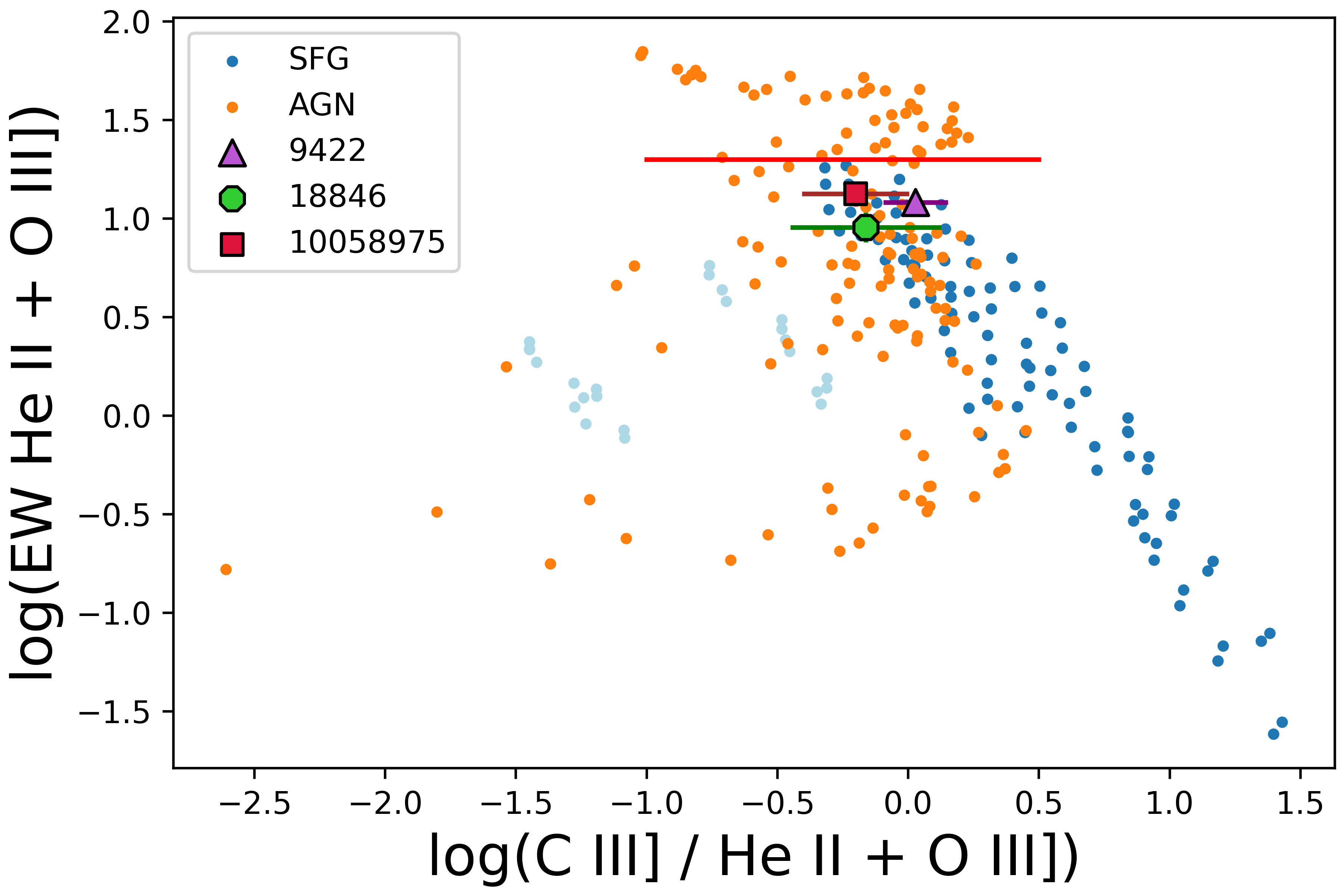}
        \label{fig:bl_b}
    \end{subfigure}\\
    \caption{The EWs of the \ion{He}{II} $\lambda$1640 and \mbox{\ion{He}{II} $\lambda$1640 + \ion{O}{III}]} lines can be seen plotted against the emission line flux ratios \mbox{\ion{C}{III}]/\ion{He}{II} $\lambda$1640} and \mbox{\ion{C}{III}]/(\ion{He}{II} $\lambda$1640 + \ion{O}{III}]) blend}. The plot to the right was created by replacing the \mbox{\ion{He}{II} $\lambda$1640} line  with the blended \mbox{\ion{He}{II} $\lambda$1640 + \ion{O}{III}]} line. The blending of \ion{He}{II} $\lambda$1640 and \ion{O}{III}] in the NIRSpec prism data reduces the diagnostic power of \ion{He}{II} $\lambda$1640 as an AGN indicator, except for AGN with the highest \mbox{\ion{He}{II} $\lambda$1640} EW.}
\label{fig:blended_EW_vs_ratio_plot}
\end{figure*}

The EWs of \mbox{\ion{He}{II} $\lambda$1640} and the blended \mbox{\ion{He}{II} + \ion{O}{III}} lines were considered and plotted against the line ratios \mbox{\ion{C}{III}]/\ion{He}{II} $\lambda$1640}, and  \mbox{\ion{C}{III}]/(\ion{He}{II} + \ion{O}{III})}, respectively. These can be seen in Figure ~\ref{fig:blended_EW_vs_ratio_plot}.

We can see on the left side  of Figure \ref{fig:blended_EW_vs_ratio_plot} that the EW of \mbox{\ion{He}{II} $\lambda$1640} creates three regions, one with only AGN which have the highest $\mathrm{EW}_{\ion{He}{II}}$, a region with only SFG possessing the lowest $\mathrm{EW}_{\ion{He}{II}}$ values, and a mixed region in the middle where the lowest metallicity SFG are located. Because of its high ionization potential, the recombination line \mbox{\ion{He}{II} $\lambda$1640}, even on its own, is a very good diagnostic parameter to separate AGN from SF galaxies.
Furthermore, the ratio \mbox{\ion{C}{III}]/\ion{He}{II} $\lambda$1640} shows a very good separation even on its own where the SFG have the highest values and the AGN the lowest. For SFG, the \mbox{\ion{He}{II} $\lambda$1640} emission line flux is smaller than for AGN, causing the ratio to get bigger.

These quantities could be used together as a diagnostic diagram, but as stated before, in the observational data \mbox{\ion{He}{II} $\lambda$1640} and \mbox{\ion{O}{III}] $\lambda$1665} are not present on their own, so these plots could not be applied here. A similar plot replacing the \mbox{\ion{He}{II} $\lambda$1640} line with the blended \mbox{\ion{He}{II} + \ion{O}{III}]} can be seen on the right side of Figure \ref{fig:blended_EW_vs_ratio_plot}, together with the original plot on the left. This shows significantly more mixing between AGN and SFG models than the plot that uses the separately detected lines. There is, however, still a region which would allow for the identification of AGN candidates, but this is much more limited. As supported by the solid red line on the right of Figure \ref{fig:blended_EW_vs_ratio_plot}, galaxies with $\log (\mathrm{EW} \, \mbox{\ion{He}{II} + \ion{O}{III}]}) > 1.3$ can be reliably identified as AGN, while those with values below this threshold fall into a mixed region that includes star-forming galaxies and AGN. The AGN above the line consist of those with an $\alpha = -1.2$.

This figure also displays our galaxies of interest 9422, 18846 and 10058975. The observed EW values from the JWST data were converted to the rest frame equivalents using a division by (1+z). This was consistently done throughout the rest of the paper. Unfortunately, the three galaxies are in the mixed region making it impossible to clearly identify them.

\subsection{EW diagrams using both optical and UV lines}\label{subsec:EW_plots}

As outlined previously, it is necessary to explore other methods for separating star-forming and AGN systems to complement diagnostics based only on flux ratios. Consequently, we also focus on the design and analysis of diagnostics based on the comparison of EWs. \cite{p12_nakajima2018} defines separating regions based on the EWs of \ion{C}{III}] and \ion{C}{IV} and \cite{p11_NakajimaMaiolino2022} on the EW of \mbox{\ion{He}{II} $\lambda$1640}. In our case, we complemented the study of these emission lines with the addition of EWs of other emission lines in the optical range (such as H$\beta$, \mbox{\ion{He}{II} $\lambda$4686}). The other diagrams we considered can be found in the Appendix on Figure \ref{fig:appendix_EWplots_all_metallicities} and show an exhaustive list of possible diagnostic diagrams using EWs on both axes. 

\begin{figure}
    \includegraphics[width=\linewidth]{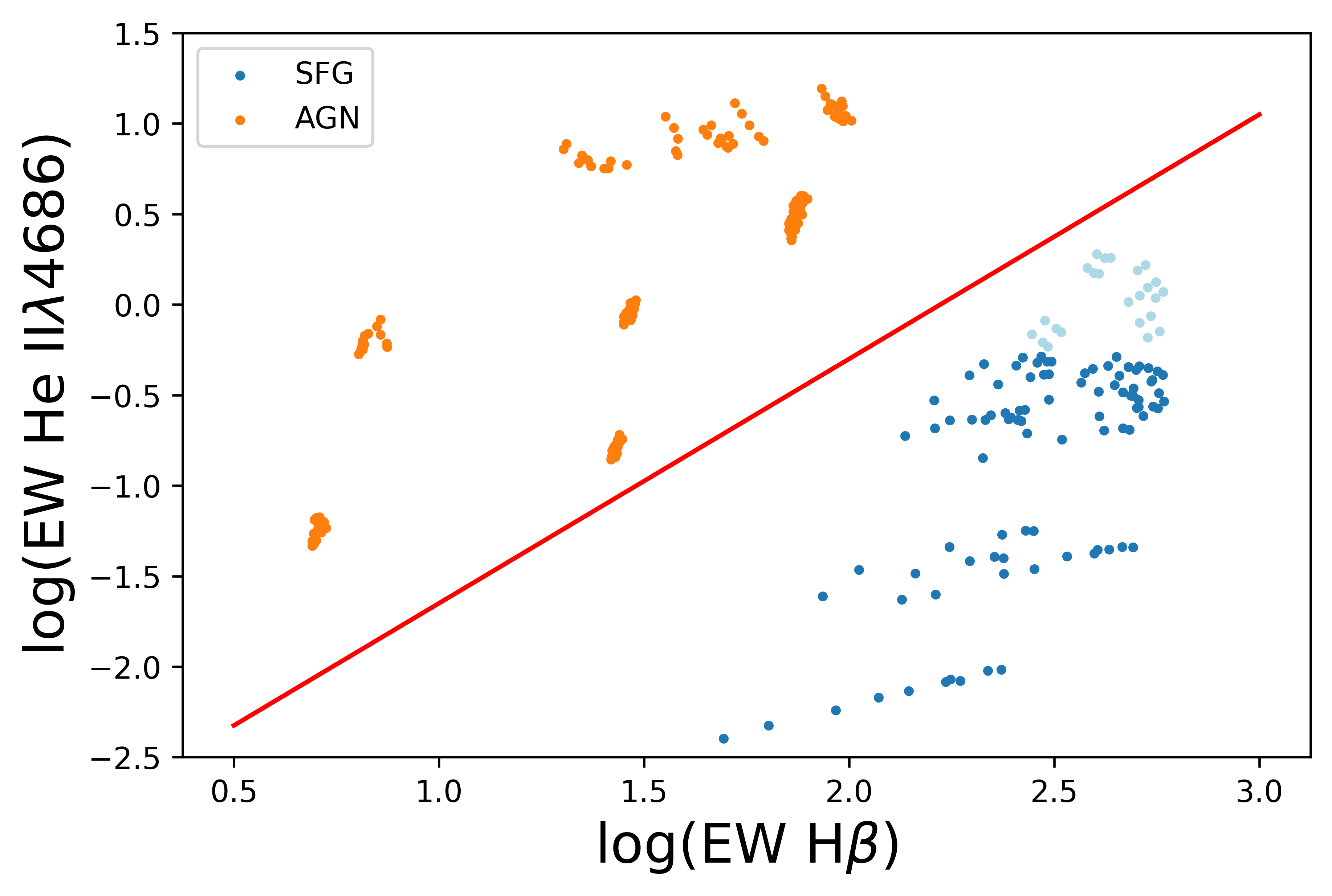}
    \caption{The modelled data on a plot displaying the EW of H$\beta$ against the EW of \mbox{He II $\lambda$4686}. The red line separates the AGN and SFG regions and is defined by Equation \ref{eq:EW_plot_separation}.}
\label{fig:EWplot_best}
\end{figure}

The plot that separates the regions the best can be seen on Figure \ref{fig:EWplot_best}. It uses the EW of H$\beta$  vs. the EW of \mbox{\ion{He}{II} $\lambda$4686}. H$\alpha$ would give a similar plot, since H$\beta$ and H$\alpha$ are very similar, but H$\beta$ is preferred due to H$\alpha$ shifting out of the JWST/NIRSpec wavelength range at high redshifts (z = 7 vs. z = 10). In the models presented by \cite{p11_NakajimaMaiolino2022}, the spectra consist of the narrow line region (NLR) emission combined with the contribution from the accretion disk. This likely leads to lower EWs of H$\beta$ in AGN models compared to those in SFG models due to the additional continuum contribution. The \mbox{\ion{He}{II} $\lambda$4686} recombination line was chosen because it appears to be a promising diagnostic due to being independent of the metallicity. It has been explored and used by  \cite{Shirazi_2012, Tozzi2023, Ubler_2023, p14_scholtz2023jades} and \cite{Dors2023}. We define a separation line to distinguish AGN from SFG as follows:
\begin{equation}
\log (\mathrm{EW} \, \mbox{\ion{He}{II} $\lambda$4686}) > 1.35 \log (\mathrm{EW} \, \mathrm{H}\beta) - 3.
\label{eq:EW_plot_separation}
\end{equation}

The problem with this graph, however, is that the \mbox{\ion{He}{II} $\lambda$4686} line is very faint, so it was not detected with a \mbox{S/N > 5} in the observational data. An upper limit calculation provides constraints on the sources containing an AGN, however, \cite{p14_scholtz2023jades} stated that this approach did not give any promising results. Even though this plot cannot be used for this data, it is  worth to consider what would be needed in order to better observe \mbox{\ion{He}{II} $\lambda$4686} (see Section \ref{subsec:sensitivity}). 

Amongst the plots that only use UV emission line EWs, the most promising is the one using the \mbox{\ion{C}{IV}} line and the \mbox{\ion{He}{II} $\lambda$1640} line if detected separately in the observational data as we discussed in Section \ref{subsubsec:UV_EW_and_ratio_plots}. The \mbox{\ion{He}{II} + \ion{O}{III}] blend} is reported in the data and will later be used in Section \ref{subsec:EW_of_three_galaxies} with our galaxies of interest.

The equivalent width of an emission line measures its strength relative to the surrounding continuum, which comes from starlight, the accretion disk, and nebular gas in both the broad and narrow line regions (BLR and NLR). In the BLR, additional continuum from the unobscured accretion disk reduces the broad line EW, while in the NLR, starlight contributes to the continuum, lowering the narrow line EW. When analyzing AGN, line ratios offer a more reliable diagnostic tool compared to equivalent widths, particularly in Type II AGN configurations. The models of \cite{p11_NakajimaMaiolino2022} consist of the NLR emission combined with the accretion disk component as previously mentioned. The presence of the accretion disk component raises the continuum level, leading to a lower EW. However, observationally we aim to identify Type II AGN, which have their BLR (and accretion disk) component obscured. This discrepancy results in artificially low EWs shown by the theoretical models, as in reality the nebular lines are measured against a diminished continuum causing higher EWs. However, the line ratios remain unaffected since they depend only on the relative strengths of the emission lines.

\subsection{EWs of galaxies 9422, 18846, and 10058975} \label{subsec:EW_of_three_galaxies}

In this section we examine real spectroscopy from galaxies observed with NIRSpec. Whilst there is extensive spectroscopy of distant galaxies with JWST, there are few galaxies observed to date that have all of the lines in emission that can be used in the diagnostics we have mentioned. We use the spectra to determine whether these galaxies are AGN and how EWs can be used to determine this.

Equivalent width measurements of some ultraviolet and optical lines in the spectra of galaxies 9422, 18846, and 10058975 with JWST observations are discussed. Two optical and three ultraviolet lines are used: H$\beta$, \mbox{[\ion{O}{III}] $\lambda$5007}, \ion{C}{IV}, \ion{C}{III}], and the blended \mbox{\ion{He}{II} + \ion{O}{III}]} line.

The continuum around each emission line in the galaxies was manually selected and fitted using the specutils Python package. The observed equivalent widths were calculated by dividing the line flux of the emission line in question by the obtained continuum level. 

Finally, using the rest frame EW measurements, the three galaxies were displayed on some of the EW vs. EW graphs we explored. Some of these plots can be seen in Figure \ref{fig:data_on_EW_plots}, and some other diagrams we considered can be found in the Appendix in Figure \ref{fig:appendix_data_on_EW_plots}.

We display solid red lines on the diagrams aiming to separate a mixed region from a region only occupied by AGN models. The first line (top left of Fig. \ref{fig:data_on_EW_plots}) follows the equation
\begin{equation}
\log (\mathrm{EW} \, [\ion{O}{III}] \lambda5007) = 30 \log (\mathrm{EW} \, \mathrm{H}\beta) - 59.
\label{eq:data_on_EW_plot_separation1}
\end{equation}
Only SFG models are to the right of this line, and mostly AGN to its left.

The second red separation line (upper right of Fig. \ref{fig:data_on_EW_plots}) follows the equation 
\begin{equation}
\log (\mathrm{EW} \, \ion{He}{II} + \ion{O}{III}] ) = 1.95 \log (\mathrm{EW} \, \ion{C}{III}]) - 0.65
\label{eq:data_on_EW_plot_separation2a}
\end{equation}
when $ 0.2 < \log (\mathrm{EW} \, \ion{C}{III}]) < 1,$
and a constant lower limit of 
\begin{equation}
\log (\mathrm{EW} \, \ion{He}{II} + \ion{O}{III}] ) = 1.3 \log (\mathrm{EW} \, \ion{C}{III}])
\label{eq:data_on_EW_plot_separation2b}
\end{equation}
when $ 1 < \log (\mathrm{EW} \, \ion{C}{III}]) < 1.8.$
Sources above this line are in a region fully occupied by AGN models.

The last separation line (bottom of Fig. \ref{fig:data_on_EW_plots}) is defined by the equation
\begin{equation}
\log (\mathrm{EW} \, \ion{C}{IV} ) = 2.3 \log (\mathrm{EW} \, \mathrm{H}\beta) - 4.2,
\label{eq:data_on_EW_plot_separation3}
\end{equation}
such that we only see AGN models above the line.

\begin{figure*}
    \begin{subfigure}{.5\textwidth}
        \centering
        \includegraphics[width=.87\linewidth]{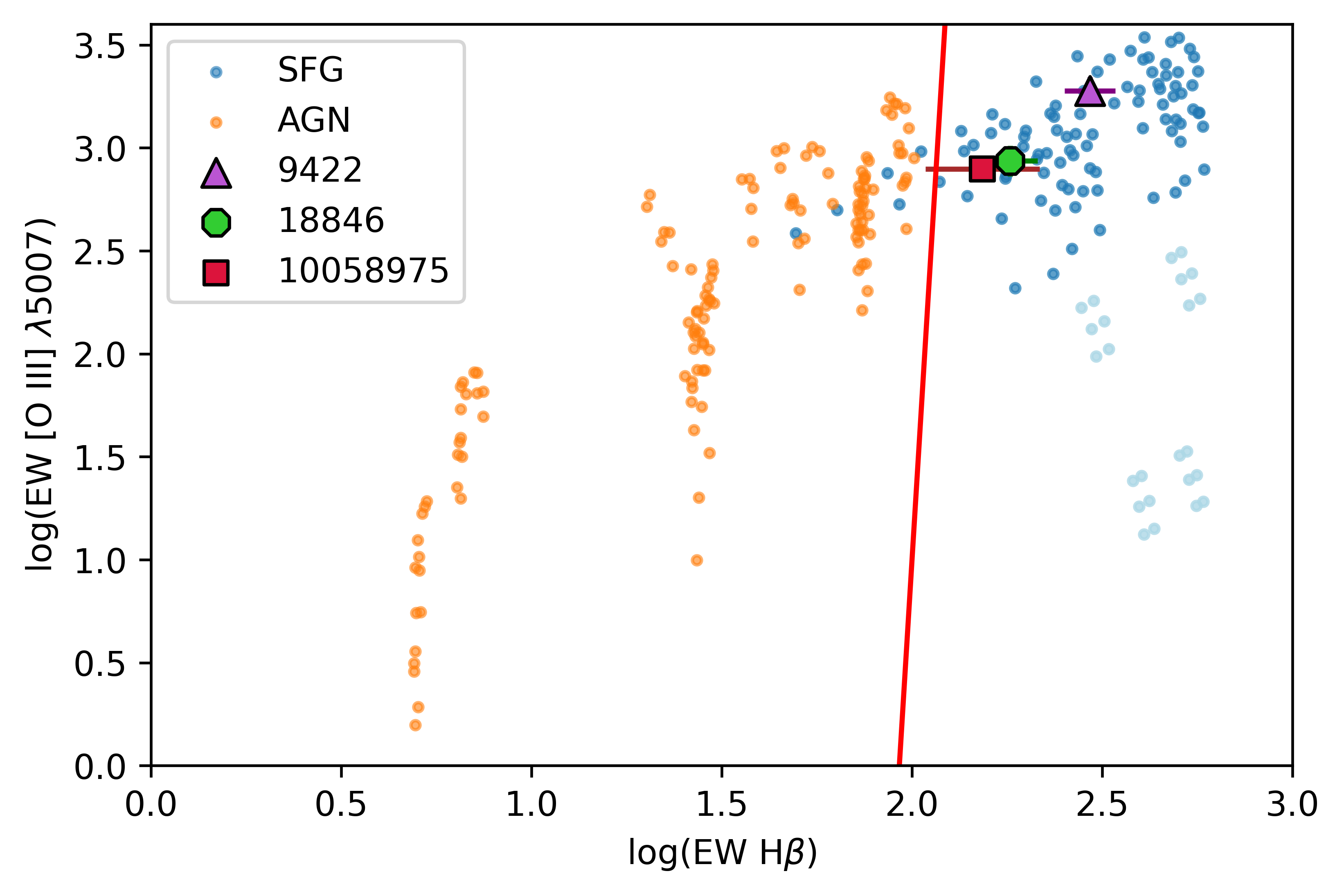}
        \label{fig:final_a}
    \end{subfigure}%
    \begin{subfigure}{.5\textwidth}
        \centering
        \includegraphics[width=.9\linewidth]{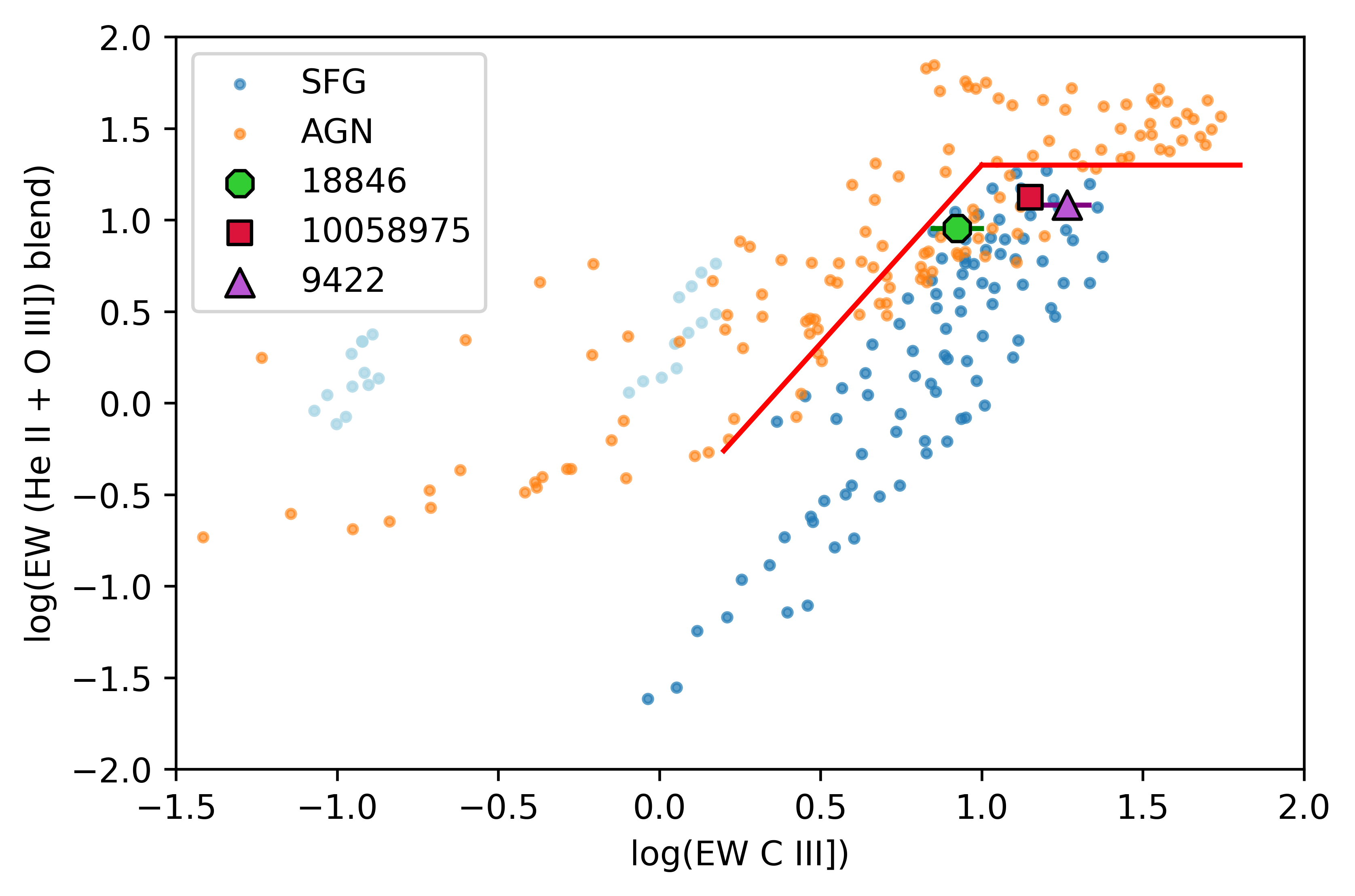}
        \label{fig:final_b}
    \end{subfigure}\\
    \begin{subfigure}{.5\textwidth}
        \centering
        \includegraphics[width=.9\linewidth]{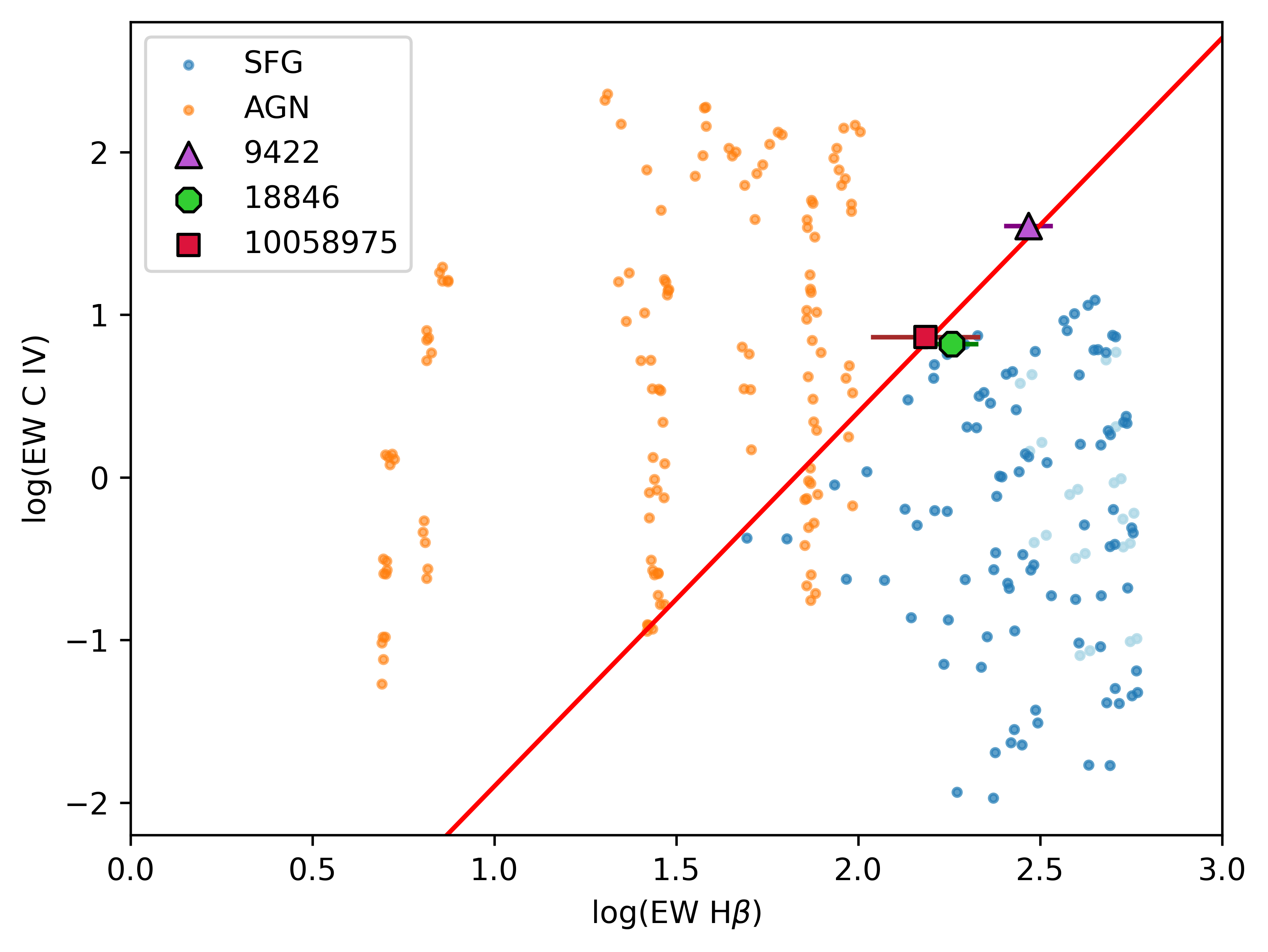}
        \label{fig:final_c}
    \end{subfigure}%
    \caption{The galaxies with NIRSpec IDs 9422, 18846, and 10058975 are displayed on some EW vs. EW plots using their rest frame EW measurements and uncertainties. The first plot (top left) only uses optical lines, the second (top right) only UV lines, and the last uses one UV and one optical line. The red solid separation lines are defined by the Equations \ref{eq:data_on_EW_plot_separation1}, \ref{eq:data_on_EW_plot_separation2a}, \ref{eq:data_on_EW_plot_separation2b}, and \ref{eq:data_on_EW_plot_separation3}.}
    \label{fig:data_on_EW_plots}
\end{figure*}

On the plot using \mbox{[\ion{O}{III}] $\lambda$5007} and H$\beta$, two optical lines, all three galaxies seem to be in the SFG region. Looking at H$\beta$ individually, values indicate that the galaxies are more similar to SFG models, especially galaxy 9422 which is the furthest from the AGN region. This is true for all other plots using H$\beta$. Galaxy 9422 exhibits the highest \mbox{[\ion{O}{III}] $\lambda$5007} EW of the three, closely aligning with the SFG models on both axes. In contrast, the other two galaxies lack a sufficient \mbox{[\ion{O}{III}] $\lambda$5007} EW to be classified as SFG based on that axis alone.

The plot using the \mbox{\ion{He}{II} $\lambda$1640 + \ion{O}{III}] blend} and \ion{C}{III}] positions the galaxies near the highest possible values of the SFG models on both axes. They also fall well within the AGN region, suggesting they may be more akin to AGN than typical star-forming galaxies.

On the \ion{C}{IV} over H$\beta$ graph, the H$\beta$ EW shows the same as previously. The \mbox{\ion{C}{IV} EW} (similarly as \mbox{\ion{C}{III}]}) places the galaxies near the highest values of the SFG models, and positions galaxy 9422 even higher, indicating the possibility of it being an AGN.

Even though these diagrams were not able to conclusively identify our three sources of interest as AGN, they clearly have the potential to be very useful due to the distinct AGN and SFG regions in the new parameter space.

\subsection{Possible use of UV-Optical ratio diagrams} \label{sec:UV_Optical_ratio_diagram}

In this section we discuss using UV-Optical ratios: \mbox{\ion{He}{II} $\lambda$1640 / H$\beta$} vs. \mbox{(\ion{C}{III}] + \ion{C}{IV}) / H$\beta$}. The utilization of ultraviolet and optical lines in the same ratio is not straightforward. Dust affects light differently depending on the wavelength (\cite{Calzetti_2000_DustAttenuation}).
To combine them in the same graph, we would need to calculate the dust correction factor ($\textnormal{K}_{\textnormal{dust}}$) and check if the factor is comparable to the separation between the models.

\begin{figure*}
    \begin{subfigure}{.49\textwidth}
        \centering
        \includegraphics[width=\linewidth]{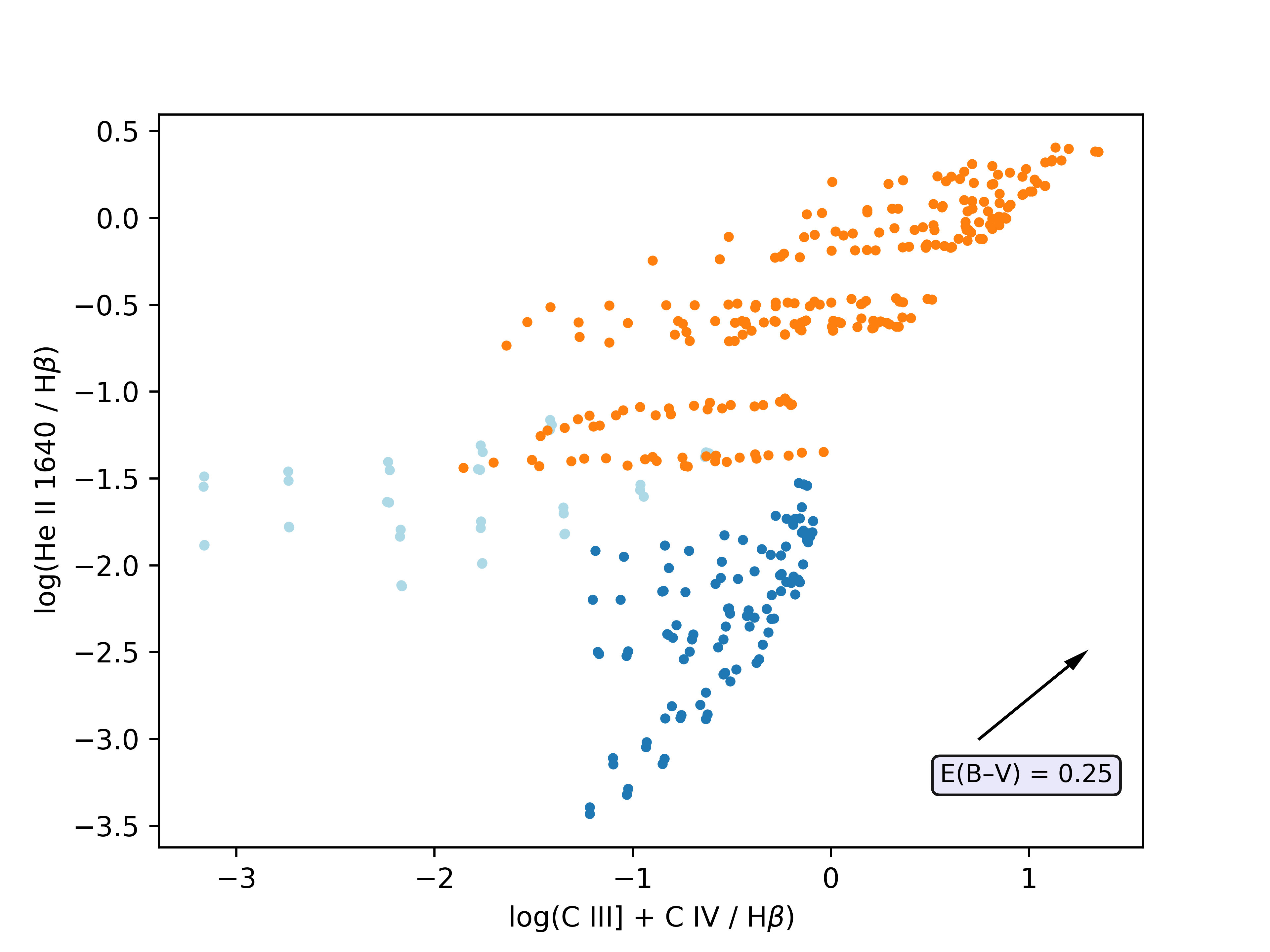}
    \end{subfigure}
    \begin{subfigure}{.49\textwidth}
        \centering
        \includegraphics[width=\linewidth]{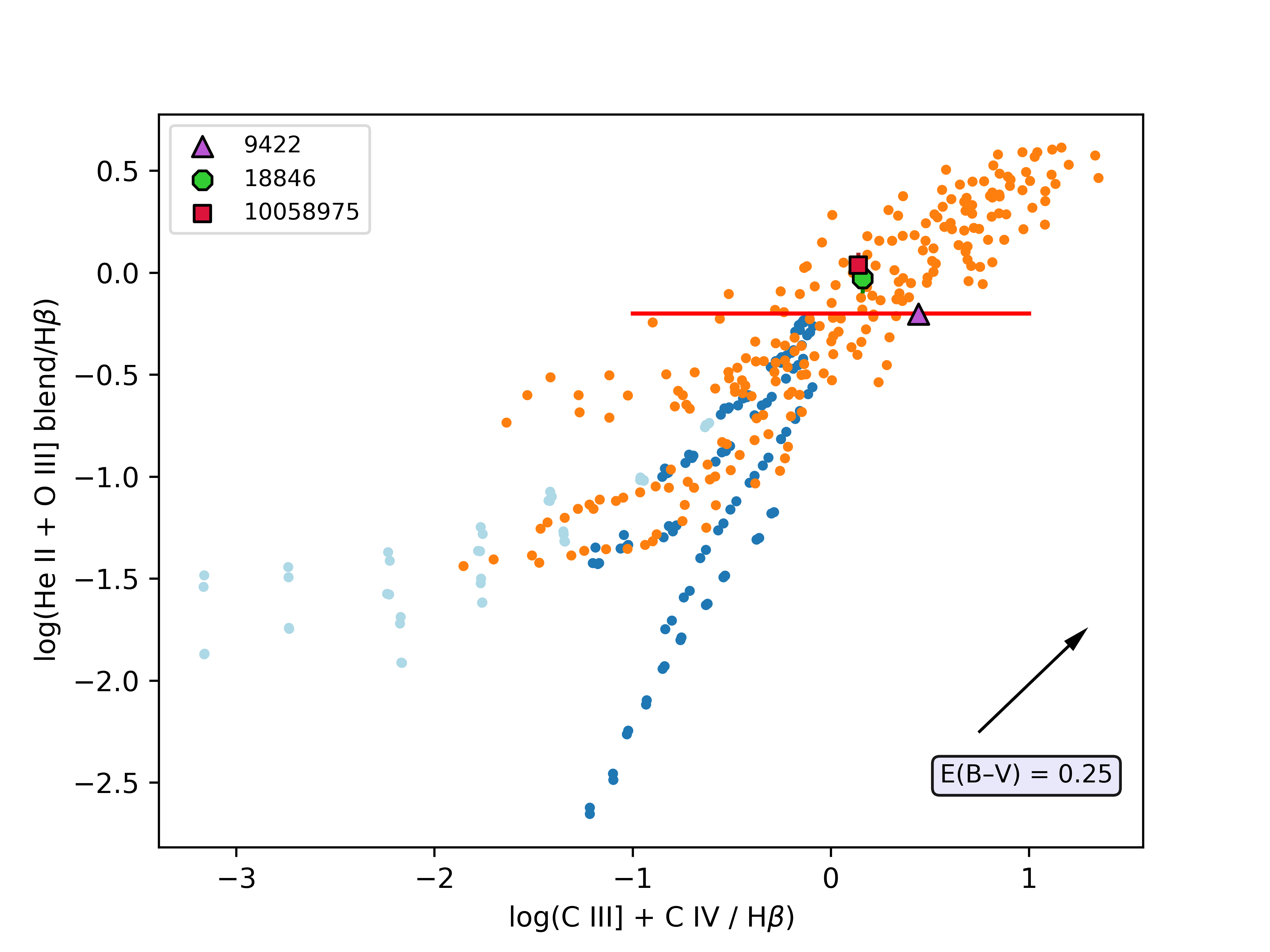}
    \end{subfigure}\\
\caption{UV-Optical ratio diagrams displaying the models of \protect\cite{p11_NakajimaMaiolino2022} (coloured as previously) and an arrow corresponding to the line ratio correction factor for E(B–V) = 0.25 assuming a Calzetti dust attenuation law. The diagram on the right was created by changing line \mbox{\ion{He}{II} $\lambda$1640} on the left diagram to the blended line \mbox{\ion{He}{II} + \ion{O}{III}]} since this is what can be detected with the NIRSpec prism.  Our sources of interest are marked with a purple triangle, a green circle and a red square.}
\label{fig:UV-Optical_ratios}
\end{figure*}

In order to estimate this dust correction factor, we used the dust attenuation curve from \cite{Calzetti_2000_DustAttenuation}. We calculated $\textnormal{K}_{\textnormal{dust}}$ assuming a value of 0.25 for the color excess. For different values of the color excess, the correction factor would scale linearly in our log-log plot. This correction factor was plotted as a black arrow in Figure \ref{fig:UV-Optical_ratios}, such that the full extent of the arrow is $\mathrm{log}(\textnormal{K}_{\textnormal{dust}})$ dex. In reality, dust effects on the spectrum are often less pronounced than shown here.

This diagram has very distinct AGN and SFG regions; however, it uses the \mbox{\ion{He}{II} $\lambda$1640} line on its own, while in the prism catalogue it appears blended with \mbox{\ion{O}{III}] $\lambda$1665}. Therefore, on the right side of Figure \ref{fig:UV-Optical_ratios}, we plot a similar diagram that uses the blended line instead. Since galaxies 18846 and 10058975 had all the necessary lines in the catalogue and we estimated the CIII] emission line flux for galaxy 9422, we mark them on the plot. We display a solid red line on the diagram aiming to separate a mixed region from a region only occupied by AGN models. Galaxies with:
\begin{equation}
\log \left( \frac{\text{\ion{He}{II} + \ion{O}{III}]}}{\text{H$\beta$}} \right) > -0.2
\label{eq:UV_OPtical_separation_line}
\end{equation}

can be reliably identified as AGN while those with values below this fall into a mixed region. This diagram shows more mixing between AGN and non-AGN compared to the one presented on the left side, but all three galaxies still fall within the AGN region. 

Since UV lines are more heavily affected by dust extinction compared to optical lines, the raw (uncorrected) line ratio will typically appear lower than the intrinsic (dust-free) value. This is because the UV flux will be reduced much more than the optical flux due to the higher attenuation at shorter wavelengths. Consequently, after applying dust corrections, the galaxies will shift to higher values on both axes, and will remain within the AGN region. We can thus identify 9422, 18846 and 10058975 as AGN candidates using this method.

Based on Figure \ref{fig:UV-Optical_ratios}, the \mbox{\ion{He}{II} $\lambda$1640 / H$\beta$} vs. \mbox{(\ion{C}{III}] + \ion{C}{IV}) / H$\beta$} diagram might be a useful diagnostic, since both axes show a good separation between the AGN and SFG models. However, it is important to note that this diagram can only be useful if the color excess is very small. This excess would need to be accurately estimated. A possible way to include this correction into the analysis would be to use the Balmer decrement method. This method consists of measuring the observed intensities of H$\alpha$ and H$\beta$ lines and comparing them to the expected intrinsic ratio.

While the Balmer decrement is a valuable method for correcting dust extinction, it has its limitations, particularly in the context of AGN. Strong ionizing radiation from the AGN's central black hole can alter the Balmer line ratios, complicating the dust extinction calculation. This makes it crucial to know the galaxy's nature before applying the method. However, since we are using this method to identify the galaxy type, it becomes more challenging to use effectively.

\section{AGN Determination from Photometry} \label{sec:photometry}
In this section we complement our analysis of the spectroscopy data with the study of the photometry data of the JADES dataset, mainly focusing on sources 9422 and 18846, following similar steps to the analysis done by \cite{p6_Juodzbalis2023}. For the identification of AGN galaxies using photometry, we use the routine EAZY (\cite{EAZY_Brammer_2008}). We apply this method to a series of spectral energy distribution (SED) fits to identify sources that are best fit by AGN SED templates. Furthermore, we present some values obtained with this method for a compilation of AGN sources from the literature.

\subsection{\texttt{EAZY} selection method} \label{subsec:EAZY}
\cite{p6_Juodzbalis2023} developed a photometric method that can identify high probability AGN candidates in the early Universe. Their goal was to create a method which would aid the search for strong AGN candidates that could then be followed up by spectroscopy. 

\texttt{EAZY} was run with two sets of templates. One of them, referred to as FSPS+Larson, consist of the 12 default Flexible Stellar Population Synthesis templates (\cite{ConroyGunn_2010_FSPS_templates_code}) with the addition of 6 templates from \cite{Larson_2023_spectral_templates}. The other set contains direct collapse black hole (DCBH) SED templates from \cite{p11_NakajimaMaiolino2022}. The accuracy of the method is highly dependent on the specific templates used with EAZY, as different templates can influence the resulting classifications and parameter estimates.

The SED fitting is done twice, once only with the FSPS+Larson templates and then another time with the Nakajima \& Maiolino set added on top of the FSPS+Larson set. The filters we use are: F090W, F115W, F150W, F200W, F277W, F356W, F444W, F335M, and F410M. We used the AGN fraction parameter, $f_{\text{AGN}}$, defined as (\cite{p6_Juodzbalis2023})
\begin{equation}
    f_{\text{AGN}} = \sum W_{i},
\end{equation}
where $W_{i}$ is the weight of each AGN template. The $f_{\text{AGN}}$ parameter, therefore, describes the relative weight of the AGN versus non-AGN templates in the best fit.

For the selection of candidates, reduced chi-squared values of the first and second fits, $\chi^{2}_{R1} \: and \: \chi^{2}_{R2}$,  are also taken into account. The second fit is the one with the added AGN templates.  The selection criteria defined by \cite{p6_Juodzbalis2023} are as follows:
\begin{enumerate}
    \item $\chi^{2}_{R2}$ < 3 ('robust' classification),
    \vspace{8pt}
    \item $f_{\text{AGN}}$ > 0.5 (source is dominantly fitted by an AGN model),
    \vspace{0.1pt}
    \item $\chi^{2}_{R2}$ < $\left(\chi^{2}_{R1} - 0.5\right)$, ensures an improved fit when the AGN models are added to the templates,

\end{enumerate}

Note that this method is more sensitive to unobscured, Type I, AGN due to those being the easiest to identify with photometry and therefore, will not give a full description of the AGN population in the data. However, this methods can identify high priority sources for future spectroscopic followup.

\subsection{The JADES dataset}\label{subsec:EAZY_on_JADES}
We applied the aforementioned \texttt{EAZY} SED fitting on the sources in the JADES NIRCam photometric catalogue, excluding the ones that do not have data in all the needed filters. This gives us 177 sources.  

\begin{figure*}
    \begin{subfigure}{.45\textwidth}
        \centering
        \includegraphics[width=\linewidth]{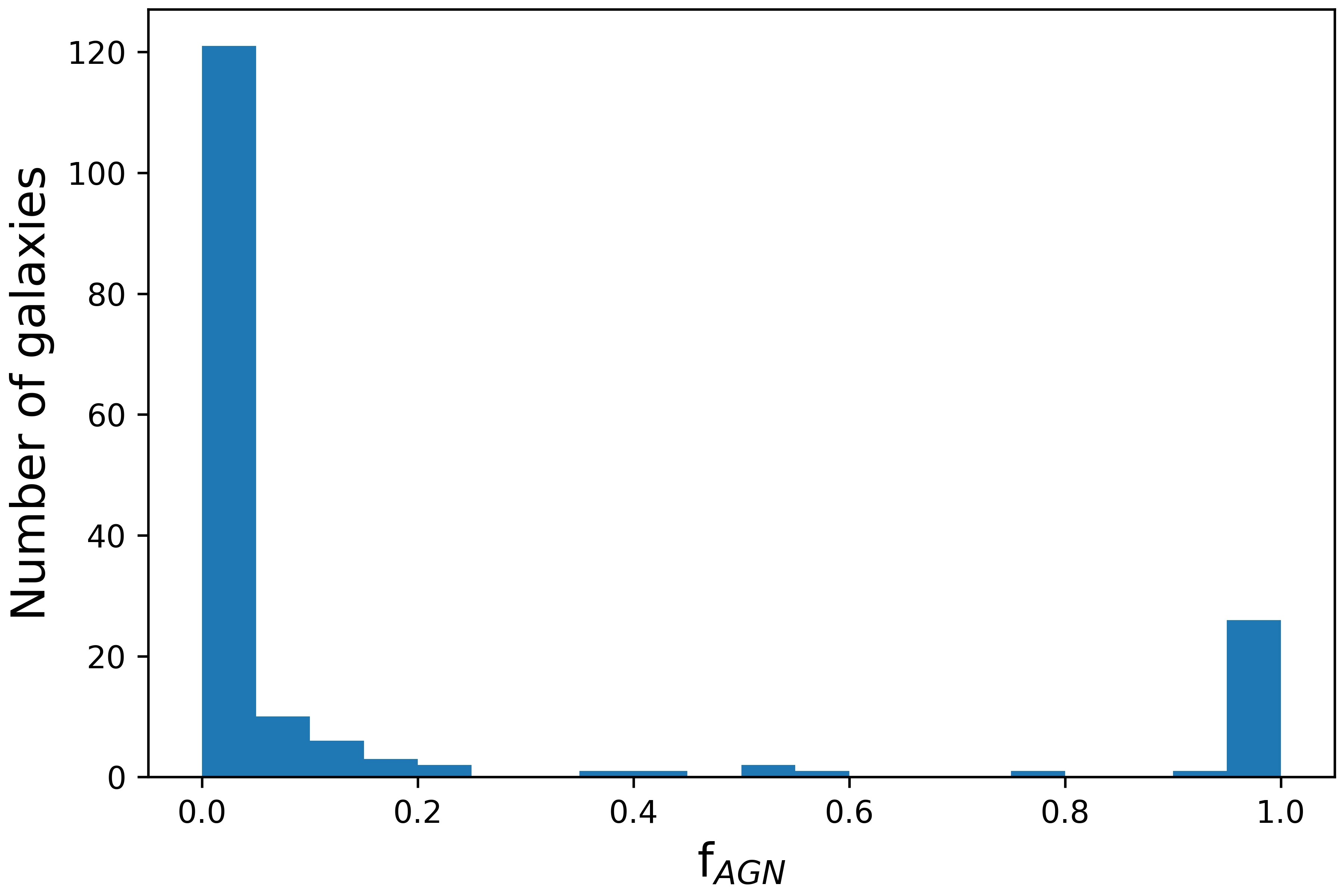}
    \end{subfigure}
    \begin{subfigure}{.45\textwidth}
        \centering
        \includegraphics[width=\linewidth]{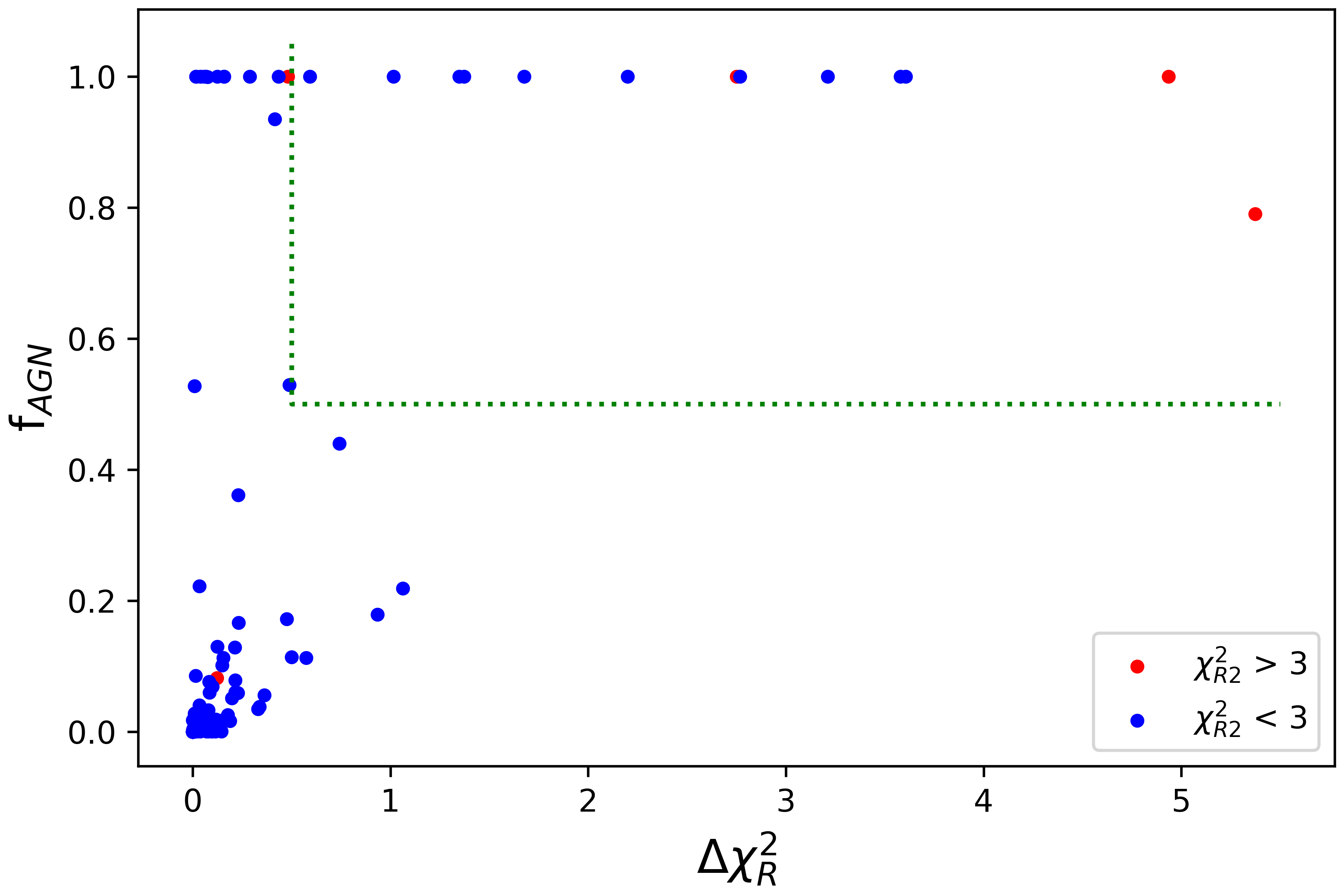}
    \end{subfigure}\\
\caption{A histogram of $f_{\text{AGN}}$ values for the JADES dataset (on the left) and a plot showing the $f_{\text{AGN}}$ values against the difference in $\chi^{2}_{R}$ between the two fits (on the right). Data points that satisfy the 'robust' classification by having their $\chi^{2}_{R2}$ value below 3, are marked with blue dots, while the the ones that do not satisfy this condition are coloured red. The green dotted lines separate the part of the parameter space where points satisfy the other two conditions: \mbox{$f_{\text{AGN}}$ > 0.5} and \mbox{$\Delta\chi^{2}$ > 0.5}.}
\label{fig:f_AGN_JADES}
\end{figure*}

On the left side of Figure \ref{fig:f_AGN_JADES}, we plot a histogram of the $f_{\text{AGN}}$ values in the JADES data. Most of the values are around 0 and with another significant fraction of them close to 1.
On the right side, we show the $f_{\text{AGN}}$ values plotted against the difference in the $\chi^{2}_{R}$ values between the two SED fits. Data points that satisfy the 'robust' classification by having their $\chi^{2}_{R2}$ value below 3, are marked with blue dots. The red data points do not satisfy this condition. The part of the parameter space where points satisfy the other two conditions: \mbox{$f_{\text{AGN}}$ > 0.5} and \mbox{$\Delta\chi^{2}$ > 0.5}, is above and to the right of the green dotted lines.

The $f_{\text{AGN}}$ values in the data are mostly concentrated close to 1 or 0, this is due to \texttt{EAZY}'s preference towards single template models. A value close to 1, therefore, does not necessarily imply that the source contains an AGN, as the conventional FSPS+Larson templates may lack the necessary versatility in their emission line characteristics to accurately replicate the observed data.

\begin{figure*}
    \begin{subfigure}{.49\textwidth}
        \centering
        \includegraphics[width=\linewidth]{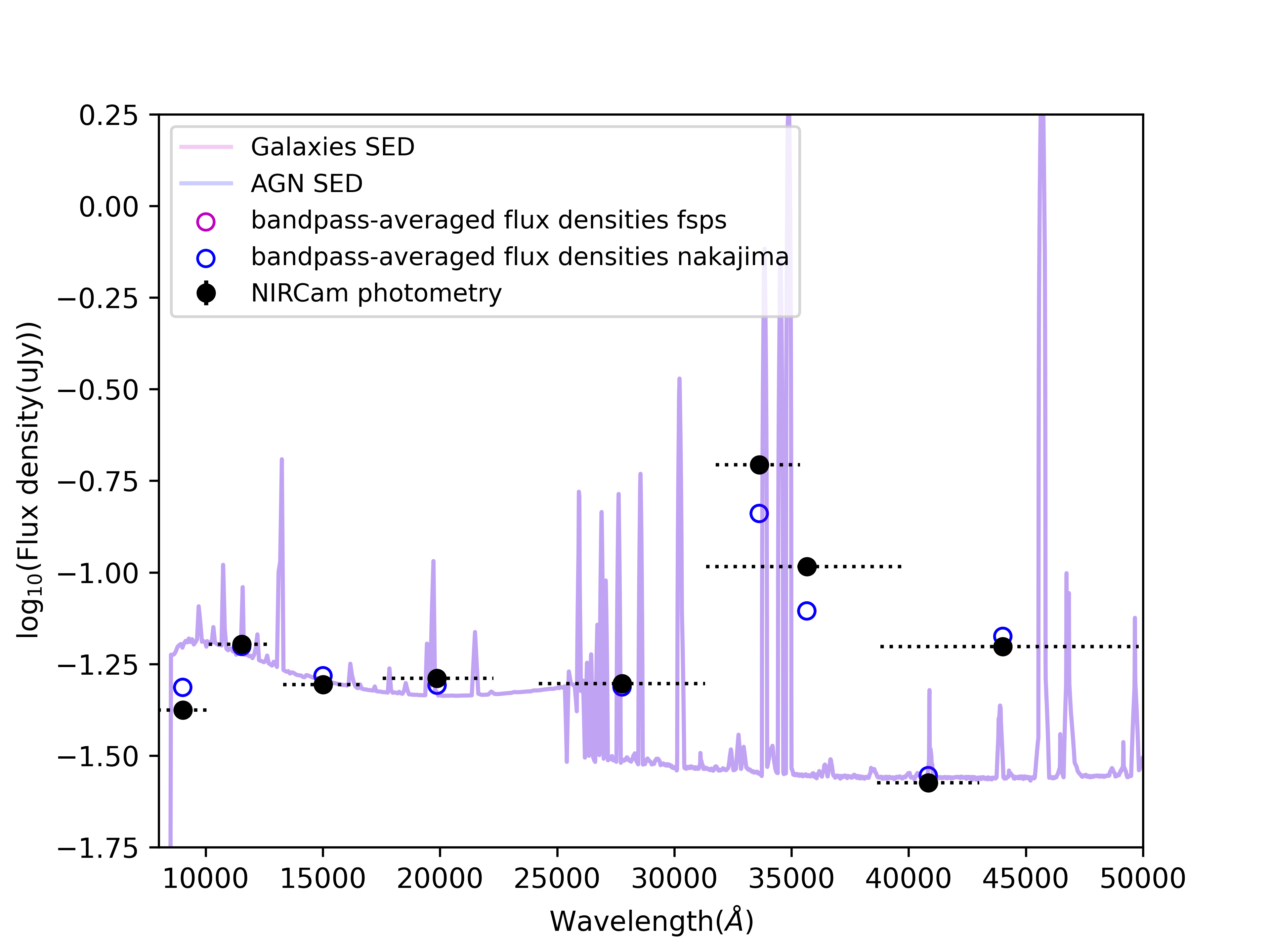}
    \end{subfigure}
    \begin{subfigure}{.49\textwidth}
        \centering
        \includegraphics[width=\linewidth]{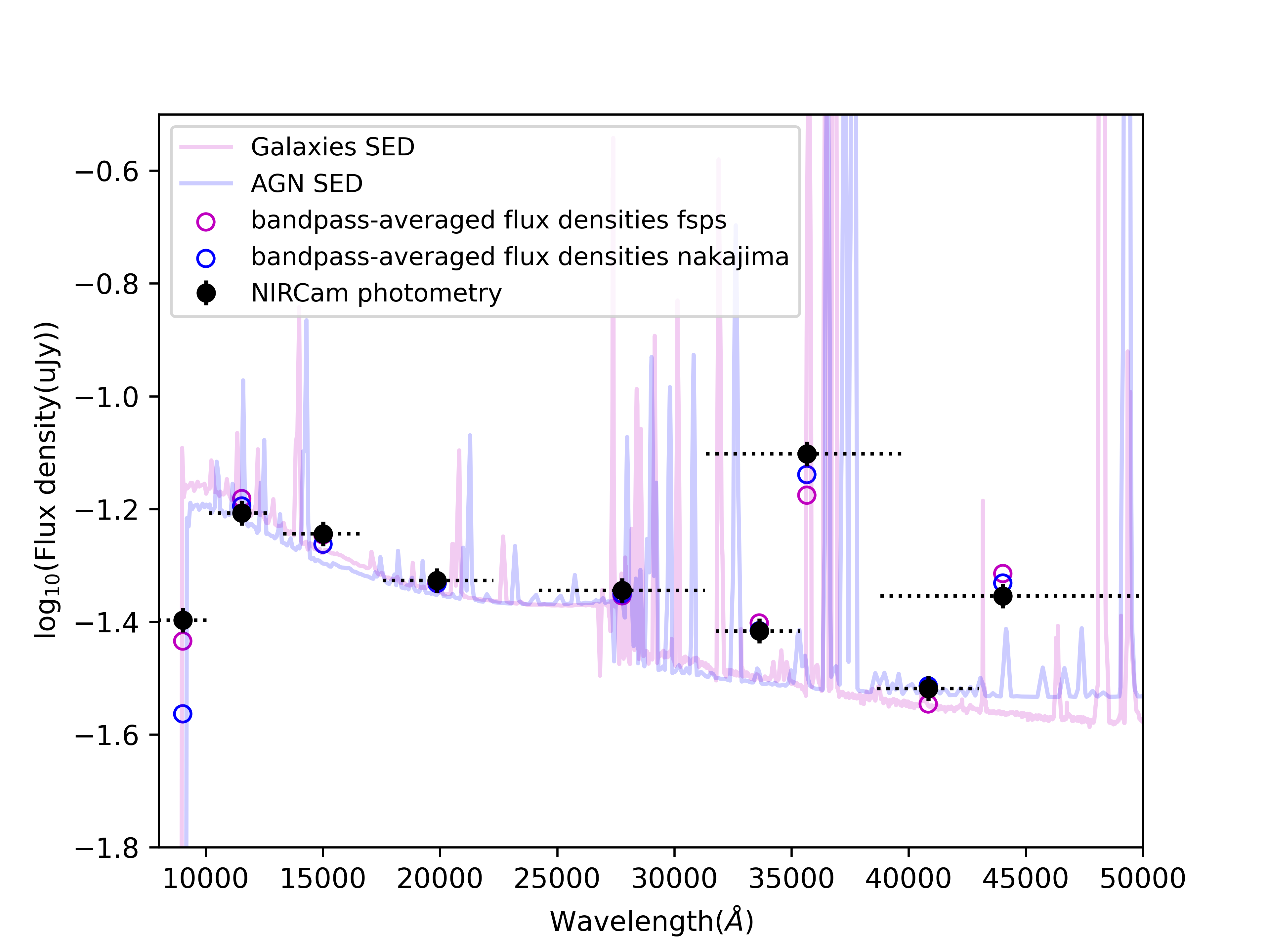}
    \end{subfigure}\\
\caption{SED fits for our two sources of interest, 9422 (on the left) and 18846 (on the right). The y axis shows the flux density in $\mu$Jy while the x axis is the observed wavelength in \text{\AA}. The FSPS+Larson and the added Nakajima \& Maiolino fits are represented by the faint magenta and blue curves. The black data points are from the NIRCam filter measurements and the black horizontal dotted lines showcase the wavelength range of the filters. The magenta and blue circles correspond to the bandpass-averaged flux densities predicted by the galaxy and the AGN fits, respectively. }
\label{fig:SED_fits}
\end{figure*}

Out of the 177 JADES sources, 31 had an $f_{\text{AGN}}$ value above 0.5. These values can be seen with the NIRSpec and NIRcam IDs of the corresponding sources in Table \ref{tab:f_AGN_JADES}. Note that 2 sources were excluded because they did not have valid $\chi^{2}_{R}$ values. Sources were grouped based on the number of criteria they satisfy out of the 3 mentioned in Section \ref{subsec:EAZY}. The first 10 galaxies satisfied all three criteria, and therefore, can be considered strong AGN candidates. Amongst these, we can see one of our sources of interest which has a NIRSpec ID of 18846 as the candidate with the lowest reduced chi squared. The second group satisfy 2 out of the 3 criteria, because the improvement of the fit is not sufficient, and the sources in the third group cannot be classified as 'robust' due to their  $\chi^{2}_{R2}$ values being above 3. 
 
\begin{table}
\centering
\begin{tabular}{ccccc}
\hline
\textbf{NIRCam ID}   & \textbf{NIRSpec ID}   & \textbf{$f_{\text{AGN}}$}   & \textbf{$\chi^{2}_{R2}$}   & \textbf{$\Delta\chi^{2}_{R}$}  \\ \hline \hline
\multicolumn{5}{c}{\textbf{$f_{\text{AGN}}$ \textgreater{} 0.5 and $\chi^{2}_{R2}$ \textless{} 3 and $\Delta\chi^{2}_{R}$ \textgreater{} 0.5}} \\ \hline
131688               & \textbf{00018846}     & 1.00            & 0.59                  & 1.68          \\
201680               & 00007892              & 1.00            & 0.85                  & 0.59          \\
110987               & 00008113              & 1.00            & 1.56                  & 2.77          \\
132780               & 00019342              & 1.00            & 1.59                  & 1.37          \\
113461               & 00009343              & 1.00            & 1.74                  & 3.61          \\
\textbf{203112}               & 00009452              & 1.00            & 1.84                  & 2.20          \\
201730               & 00007938              & 1.00            & 1.86                  & 1.02          \\
107462               & 00006519              & 1.00            & 2.00                  & 1.35          \\
114213               & 00009743              & 1.00            & 2.25                  & 3.58          \\
210963               & 00019606              & 1.00            & 2.74                  & 3.21          \\ \hline \hline
\multicolumn{5}{c}{\textbf{$f_{\text{AGN}}$ \textgreater{} 0.5 and $\chi^{2}_{R2}$ \textless{} 3, but $\Delta\chi^{2}_{R}$ \textless{} 0.5}}    \\ \hline
98452                & 00003322              & 0.53            & 0.85                  & 0.49          \\
197791               & 00004009              & 1.00            & 0.06                  & 0.43          \\
113523               & 10001946              & 1.00            & 0.15                  & 0.16          \\
111668               & 10009320              & 1.00            & 1.09                  & 0.12          \\
198974               & 00005040              & 1.00            & 0.79                  & 0.07          \\
139168               & -                     & 1.00            & 0.56                  & 0.07          \\
109645               & 00007507              & 1.00            & 0.71                  & 0.06          \\
128470               & 00017251              & 1.00            & 0.99                  & 0.06          \\
96216                & -                     & 1.00            & 0.98                  & 0.04          \\
131067               & 10014177              & 1.00            & 0.64                  & 0.02          \\
103332               & 10000618              & 0.53            & 0.71                  & 0.01          \\
105635               & 00005759              & 0.94            & 1.31                  & 0.42          \\
107324               & -                     & 1.00            & 1.56                  & 0.16          \\
130158               & 10014220              & 1.00            & 1.89                  & 0.29          \\
128771               & -                     & 1.00            & 1.94                  & 0.02          \\ \hline \hline
\multicolumn{5}{c}{\textbf{$f_{\text{AGN}}$ \textgreater{} 0.5, but $\chi^{2}_{R2}$ \textgreater{} 3}}                       \\ \hline
\textbf{103349}              & 10000626              & 1.00            & 3.16                  & 0.48          \\
209707               & 00018090              & 1.00            & 9.55                  & 4.93          \\
\textbf{114790}             & 00010073              & 1.00            & 9.99                  & 2.75          \\
\textbf{110905}             & 00008083              & 0.79            & 10.62                 & 5.37          \\ \hline
\end{tabular}
\caption{Sources with $f_{\text{AGN}}$ values above 0.5 in the JADES photometric catalogue. The first 10 sources satisfy the 3 criteria discussed in Section \ref{subsec:EAZY}, and are therefore strong AGN candidates. One of our sources of interest, 18846, is marked with a bold NIRSpec ID. The galaxies with bold NIRCam IDs have already been identified as AGN and more information can be found about them in Table \ref{tab:AGN_list}. Note that the $f_{\text{AGN}}$ values are written to a precision of 3 significant digits.}
\label{tab:f_AGN_JADES}
\end{table}

\subsection{NIRSpec Sources 9422 and 18846}\label{subsec:EAZY_9422_18846}

The $f_{\text{AGN}}$ values found for the sources with NIRSpec spectra, including those with IDs 9422 and 18846 (NIRCam IDs 113585 and 131688) are 0.0 and 0.9999461. This shows that 18846 is an excellent AGN candidate. We note again that the 0.0 value for galaxy 9422 does not mean that it does not have an AGN, only that this method was not able to identify it as a strong candidate.

For the specific case of galaxy 9422, \cite{p15_cameron2023} analyses its spectrum. It shows a steep turnover in the ultraviolet continuum, which is indicative of two-photon emissions from neutral hydrogen and it is not typically seen in AGN or X-ray binaries. On the other hand, \cite{p14_scholtz2023jades} classifies it as an AGN using the two Helium lines, \cite{Tacchella_2024_GS9422} classifies it as a galaxy with an obscured type-2 AGN, and \cite{Li_2024_GS9422}) states that it is combination of young metal-poor stars and a low-luminosity AGN.

For 18846, the first SED fit with the FSPS+Larson templates gave a $\chi^{2}_{R}$ of 2.25, while the second fit with the added Nakajima\&Maiolino templates gave a better value of 0.59. The estimated photometric redshifts were 6.34, and 6.5 for the two fits, respectively. These compare well to the actual spectroscopic redshift of 6.342.
Therefore, 18846 satisfies all the criteria in Section \ref{subsec:EAZY} for being a strong AGN candidate. This result gives motivation to further investigate galaxy 18846. Especially considering its spectral resemblance to sources 9422 and 10058975 which have already been selected as AGN by multiple studies (\cite{p14_scholtz2023jades}, \cite{Tacchella_2024_GS9422}, \cite{Li_2024_GS9422}).

The flux values in the 8 filters used for the SED fitting for our two sources can be seen in Table \ref{tab:filters_9422_18846}. Note, however, that the fitting was run with a minimum flux error of 5\%, accounting for uncertainties in the calibration.

\begin{table}
\centering
\begin{tabular}{ccc}
\hline
\textbf{Filter} & \textbf{9422 [$\mu$Jy]}   & \textbf{18846 [$\mu$Jy]}  \\ \hline \hline
\textbf{F090W}  & 0.0421 $\pm$ 0.0021 & 0.0401 $\pm$ 0.0020 \\
\textbf{F115W}  & 0.0636 $\pm$ 0.0032 & 0.0620 $\pm$ 0.0031 \\
\textbf{F150W}  & 0.0494 $\pm$ 0.0025 & 0.0570 $\pm$ 0.0028 \\
\textbf{F200W}  & 0.0514 $\pm$ 0.0026 & 0.0471 $\pm$ 0.0024 \\
\textbf{F277W}  & 0.0497 $\pm$ 0.0025 & 0.0453 $\pm$ 0.0023 \\
\textbf{F356W}  & 0.1037 $\pm$ 0.0052 & 0.0790 $\pm$ 0.0040 \\
\textbf{F444W}  & 0.0627 $\pm$ 0.0031 & 0.0442 $\pm$ 0.0022 \\
\textbf{F335M}  & 0.1965 $\pm$ 0.0098 & 0.0384 $\pm$ 0.0019 \\
\textbf{F410M}  & 0.0266 $\pm$ 0.0013 & 0.0303 $\pm$ 0.0015 \\ \hline
\end{tabular}
\caption{Average flux densities for our two sources in units of $\mu$Jy across the filters used in the SED fitting with \texttt{EAZY}. Errors in the flux values are displayed, however, note that the fitting was run with a minimum flux error of 5\% to account for equipment calibration uncertainties.}
\label{tab:filters_9422_18846}
\end{table}

The SED fits for the two galaxies of interest can be seen on Figure \ref{fig:SED_fits}. Note that the units and the scale of the y axes are different than in Figure \ref{fig:spectra}. The black data points are the NIRCam photometry values for the mentioned filters and the black horizontal dotted lines showcase the wavelength range of these filters. The circles represent the bandpass-averaged flux densities predicted by the SED fits. The blue line and circles represent the result of the second fit done with the added Nakajima\&Maiolino templates, while the magenta line and circles show the results of the first fit.

We can see that for 9422, the two fits are overlapping, meaning that adding the AGN templates did not change the best fit, hence the low $f_{\text{AGN}}$ value. For galaxy 18846, the addition of the Nakajima\&Maiolino set improved the fit, although the improvement is marginal, primarily due to the filters affected by emission lines like F356W and F444W. As a result, it has a high $f_{\text{AGN}}$ value.

\subsection{Previously identified AGN}\label{subsec:known_AGN_literature}

Furthermore, we compiled a list of AGN that have been identified using various methods (such as the BPT and the VO87 diagrams or the presence of broad lines) and obtained $f_{\text{AGN}}$ values for them. These are presented in Table \ref{tab:AGN_list}. The sources with high $f_{\text{AGN}}$ values are written in bold. 

We can see that only 4 of the 32 confirmed AGN sources have high $f_{\text{AGN}}$ values. This shows once more that the $f_{\text{AGN}}$ does not correspond to a physical quantity, and galaxies with low values can still be identified as AGN. As is the case of galaxy 9422 which, though has a very low $f_{\text{AGN}}$ value, is claimed to be an obscured AGN by \cite{Tacchella_2024_GS9422} and \cite{Li_2024_GS9422}, as already mentioned. This method might be able to identify some sources which have a high probability of containing an AGN, but it also disregards other possible candidates.

In conclusion, although this photometry analysis cannot be used as a completely independent diagnostic tool for finding AGN in galaxies, it can still be a good resource to accompany spectroscopic analysis. By looking first at the spectroscopy of the galaxies that pass our criteria, we can prioritize the galaxies that have a higher probability of being an AGN. Nonetheless, we should not claim that a galaxy is not an AGN just by the results of the photometric analysis, but would need further confirmation from spectroscopy to back up the finding of AGN in these systems.

\begin{table*}
\begin{tabular}{ccccc}
\hline
\multicolumn{2}{c}{\textbf{IDs}}               & \multirow{2}{*}{\textbf{$f_{\text{AGN}}$}} & \multirow{2}{*}{\textbf{Selection method}}                                       & \multirow{2}{*}{\textbf{References}}                                                \\
\textbf{NIRCam ID}    & \textbf{NIRSpec ID}    &                                  &                                                                                  &                                                                                     \\ \hline
113585                & 00009422               & 0.00                             & HeII1640 and HeII4686                                                            & \cite{p14_scholtz2023jades}                                                                 \\
103366                & 00004902               & 0.11                             & S2-VO87*                                                                         & \cite{p14_scholtz2023jades}                                                                 \\
108715                & 00007099               & 0.02                             & S2-VO87*                                                                         & \cite{p14_scholtz2023jades}                                                                 \\
110238                & 00007762               & 0.18                             & high ion.                                                                        & \cite{p14_scholtz2023jades}                                                                 \\
\textbf{110905}       & \textbf{00008083}      & \textbf{0.79}                    & \begin{tabular}[c]{@{}c@{}}high ion. \& HeII 4686 \\ \& broad lines\end{tabular} & \begin{tabular}[c]{@{}c@{}}\cite{p14_scholtz2023jades} \&\\ \cite{p4_maiolino2023jades}\end{tabular} \\
111661                & 00008456               & 0.01                             & HeII4686                                                                         & \cite{p14_scholtz2023jades}                                                                 \\
112500                & 00008880               & 0.00                             & N2-BPT                                                                           & \cite{p14_scholtz2023jades}                                                                 \\
\textbf{203112}       & \textbf{00009452}      & \textbf{1.00}                    & N2-BPT                                                                           & \cite{p14_scholtz2023jades}                                                                 \\
\textbf{114790}       & \textbf{00010073}      & \textbf{1.00}                    & HeII4686                                                                         & \cite{p14_scholtz2023jades}                                                                 \\
208642                & 00016745               & 0.11                             & S2-VO87*                                                                         & \cite{p14_scholtz2023jades}                                                                 \\
208919                & 00017072               & 0.05                             & HeII1640                                                                         & \cite{p14_scholtz2023jades}                                                                 \\
209348                & 00017670               & 0.00                             & HeII4686                                                                         & \cite{p14_scholtz2023jades}                                                                 \\
137667                & 00021842               & 0.00                             & high ion                                                                         & \cite{p14_scholtz2023jades}                                                                 \\
\textbf{103349}       & \textbf{10000626}      & \textbf{1.00}                    & high ion. \& HeII4686                                                            & \cite{p14_scholtz2023jades}                                                                 \\
198071                & 10035295               & 0.00                             & HeII1640                                                                         & \cite{p14_scholtz2023jades}                                                                 \\
200002                & 10036017               & 0.01                             & N2-BPT \& S2-VO87                                                                & \cite{p14_scholtz2023jades}                                                                 \\
197348                & 10013704               & 0.00                             & broad lines                                                                      & \cite{p4_maiolino2023jades}                                                                \\
\multicolumn{2}{c}{\textbf{GLASS 160133}}               & \textbf{1.00}                    & broad lines                                                                      & \cite{Harikane2023}                                                                \\
\multicolumn{2}{c}{GLASS 150029}               & 0.00                             & broad lines                                                                      & \cite{Harikane2023}                                                                \\
\multicolumn{2}{c}{CEERS 00746 or  CEERS 3210} & 0.00                             & broad lines                                                                      & \begin{tabular}[c]{@{}c@{}}\cite{Harikane2023} \\ \& \cite{Kocevski2023}\end{tabular}                      \\
\multicolumn{2}{c}{CEERS 00672}                & 0.00                             & broad lines                                                                      & \cite{Harikane2023}                                                                \\
\multicolumn{2}{c}{CEERS 02782 or  CEERS 1670} & 0.01                             & broad lines                                                                      & \begin{tabular}[c]{@{}c@{}}\cite{Harikane2023} \\ \& \cite{Kocevski2023}\end{tabular}                      \\
\multicolumn{2}{c}{CEERS 00397}                & 0.01                             & broad lines                                                                      & \cite{Harikane2023}                                                                \\
\multicolumn{2}{c}{CEERS 01236}                & 0.00                             & broad lines                                                                      & \cite{Harikane2023}                                                                \\
\multicolumn{2}{c}{GOODS-S-13971}              & 0.00                             & broad lines                                                                      & \cite{Matthee_2024}                                                                 \\
\multicolumn{2}{c}{MSA 10686}                  & 0.00                             & broad lines                                                                      & \cite{Greene_2023}                                                                  \\
\multicolumn{2}{c}{MSA 13821}                  & 0.00                             & broad lines                                                                      & \cite{Greene_2023}                                                                  \\
\multicolumn{2}{c}{MSA 23608}                  & 0.00                             & broad lines                                                                      & \cite{Greene_2023}                                                                  \\
\multicolumn{2}{c}{MSA 32265}                  & 0.00                             & broad lines                                                                      & \cite{Greene_2023}                                                                  \\
\multicolumn{2}{c}{MSA 33437}                  & 0.00                             & broad lines                                                                      & \cite{Greene_2023}                                                                  \\
\multicolumn{2}{c}{MSA 35488}                  & 0.00                             & broad lines                                                                      & \cite{Greene_2023}                                                                  \\
\multicolumn{2}{c}{MSA 38108}                  & 0.00                             & broad lines                                                                      & \cite{Greene_2023}                                                                  \\ \hline
\end{tabular}
\caption{List of identified AGN with the obtained $f_{\text{AGN}}$ values (to 3 significant digits), identification methods, and references. Sources that satisfy the $f_{\text{AGN}}$ > 0.5 criterion are written in bold. The first three of these also appear in Table \ref{tab:f_AGN_JADES}.}
\label{tab:AGN_list}
\end{table*}

\section{Sensitivity of diagrams} \label{subsec:sensitivity}
In this section we aim to further discuss the use of the diagrams proposed in the previous sections (Sections \ref{subsec:UV_alternatives}, \ref{subsec:EW_plots}, and \ref{subsec:EW_of_three_galaxies}) by exploring their sensitivity. We do this by evaluating the needed signal-to-noise ratio (SNR) that would make the parameter space a useful diagnostic, then we calculate the needed source brightness required for detections with such precision. Finally, we make assumptions related to the exposure time dependence of this required source brightness.

\subsection{Required Signal to Noise Ratio Values}\label{subsec:SNR}
 
We found that a SNR of 5 on the line fluxes and the continuum levels is sufficient to motivate the use of all previously discussed diagrams, and that for many, 3$\sigma$ line detections could still give useful insight. Data points with these uncertainties provide useful information about the ionization sources of the galaxies in question. This is because the error bars of the diagnostic quantities are sufficiently small with respect to the separation between SFG and AGN in the diagrams.

The actual UV SNRs for our three target sources are listed in Table \ref{tab:SNRs}. As shown, each galaxy has an SNR above 5 for several lines, making them well-suited to provide meaningful information.

\begin{table*}
\centering
\begin{tabular}{c|cc|cc|cc} \hline
\textbf{}              & \multicolumn{2}{c|}{\textbf{9422}} & \multicolumn{2}{c|}{\textbf{18846}} & \multicolumn{2}{c}{\textbf{10058975}} \\  \hline \hline
\textbf{Emission line} & \textbf{Line flux SNR}   & \textbf{Continuum SNR}  & \textbf{Line flux SNR}   & \textbf{Continuum SNR}   & \textbf{Line flux SNR}    & \textbf{Continuum SNR}    \\ \hline 
\textbf{He II + O III]} & 14.9 &  21.2 &  6.0  &  20.3 & 7.2 & 23.6 \\
\textbf{C IV}           & 12.8 &  24.3 & 6.7  & 23.2 &  6.0 &  24.0 \\
\textbf{C III]}         &  - &  29.4  &  5.3  &  40.9  &  17.1  &  19.8 \\ \hline
\end{tabular}
\caption{Signal-to-noise ratios of the ultraviolet line fluxes and their continuum level in our three sources of interest (NIRSpec IDs 9422, 18846, and 10058975). The line flux SNRs were calculated using the JWST/Prism line flux catalogue, while the continuum level SNRs were calculated by dividing our continuum level estimate with the standard deviation of the continuum sample points from our continuum fit. The \mbox{\ion{C}{III}]} line of galaxy 9422 was not reported in the JWST/Prism line flux catalogue.}
\label{tab:SNRs}
\end{table*}

\subsection{Required brightness vs. exposure time}\label{subsec:brightness_vs_exp_time}

For a correct assessment of our diagnostics in the spectroscopic analysis, it is important to have an evaluation of the number of hours needed for a correct detection of the emission lines. We calculated the minimum continuum brightness ($C_{\text{min}}$) a source must have for its emission line with a rest frame EW of $\mathrm{EW}_{\text{rest}}$ to be 5$\sigma$ detected by using the 10$\sigma$ line sensitivity (in units of $\mathrm{erg/s/cm^{2}}$) provided by \cite{Eisenstein_2023_JADESoverview}.

We used the 5$\sigma$ limits relying on the argument on the needed SNRs in Section \ref{subsec:SNR}. These values can be found in Table \ref{tab:F_5sigma_values} where $F_{5\sigma}(z)$ is the 5$\sigma$ detection limit obtained from \cite{Eisenstein_2023_JADESoverview} assuming a redshift z = 6. In order to estimate the source brightness needed for a 5$\sigma$ detection with a different exposure time observation, we assumed that the line sensitivity scales as $1/\sqrt{t}$. This is a reasonable assumption since for background limited observations, like these faint high redshift galaxies, the SNR scales as $\sqrt{t}$ and is proportional to the flux.

\begin{table}
\centering
    \begin{tabular}{lc}
    \textbf{Emission line} & \textbf{$F_{5\sigma}$ ($10^{-18} \mathrm{erg/s/cm^{2}}$)} \\ \hline \hline
    \textbf{\mbox{He II $\lambda$1640}} &  1.35  \\ 
    \textbf{\mbox{C IV}} & 1.55  \\ 
    \textbf{\mbox{C III]}} & 0.85 \\ 
    \textbf{\mbox{O III] $\lambda$1665}}  & 1.25 \\ 
    \textbf{\mbox{He II $\lambda$4686}}  &  0.15  \\ 
    \textbf{\mbox{H$\beta$}}   &  0.15 \\ 
    \textbf{\mbox{[O III] $\lambda$5007}}   &  0.14  \\ \hline              
    \end{tabular}
\caption{ 5$\sigma$ NIRSpec prism detection limits for various emission lines assuming a source at redshift \mbox{z = 6} and an exposure time of 28 hours. These values were calculated using the prism sensitivity line in \protect\cite{Eisenstein_2023_JADESoverview}, and they represent the minimum line fluxes needed for a 5$\sigma$ detection.}
\label{tab:F_5sigma_values}
\end{table}

\begin{figure}
 \includegraphics[width=.48\textwidth, height=.9\textheight, keepaspectratio]{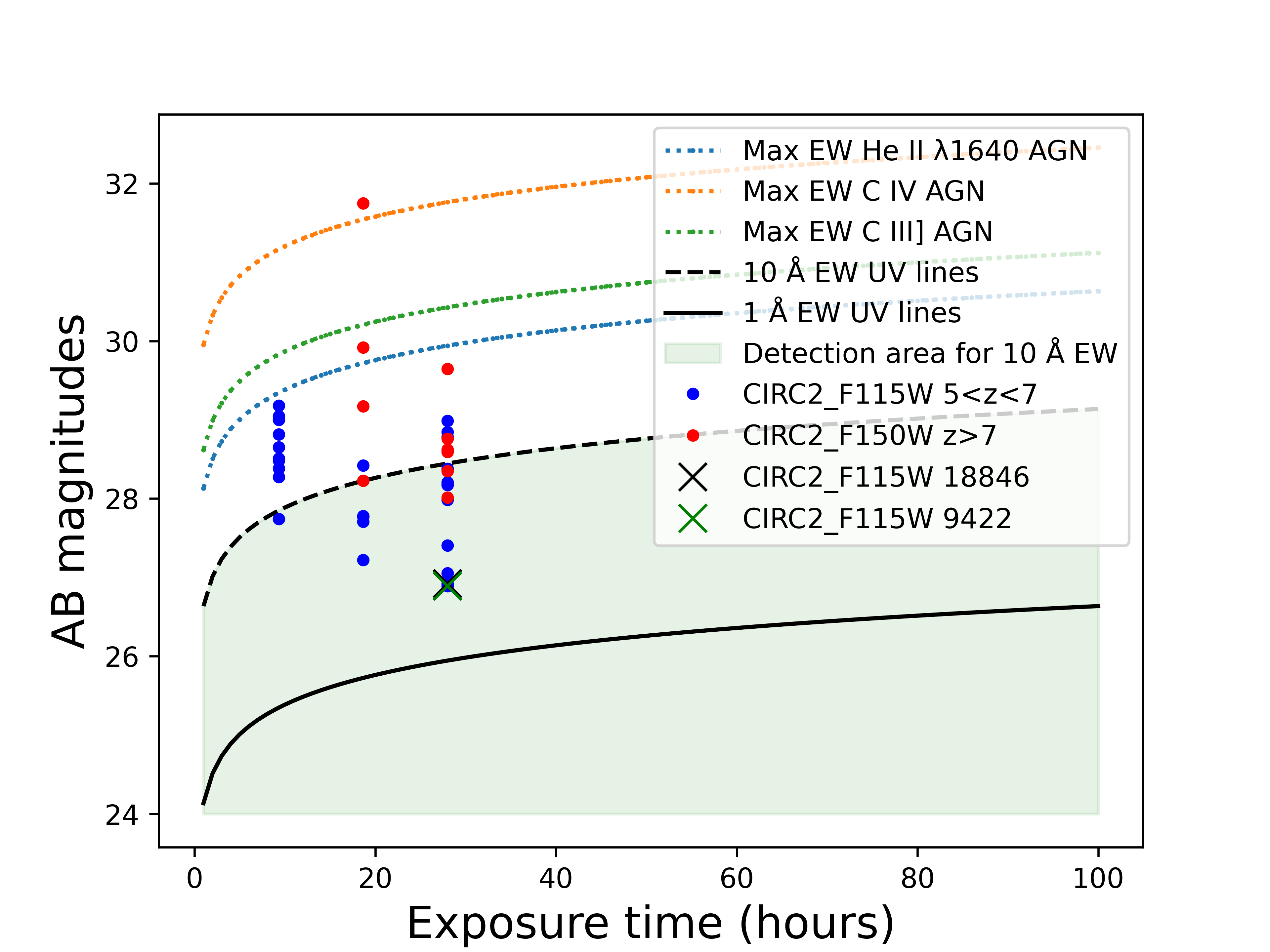}
    \caption{Diagrams displaying the exposure time dependence of the required brightness (in AB mag) a source at \mbox{z = 6} needs to have in order to achieve 5$\sigma$ line flux detections in the case of restframe EWs of 1 $\text{\AA}$, 10 $\text{\AA}$, or the maximum EW value of the AGN models. The 1 $\text{\AA}$ UV average curve is displayed as a black line. The 10 $\text{\AA}$ UV average curve is displayed as a black dashed line, and the dotted coloured lines correspond to the maximum EWs in the AGN models for some of the lines in question. These maximum EW values are around 228, 55, and 42 $\text{\AA}$ for \mbox{\ion{C}{IV}}, \mbox{\ion{C}{III}]}, and  \mbox{\ion{He}{II} $\lambda$1640}, respectively. The green-shaded region represents the 5$\sigma$ line flux detection area for UV lines with an EW of 10 $\text{\AA}$. Blue and red data points show the CIRC2 NIRCam measurements for the JADES sample, where the F115W filter was used for sources at \mbox{5 < z < 7}, and F150W for sources at \mbox{z > 7}. Two of our galaxies of interest, 9422 and 18846 (with NIRCam IDs 113585 and 131688, respectively), are located by overlapping green and black crosses, and are amongst the brightest sources.)}
\label{fig:required_ABmag_vs_time}
\end{figure}

The required brightness vs. exposure time diagram for the 5$\sigma$
detection can be seen on Figure \ref{fig:required_ABmag_vs_time}. The 1 $\text{\AA}$ UV average curve is displayed as a black line. The 10 $\text{\AA}$ UV average curve is displayed as a black dashed line. The coloured dotted lines correspond to the maximum EW values of the AGN models for some of the emission lines. These maximum EW values are around 228, 55, and 42 $\text{\AA}$ for \mbox{\ion{C}{IV}}, \mbox{\ion{C}{III}]}, and  \mbox{\ion{He}{II} $\lambda$1640}, respectively. The green-shaded region represents the 5$\sigma$ line flux detection area for UV lines with an EW of 10 $\text{\AA}$. We can see that with longer exposure times, the sources need to be less bright to achieve the same precision. 

We have considered exposure times from 1 hour (quite commonly adopted with NIRSpec) to 100 hours (the maximum that anyone will probably ever get). Increasing the time exposure in this time range can reduce the required brightness by two magnitudes. The more intense reduction in brightness would be reflected in the first 10 hours of exposition. Additionally, to resolve an EW of 1 $\text{\AA}$ requires orders of brightness that are 2.5 AB magnitudes lower than for EW of 10 $\text{\AA}$. 

On Figure \ref{fig:required_ABmag_vs_time}, we also display JADES sources using the NIRCam CIRC2 measurement associated with a 0.15 arcsec radius circular aperture. In order for the UV emission lines to fall in  the appropriate region of the observed spectrum, the F115W filter was used for sources at \mbox{5 < z < 7}, and F150W for sources at \mbox{z > 7}. These are represented with blue and red data points, respectively. Two of our sources of interest, 9422 and 18846 (NIRCam IDs 113585 and 131688, respectively), can be seen marked with overlapping green and black crosses. 9422 has a value of 26.89 AB magnitude and 18846 has a value of 26.92 AB magnitude. Unfortunately, photometry was not available for galaxy 10058975.  We can see that galaxies 9422 and 18846 are amongst the brightest sources in the data.

Note that many of the observational data points lie just below the maximum curve for the \mbox{\ion{He}{II} $\lambda$1640} line, meaning they would only be bright enough to exhibit this line to a sufficient SNR if they were extremely powerful AGN. For better observations of these extremely faint lines, such as \mbox{\ion{He}{II} $\lambda$1640} or \mbox{\ion{He}{II} $\lambda$4686}, longer exposure times are required on brighter sources. With the current data, primarily targeting faint galaxies, we cannot effectively distinguish between AGN and SFG using UV-diagnostics in most cases.
The need for the observation of these faint lines is supported by the analysis of \cite{p14_scholtz2023jades} who identified galaxies 9422 and 10058975 as AGN using \mbox{\ion{He}{II} $\lambda$1640} and  \mbox{\ion{He}{II} $\lambda$4686} diagnostics. 

We saw in Figure \ref{fig:required_ABmag_vs_time} that JADES sources are well above the 1 $\text{\AA}$ line. This shows that it would be unreasonable to expect detections of 1 $\text{\AA}$ rest frame EWs to a sufficient precision for these diagrams. This reduces the parameter space of the proposed UV EW diagrams significantly. On Figures \ref{fig:blended_EW_vs_ratio_plot}-\ref{fig:data_on_EW_plots}, we cannot expect to see observed data points with EW values less than  1 $\text{\AA}$, meaning no observations below a value 0 on the EW axes of these diagrams.

In some cases this means that the expected detection region of the diagram completely excludes the SFG models. This happens for diagrams using the EW of \mbox{\ion{He}{II} $\lambda$1640} or \mbox{\ion{He}{II} $\lambda$4686}, further showing why the detection of these two lines indicate the presence of an AGN.

\section{Conclusions}\label{sec:conclusions}

In this paper we explore how to identify active galactic nuclei (AGN) within galaxies with data from the JWST.  We show that existing optical diagrams such as the BPT diagram and the VO87 diagram pose limitations when analysing AGN at high redshifts. These limitations include significant regions where AGN and SFG overlap, and missing lines outside the JWST's detection range, rendering analysis impossible.  Therefore, new methods and ideas are needed to make progress within this very important topic. 

We thus test and use a variety of new potential diagnostics, that can be used to distinguish active galactic nuclei and star-forming galaxies in the early universe. We use the photoionization models created by \cite{p11_NakajimaMaiolino2022} and the JWST/NIRSpec prism data (\cite{p1_bunker2023jades}). Amongst the diagrams are some that only use ultraviolet emission line ratios, others use equivalent widths and line ratios, and some solely use equivalent widths of both optical and ultraviolet lines. We examine where the galaxies with JADES NIRSpec IDs 9422, 18846, and 10058975 fall on our new equivalent width plots, to determine their nature. We then evaluate the sensitivity and applicability of these possible diagnostics with the current observational strategy and also extend our analysis following a photometric selection method for which we use the JWST NIRCam photometry catalogue (\cite{NIRCam_Rieke_2023}).

Our conclusions about the explored spectroscopic diagnostics are the following:
\begin{enumerate}
    \item The standard diagnostic diagrams (BPT and VO87) are inefficient for the identification of AGN in the Early Universe, therefore new diagnostic methods need to be considered.
    \item The maximum starburst line on the N2-BPT and S2-VO87 diagrams agree well with the presented SFG and AGN photoionization models, however the line thought to mark the lower limit of the mixed region does not.
    \item The AGN/SFG demarcation line defined by \cite{Backhaus_2022} on the OHNO diagram  is inconsistent with the presented photoionization models. However, with the introduction of newer separation curves, including both our own and the one from \cite{Feuillet_2024}, the OHNO diagram shows great potential for effectively selecting AGN candidates. We list a few potential candidates from the JADES prism and grating catalogues. 
    \item The ratios (\ion{C}{III}] + \ion{C}{IV}) / \ion{He}{II} $\lambda$1640, \mbox{\ion{O}{III}] $\lambda\lambda$1665/\ion{He}{II} $\lambda$1640}, and \mbox{\ion{C}{III}]/\ion{He}{II} $\lambda$1640} ratios, as well as the equivalent widths of the \mbox{\ion{He}{II} $\lambda$1640} and \mbox{\ion{He}{II} $\lambda$4686} recombination lines, are the most efficient quantities in separating AGN and SF galaxies based on the models. Combining these quantities can create powerful diagnostics. The ratio \mbox{\ion{C}{IV} / \ion{C}{III}}  can also be useful when combined with one of the previous quantities.
    \item Diagrams using the \mbox{\ion{He}{II} $\lambda$1640 + \ion{O}{III}]} lines can provide some information about the nature of the galaxy. However, for robust diagnostics, the \mbox{\ion{He}{II} $\lambda$1640} and \mbox{\ion{O}{III}] $\lambda\lambda$1665} lines are better if observed separately.
    \item The optical EW diagram using the lines \mbox{\ion{He}{II} $\lambda$4686} vs. either H$\alpha$ or H$\beta$ provides an exceptionally good separation between AGN and SF galaxies. However, the \mbox{\ion{He}{II} $\lambda$4686} line is extremely hard to detect and the Balmer lines might not be reliable diagnostic quantities at high redshifts due to the limited ability of JWST to probe all spectral wavelengths and resolutions. 
    \item The plots that overlay the equivalent widths of galaxies 9422, 18846, and 10058975 on the modelled datasets yield  ambiguous results. Further analysis with more ultraviolet diagrams (perhaps using \mbox{\ion{He}{II} $\lambda$1640} and \mbox{\ion{O}{III}] $\lambda$1665} as separate lines) are needed to obtain conclusive results. 
    \item In many diagrams, the only star forming galaxies that are located in the AGN region have metallicities of 0.0007 Z$_{\odot}$ and 0.007 Z$_{\odot}$. These values are extremely low and would be rarely detected, thus the diagrams presented could still be very effective in separating AGN and SF galaxies.
    \item The future observational strategy should adapt to the need for longer exposure times in order to make the new diagnostics effective. With the current data, primarily targeting faint galaxies, we cannot effectively distinguish between AGN and SFG using UV-diagnostics in most cases. 
    \item 5$\sigma$ line flux observations are sufficient for the effective use of the proposed diagrams (in Sections \ref{subsec:UV_alternatives}, \ref{subsec:EW_plots}, and \ref{subsec:EW_of_three_galaxies}), and in many cases, observations with a precision of 3$\sigma$ can still provide useful insight. 
    \item Some UV-Optical line ratios diagnostics, such as the \mbox{\ion{He}{II} $\lambda$1640/H$\beta$} vs. \mbox{(\ion{C}{III}] + \ion{C}{IV}) / H$\beta$} diagram, may be useful in the case of a low dust attenuation. When working with line flux ratios, it is important to be careful of dust attenuation since it affects light differently depending on the wavelength.  The \mbox{\ion{He}{II} $\lambda$1640/H$\beta$} ratio shows a very clear separation between the AGN and SFG photoionization models, while high values of the \mbox{(\ion{C}{III}] + \ion{C}{IV}) / H$\beta$} ratio also create a region solely occupied by AGN models. 9422, 18846 and 10058975 are identified as AGN candidates using this method.  
\end{enumerate}

Furthermore, our conclusions about the photometric selection method are:
\begin{enumerate}
    \item Galaxy 18846 is a strong AGN candidate supported by the photometric selection method relying on the work of \cite{p6_Juodzbalis2023}, as well as its strong UV spectra resembling some already identified AGN. 
    \item We present a sample of similarly strong AGN candidates in the JADES data supported by the same method in Table \ref{tab:f_AGN_JADES}.   
    \item Photometric preselection methods, such as the one used in this project, can become useful tools for selecting a few strong candidates for spectroscopic followup from large samples of galaxies. However, it should  be noted that they can be biased and should therefore be used together with other selection methods. 
    \item Some templates are more biased towards a specific type of AGN which can lead to low values of the indicator of the fraction of AGN and an incorrect identification of AGN. That is why it is important to consider in the analysis an extensive and accurate database of photoionization models that would represent the most complete description of AGN and SFG. 
\end{enumerate}

We focused on galaxies 9422, 18846 and 10058975 due to their strong UV lines, which are potentially indicative of AGN activity. However, the result we find are inconclusive for these particular systems, showing the difficulty in knowing if a system has an AGN or not. In the UV line ratio plots, galaxies 18846 and 10058975 fell into the mixed region, while galaxy 9422 appeared in the AGN region. On the other hand, all galaxies were placed in either the SFG or mixed region in the EW plots. Although the potential underestimation of EW values in the models suggests that line ratios diagnostics are more reliable. Additionally, we identified an intriguing UV-optical plot using \mbox{\ion{He}{II} $\lambda$1640/H$\beta$} vs. \mbox{(\ion{C}{III}] + \ion{C}{IV}) / H$\beta$} placing the galaxies 9422, 18846 and 10058975 in the AGN region. Any dust correction would further push them into the AGN region of the diagram.

Overall, our results provide a way forward to find AGN in distant galaxies. It is clear that even with detailed spectroscopy it will not always be easy to find these systems, although they are critical for understanding how galaxy evolution and supermassive black hole evolution have occurred in the distant past. Our results also show how AGN can be found through different new approaches using the full wavelength span available to JWST.  These approaches can be used in new surveys with spectroscopy.

\section*{Acknowledgements} 

We acknowledge support from the ERC Advanced Investigator Grant EPOCHS (788113), as well as two studentships from the STFC. JT also acknowledges support from the Simons Foundation and JWST program 3215. Support for program 3215 was provided by NASA through a grant from the Space Telescope Science Institute, which is operated by the Association of Universities for Research in Astronomy, Inc., under NASA contract NAS 5-03127. IJ acknowledges support by the Huo Family Foundation through a P.C. Ho PhD Studentship, by the Science and Technology Facilities Council
(STFC), by the ERC through Advanced Grant 695671 ‘QUENCH’,
and by the UKRI Frontier Research grant RISEandFALL. KN acknowledges support from
JSPS KAKENHI Grant: JP20K22373 and JP24K07102. This work is based on observations made with the NASA/ESA \textit{Hubble Space Telescope} (HST) and NASA/ESA/CSA \textit{James Webb Space Telescope} (JWST) obtained from the \texttt{Mikulski Archive for Space Telescopes} (\texttt{MAST}) at the \textit{Space Telescope Science Institute} (STScI), which is operated by the Association of Universities for Research in Astronomy, Inc., under NASA contract NAS 5-03127 for JWST, and NAS 5–26555 for HST. The authors thank all involved with the construction and operation of JWST, without whom this work would not be possible.

This work is based on the photoionization models created by \cite{p11_NakajimaMaiolino2022} and uses observations made by the James Webb Space Telescope Near Infrared Spectrograph (NIRSpec, \cite{NIRSpec_Jakobsen_2022}, \cite{NIRSpec_Ferruit_2022}, \cite{Eisenstein_2023_JADESoverview}) as well as the Near Infrared Camera (NIRCam, \cite{NIRCam_Rieke_2023}, \cite{Eisenstein_2023_JADESoverview}). We thank Professor Roberto Maiolino for his valuable feedback, which contributed to the improvement of this work.

Furthermore, the following Python libraries were used: \texttt{ASTROPY} (\cite{astropy:2022}); 
\texttt{SPECUTILS} (\cite{specutils2023}); \texttt{ MATPLOTLIB} (\cite{matplotlib2007}); \texttt{SCIKIT-LEARN} (\cite{scikit-learn}); \texttt{NUMPY} (\cite{numpy2020}); \texttt{PANDAS} (\cite{pandas2020}); \texttt{LIME} (\cite{Fern_ndez_2024}).




\bibliographystyle{mnras}
\bibliography{paper.bib} 





\appendix

\section{Additional diagrams}

The following diagrams provide additional support for the analysis discussed in the main text by presenting an extended set of diagnostic plots. These include further diagrams of EW and line ratios for both UV and optical lines, as well as data for the three galaxies of interest.

We do not include these plots in the main paper as we did not find that these plots and relationships between these quantities are superior to the plots we consider in the main paper.  We include these here for completeness and in case these plots may be useful in the future for those who are interested in particular lines and combinations of lines for distinguishing AGN from star forming galaxies.

\begin{figure*}
    \includegraphics[width=.9\textwidth, height=.9\textheight, keepaspectratio]{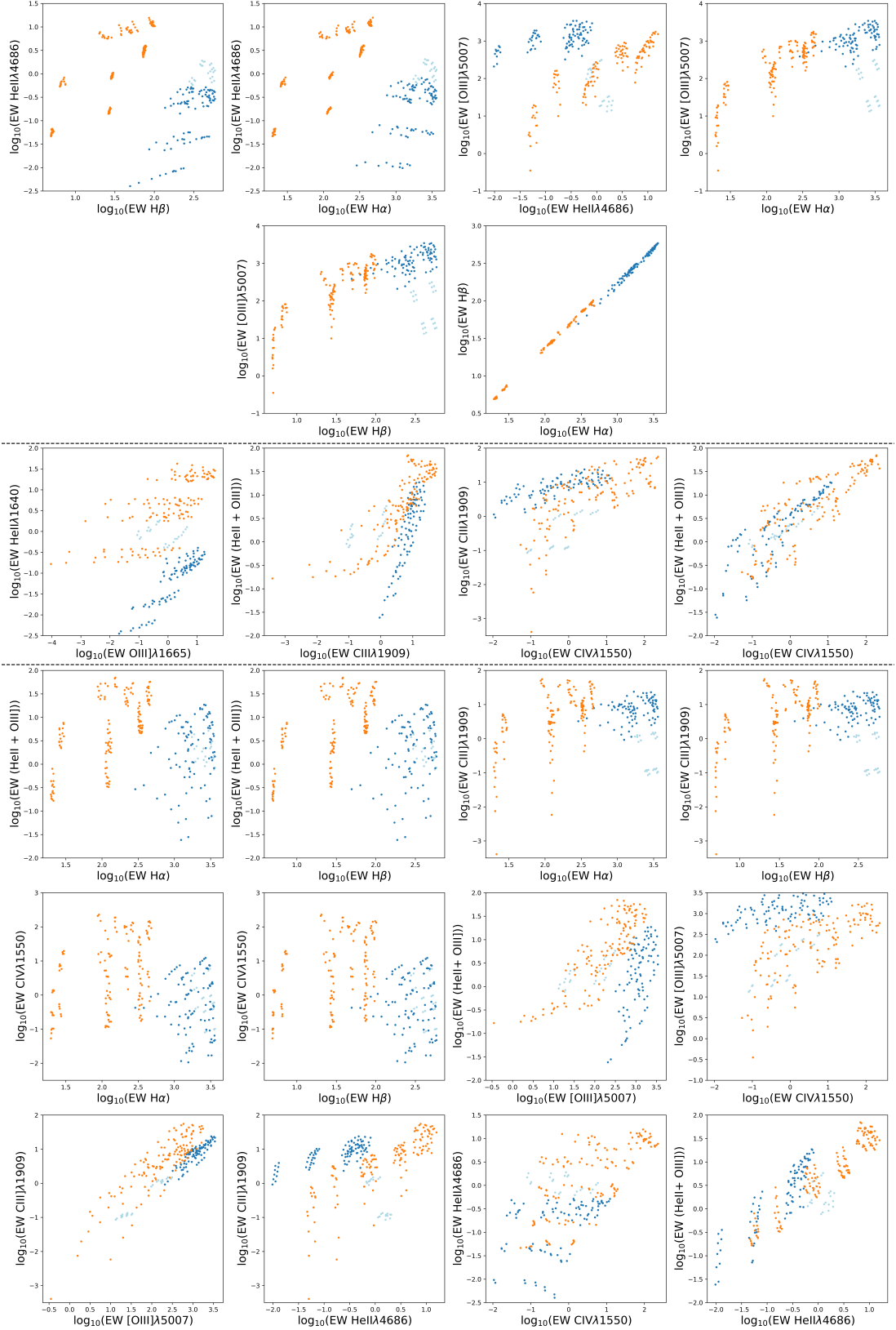}
    \caption{Plots displaying EWs calculated from the modelled data on both axes are separated into three groups: ones only using optical lines (top), ones only using UV lines (middle), and the ones using one of each (bottom). The colour coding is the same as in previous figures. Note that the first plot in the first row (upper left) also appears in Figure \ref{fig:EWplot_best} and it is included here for display purposes.}
\label{fig:appendix_EWplots_all_metallicities}
\end{figure*}

\begin{figure*}
    \begin{subfigure}{.5\textwidth}
        \centering
        \includegraphics[width=.9\linewidth]{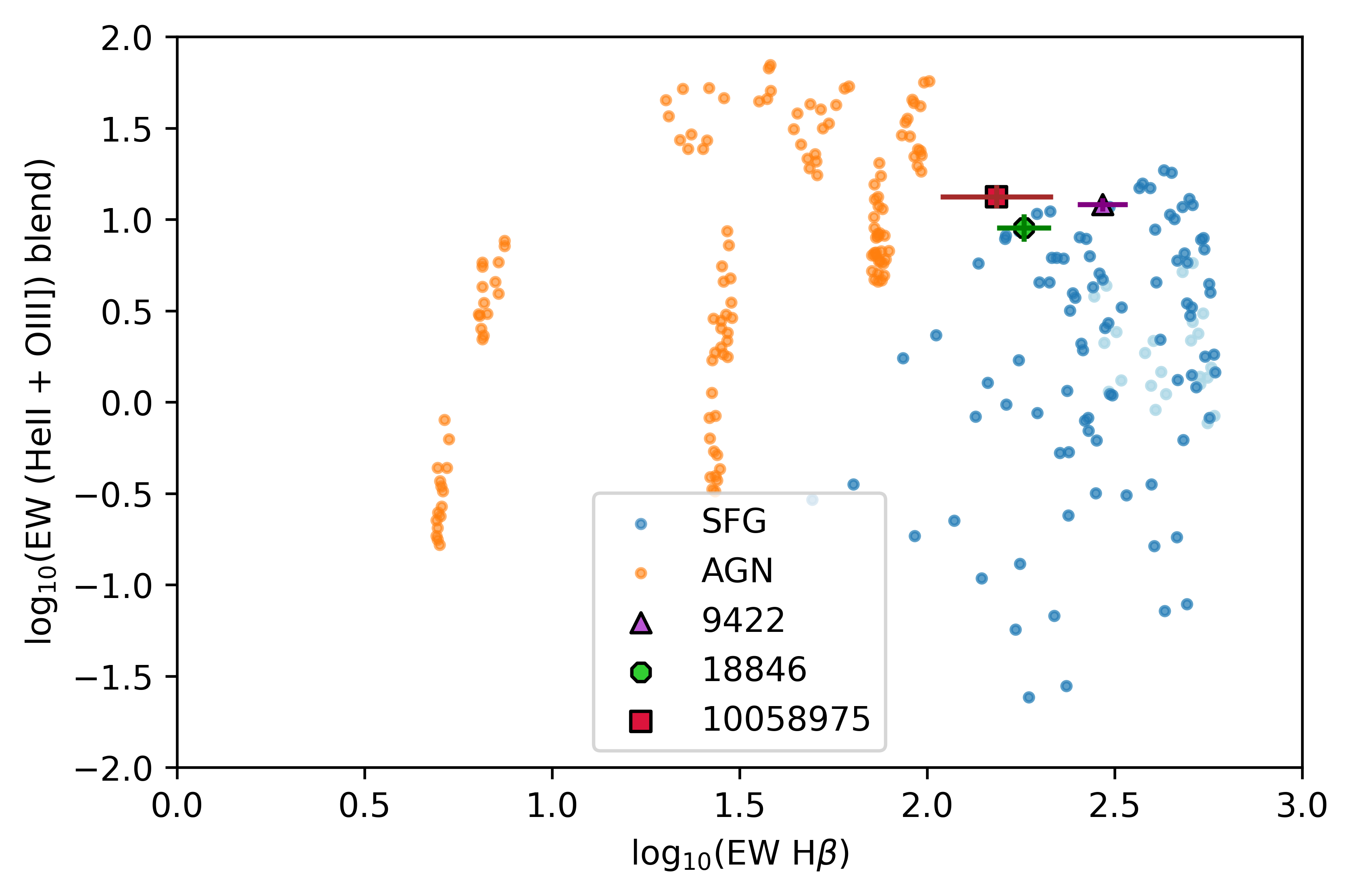}
    \end{subfigure}%
    \begin{subfigure}{.5\textwidth}
        \centering
        \includegraphics[width=.9\linewidth]{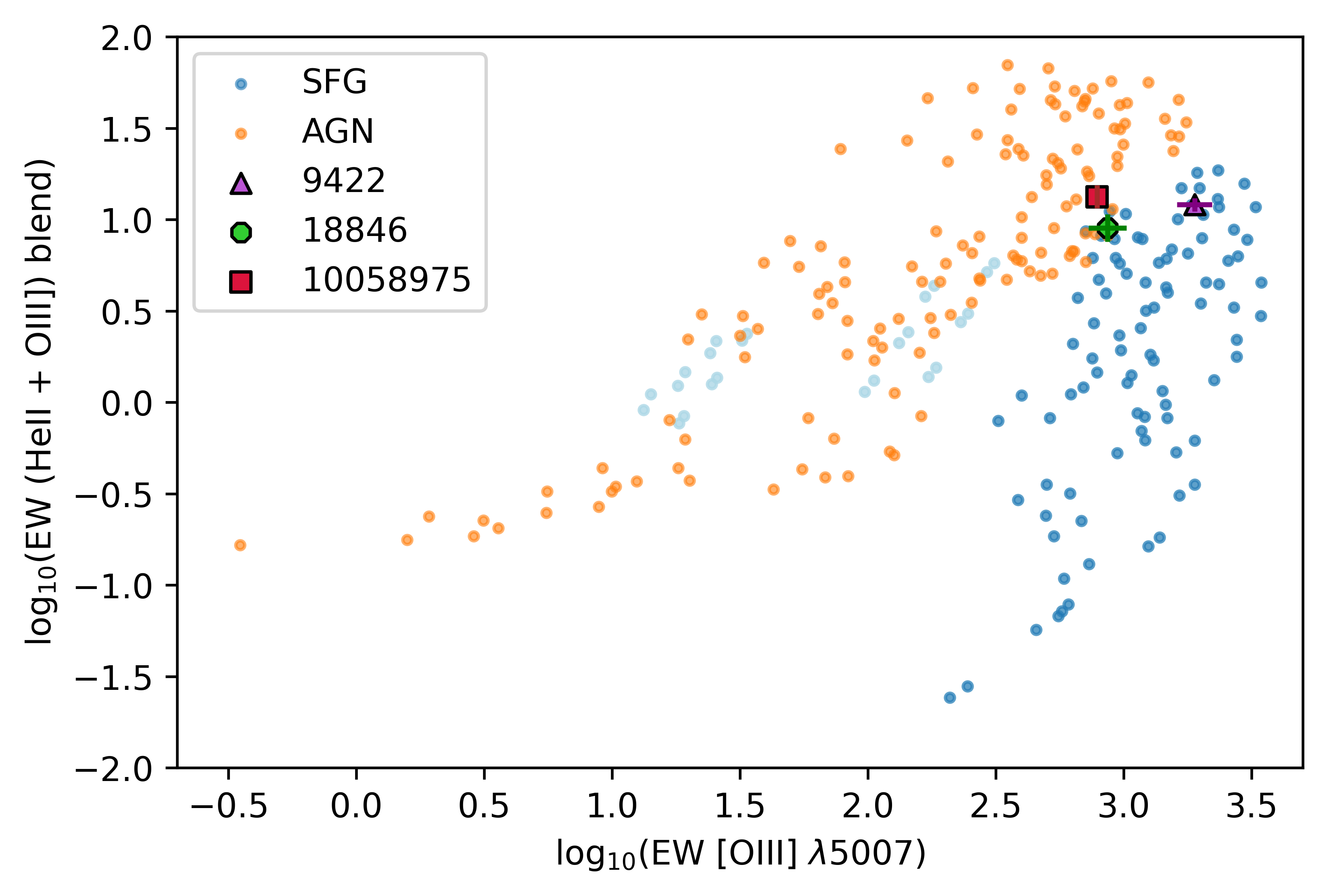}
    \end{subfigure}\\
    \begin{subfigure}{.5\textwidth}
        \centering
        \includegraphics[width=.9\linewidth]{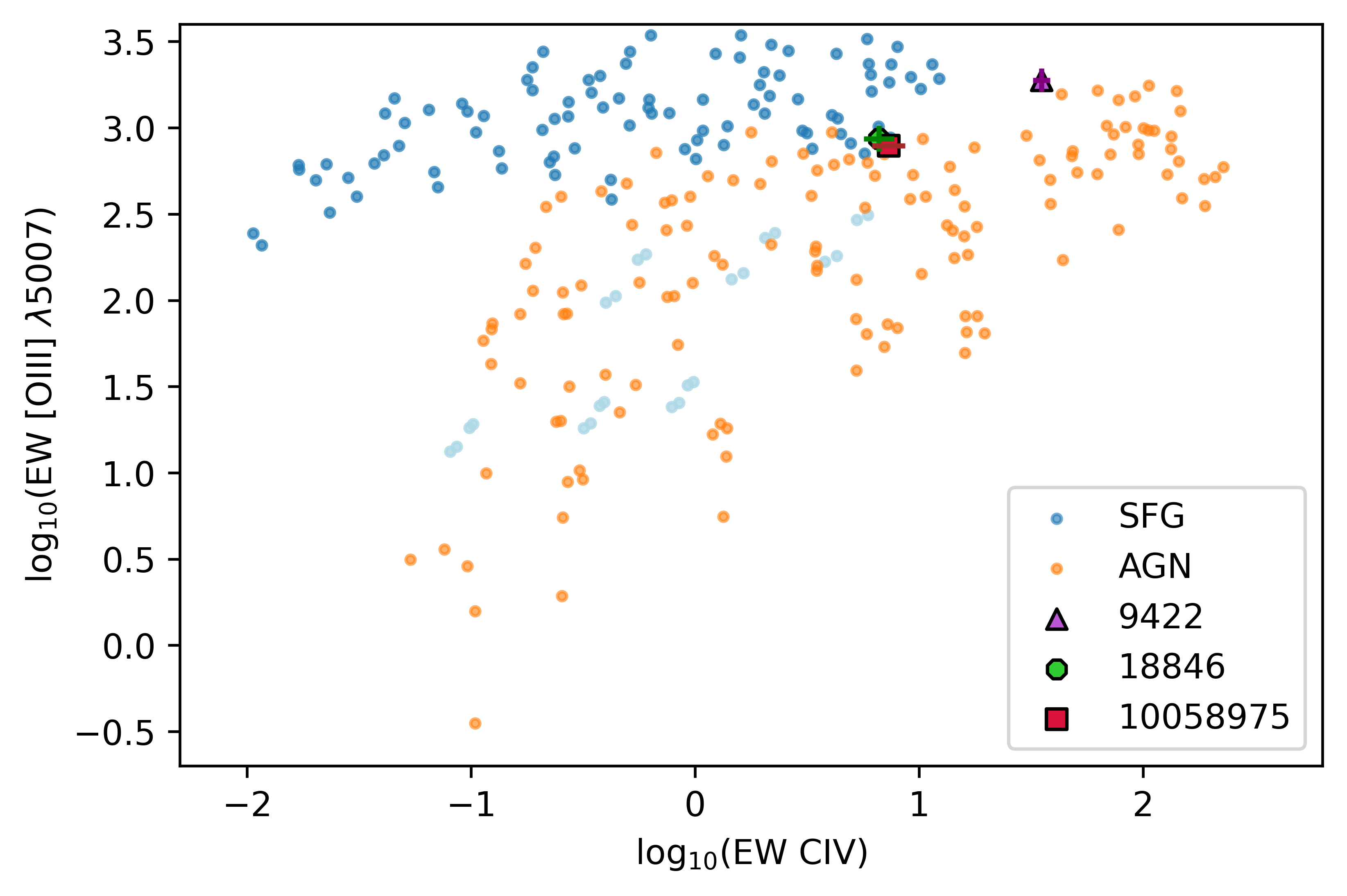}
    \end{subfigure}\\
    \caption{The galaxies with NIRSpec IDs 9422, 18846, and 10058975 are displayed on some of the EW vs. EW plots from Figure \ref{fig:appendix_EWplots_all_metallicities} using their rest frame EW measurements and uncertainties. The plots use one UV and one optical line. This figure is an extension of Figure \ref{fig:data_on_EW_plots}.}
    \label{fig:appendix_data_on_EW_plots}
\end{figure*}
 
\begin{figure*}
    \begin{subfigure}{.33\textwidth}
        \centering
        \includegraphics[width=.8\linewidth]{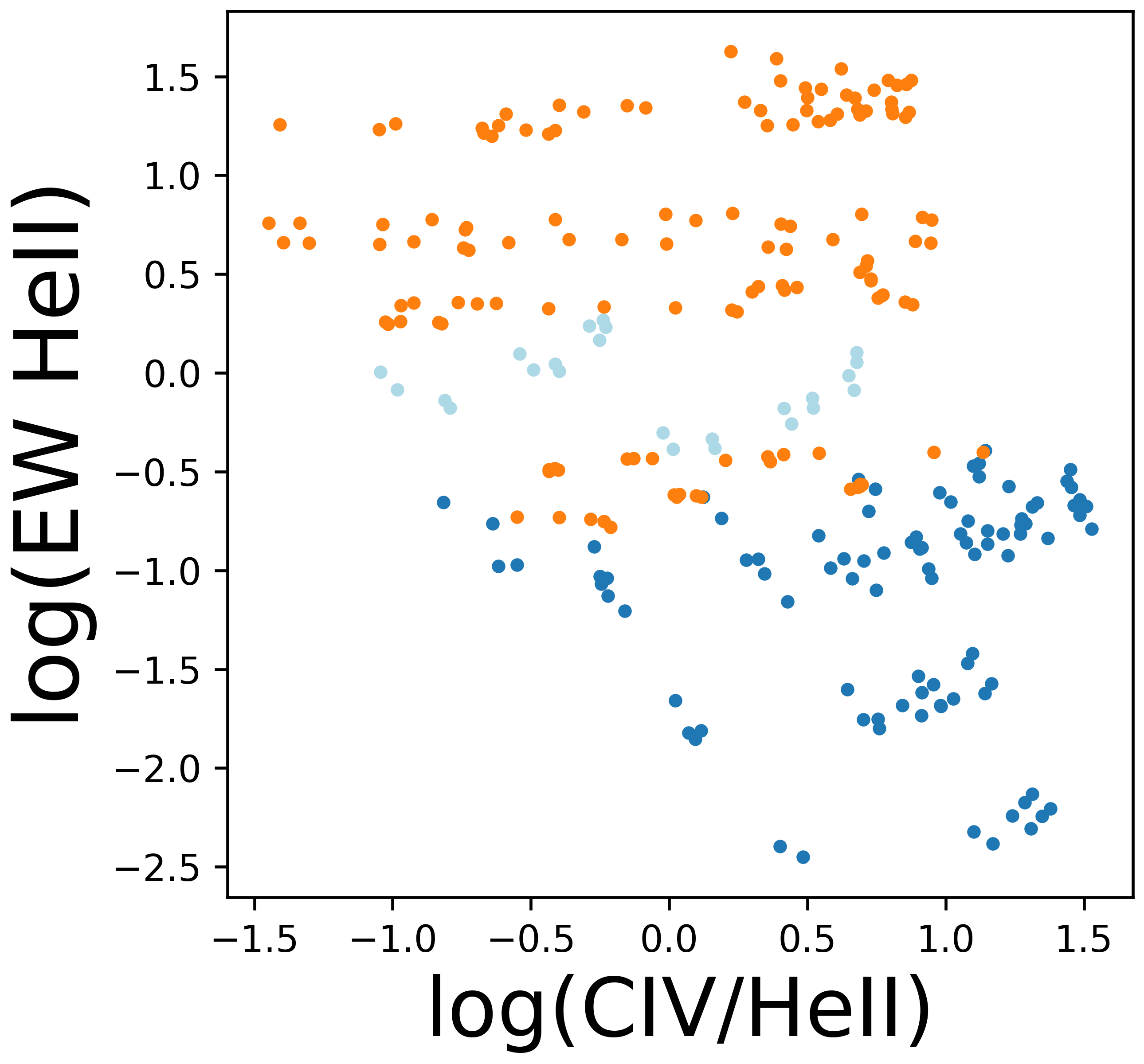}
        \label{fig:final_a_ap}
    \end{subfigure}%
    \begin{subfigure}{.33\textwidth}
        \centering
        \includegraphics[width=.8\linewidth]{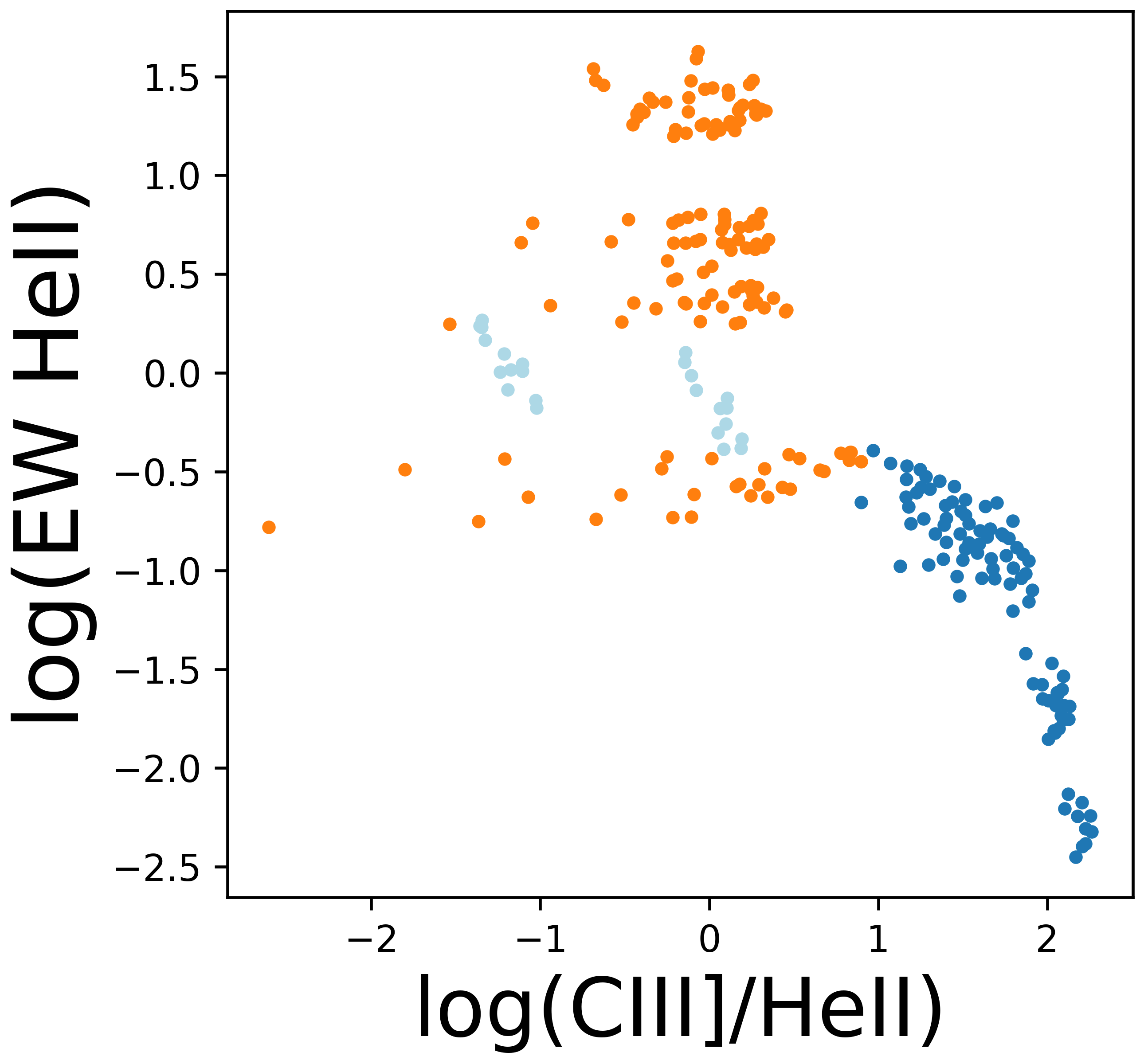}
        \label{fig:final_b_ap}
    \end{subfigure}%
    \begin{subfigure}{.33\textwidth}
        \centering
        \includegraphics[width=.8\linewidth]{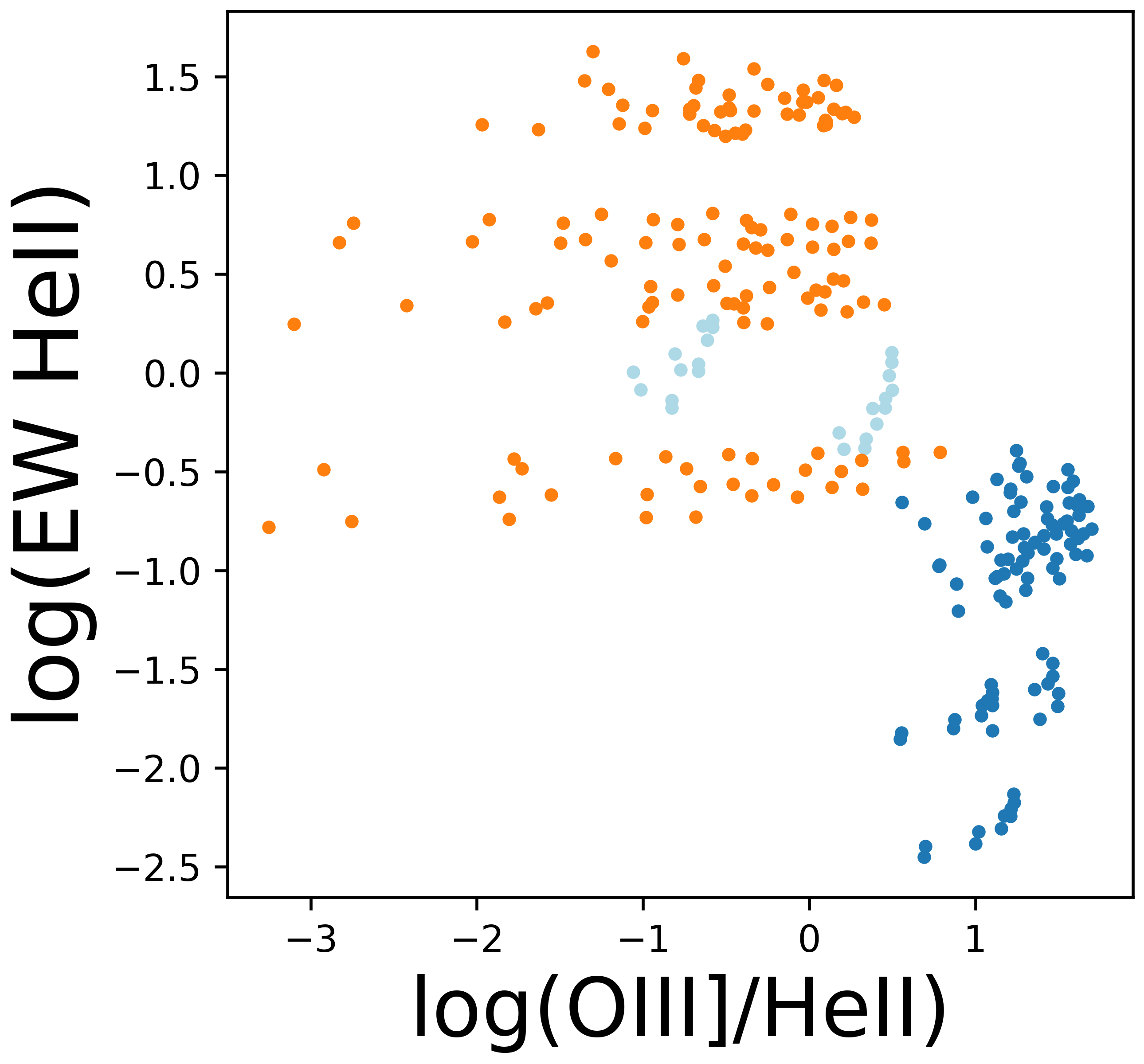}
        \label{fig:final_c_ap}
    \end{subfigure}\\
    \begin{subfigure}{.33\textwidth}
        \centering
        \includegraphics[width=.8\linewidth]{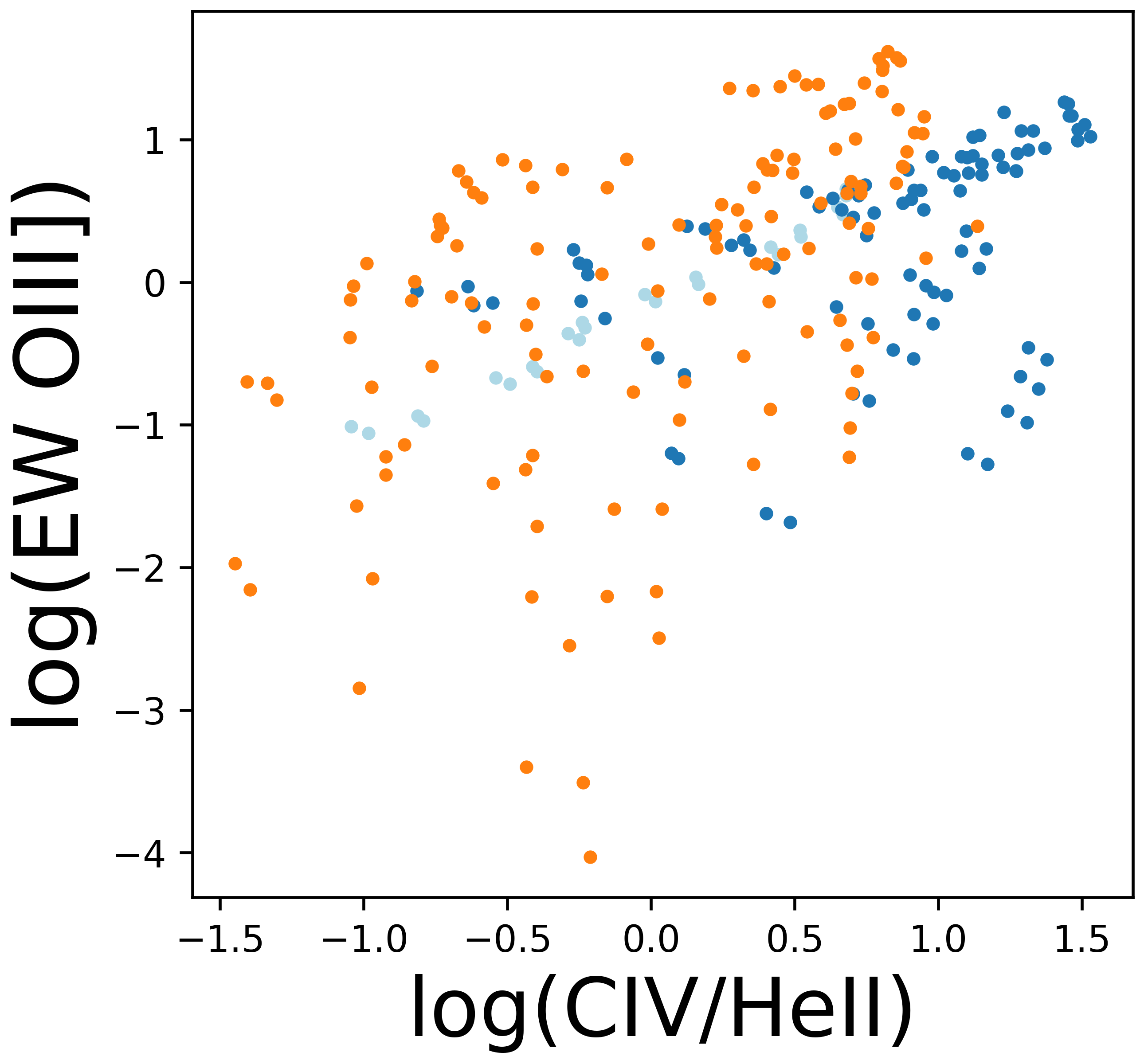}
        \label{fig:final_d}
    \end{subfigure}%
    \begin{subfigure}{.33\textwidth}
        \centering
        \includegraphics[width=.8\linewidth]{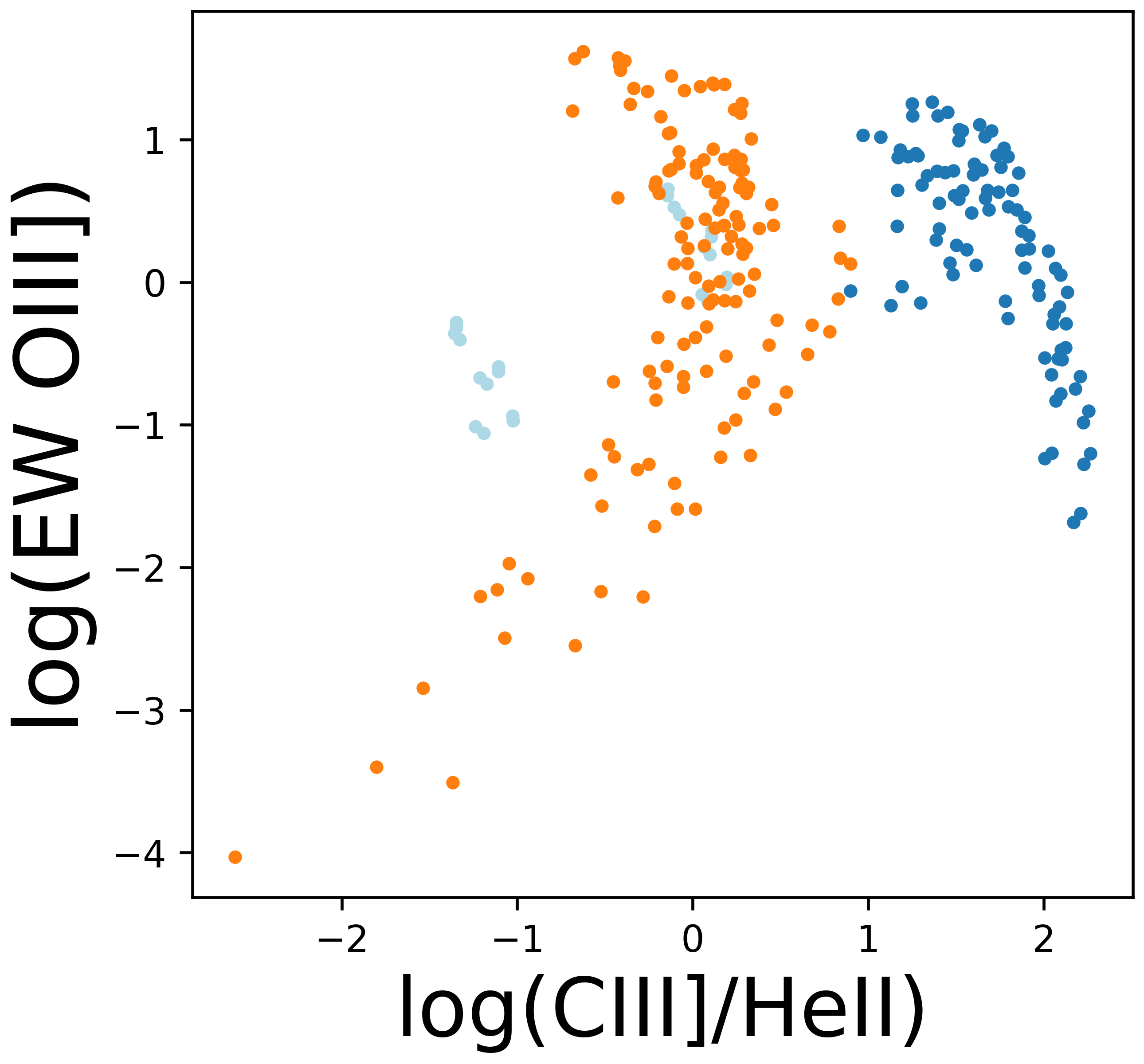}
        \label{fig:final_e}
    \end{subfigure}%
    \begin{subfigure}{.33\textwidth}
        \centering
        \includegraphics[width=.8\linewidth]{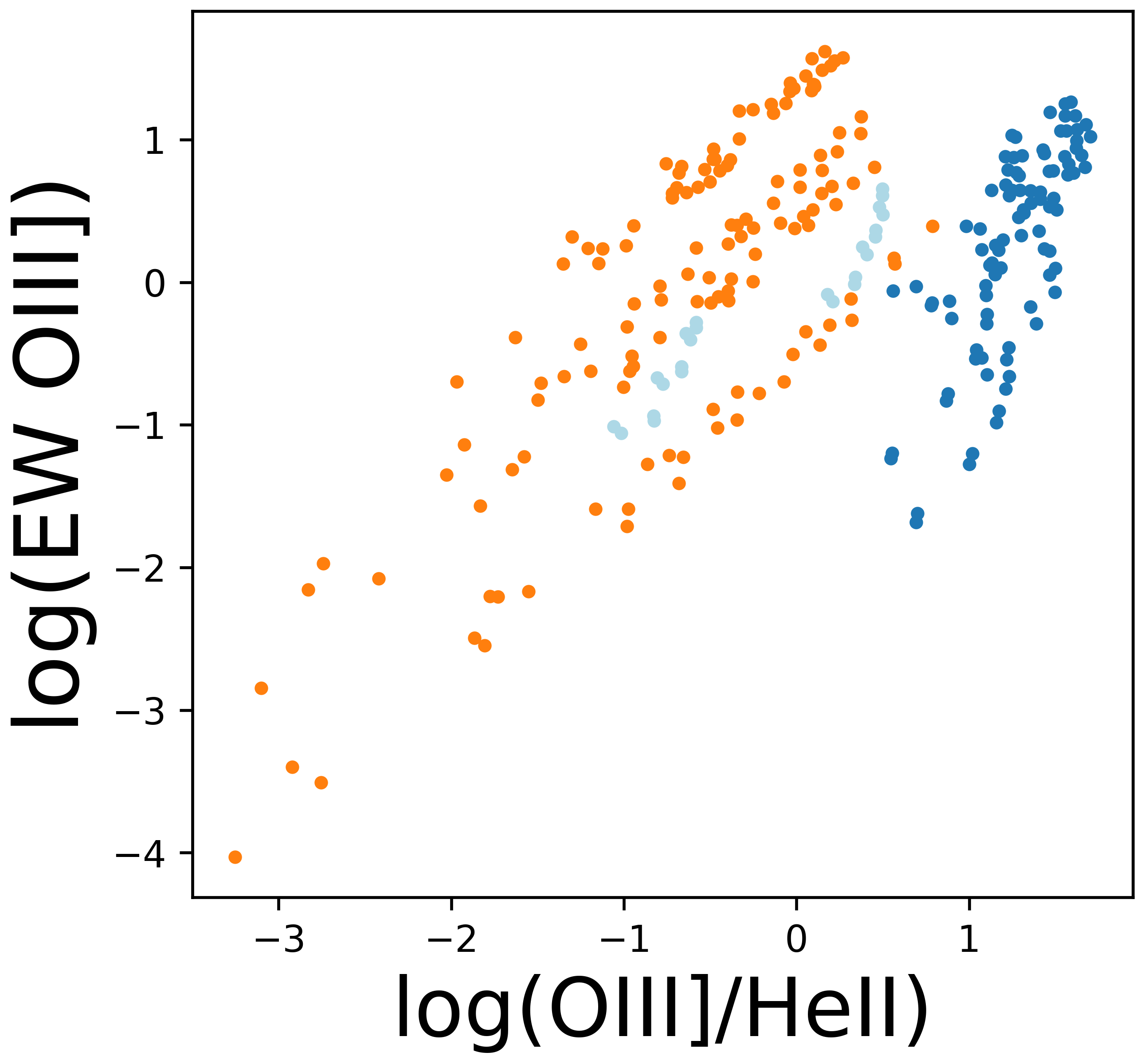}
        \label{fig:final_f}
    \end{subfigure}\\
    \begin{subfigure}{.33\textwidth}
        \centering
        \includegraphics[width=.8\linewidth]{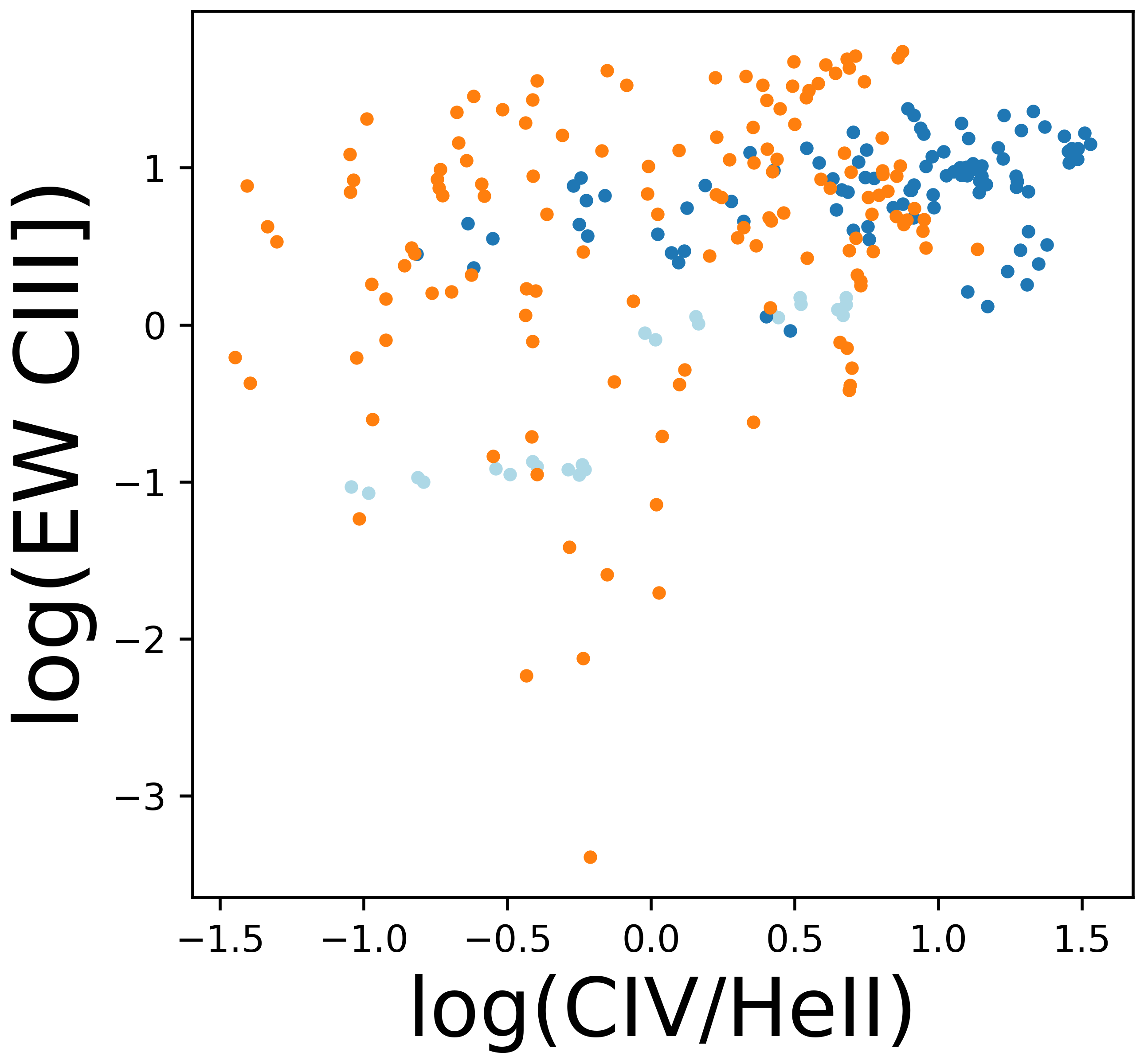}
        \label{fig:final_g}
    \end{subfigure}%
    \begin{subfigure}{.33\textwidth}
        \centering
        \includegraphics[width=.8\linewidth]{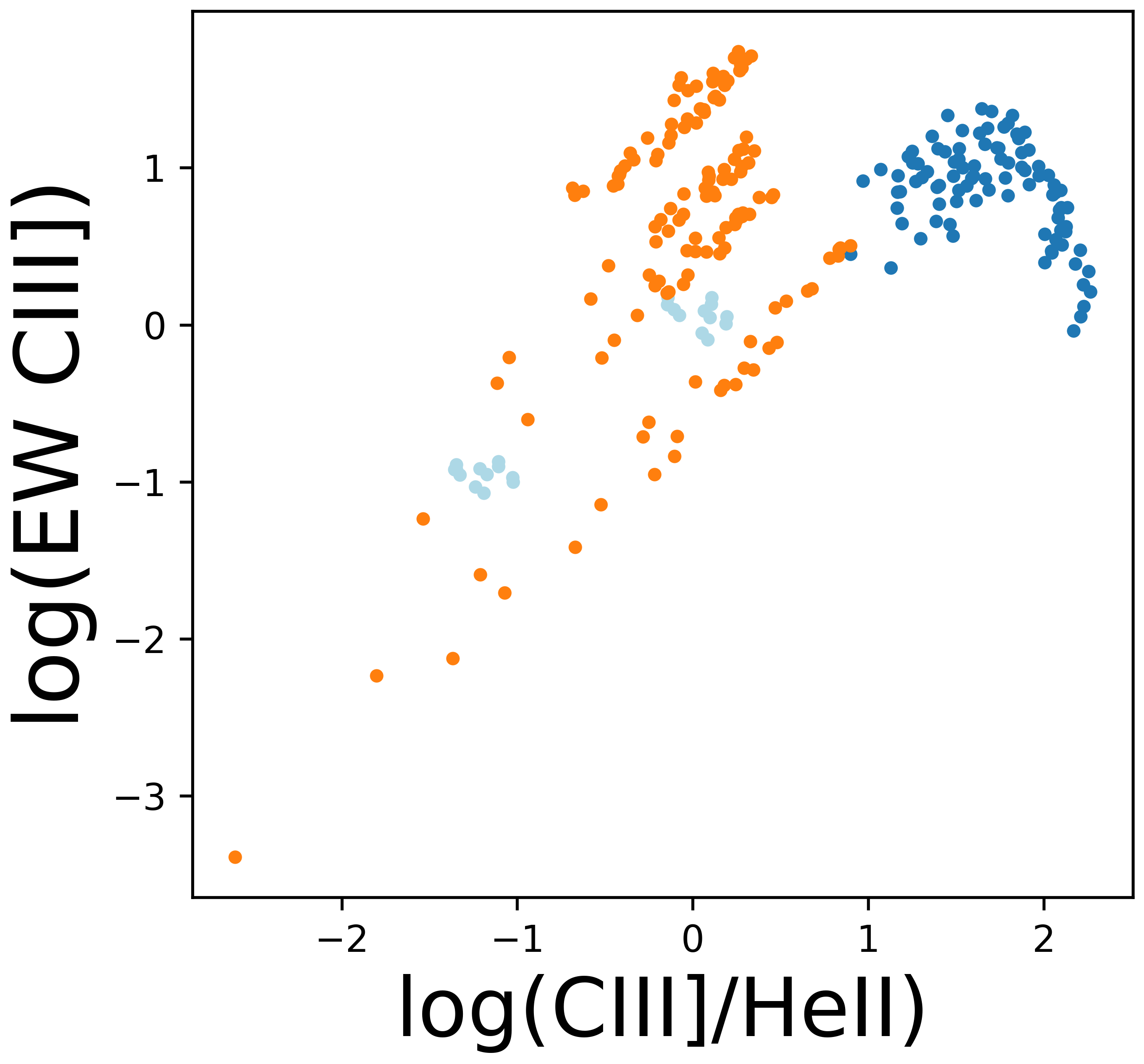}
        \label{fig:final_h}
    \end{subfigure}%
    \begin{subfigure}{.33\textwidth}
        \centering
        \includegraphics[width=.8\linewidth]{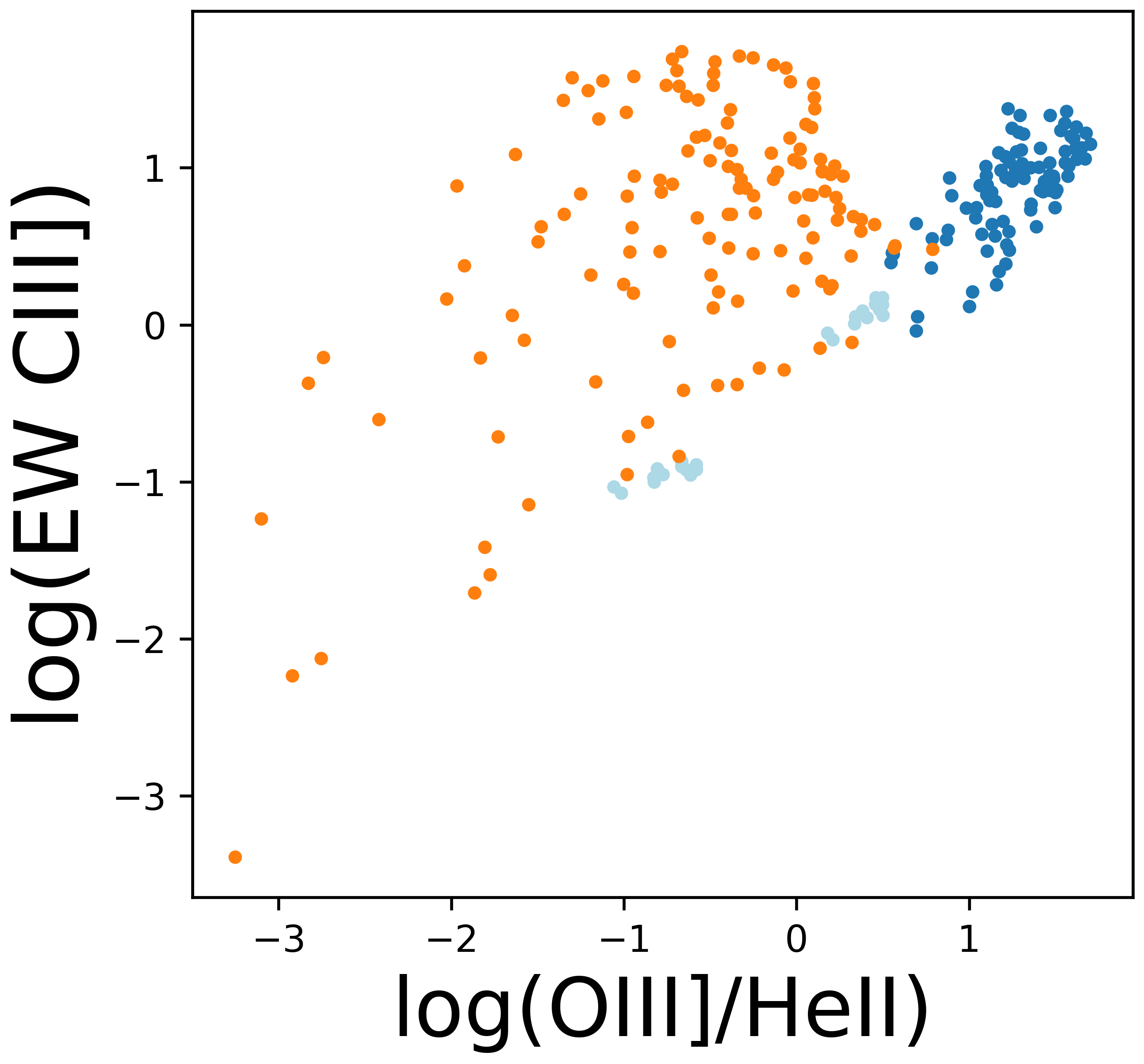}
        \label{fig:final_i}
    \end{subfigure}
    \caption{The (base 10) logarithmic EWs of the ultraviolet emission lines \mbox{\ion{He}{II} $\lambda$1640}, \mbox{\ion{O}{III}] $\lambda$1665}, and \ion{C}{III}] plotted against the line ratios \mbox{\ion{C}{IV}/\ion{He}{II} $\lambda$1640}, \mbox{\ion{C}{III}]/\ion{He}{II} $\lambda$1640}, and \mbox{\ion{O}{III}] $\lambda$1665/\ion{He}{II} $\lambda$1640}. The colour coding is the same as in previous figures. In terms of equivalent widths, \mbox{\ion{He}{II} $\lambda$1640} shows the most effective separation between AGN and SFG. Note that the middle plot in the upper row also appears in Figure \ref{fig:blended_EW_vs_ratio_plot} and it is included here for display purposes.}
    \label{fig:EW_vs_ratio_plots}
\end{figure*}

\bsp	
\label{lastpage}
\end{document}